\documentclass[11pt,twoside,english,listtotoc, tablecaptionabove, bibtotoc, parskip=full, headings=small, numbers=noenddot,notitlepage]{report}

\usepackage[T1]{fontenc}
\usepackage[latin9]{inputenc}
\usepackage[a4paper]{geometry}
\usepackage{fancyhdr}
\usepackage{float}
\usepackage{amsmath}
\usepackage{amssymb}
\usepackage{graphicx}

\makeatletter
\title{\textbf{Mass generation and deconfinement in pure\\Yang-Mills theory: a variational study}\\ \ }
\author{Giorgio Comitini\thanks{E-mail: giorgio.comitini@studium.unict.it}\\ \\ \textit{Dipartimento di Fisica e Astronomia dell'Universit\`{a} di Catania,}\\\textit{Via S. Sofia 64, I-95123 Catania, Italy}\\ \\}
\usepackage{fancyhdr}
\renewcommand{\footnoterule}{\kern+5\p@ \hrule \@width 5.9in \kern2.6\p@}

\@ifundefined{showcaptionsetup}{}{%
 \PassOptionsToPackage{caption=false}{subfig}}
\usepackage{subcaption}
\makeatother

\usepackage{babel}

\usepackage{braket}
\usepackage{mathrsfs}

\newcommand{\D}{\mathcal{D}}
\newcommand{\F}{\mathcal{F}}
\newcommand{\Z}{\mathcal{Z}}
\newcommand{\BS}{\mathcal{S}}
\newcommand{\cbar}{\overline{c}}
\newcounter{count}
\numberwithin{equation}{count}

\begin{document}

\pagenumbering{gobble}
\maketitle
\
\begin{abstract}
A simple variational argument based on the Gaussian Effective Potential (GEP) is put forward to give evidence for mass generation and deconfinement in pure Yang-Mills SU(N) theory. The GEP analysis shows that the massless gaussian vacuum of Yang-Mills theory is perturbatively unstable towards a massive gaussian vacuum, indicating that a massive expansion is the natural choice for computations in YMT. At finite temperature, the GEP provides an optimal temperature-dependent mass parameter for the expansion and signals the occurrence of a weakly first-order phase transition at $T_{c}\approx 255$ MeV for $N=3$. The equation of state is found to be in good agreement with the lattice data. This work is complemented with review material on the standard and massive expansions of Yang-Mills theory and on the formalism of quantum field theory at finite temperature. Comparisons are made with lattice results and numerical tables are provided to support our findings.
\end{abstract}
\clearpage

\thispagestyle{empty}
\
\\
\\
\\
\\
\\
\\
\\
\begin{flushright}\textit{
To the memory of Professor Gaetano Giaquinta
}\end{flushright}

\clearpage
\thispagestyle{empty}
\
\clearpage

\pagenumbering{Roman}

\chapter*{Preface to the ArXiv version\index{Preface to the ArXiv version}}
This work has been presented on October 5, 2017 as the final thesis for the degree in Physics at the University of Catania, under the supervision of Prof. Fabio Siringo. Its main results, namely evidence of variational nature for mass generation and deconfinement in pure Yang-Mills SU(N) theory, were obtained by expanding the action of the theory around a free (kinetic) term which treats the transverse gluon modes, as well as the longitudinal gluon and ghost modes, as massive excitations of the fields, while leaving the original Lagrangian unchanged. These results were first presented in ref.\cite{comitini}. Recently, a second solution to the variational problem has been proposed \cite{comitini2} with the aim of making connection with the method employed in ref.\cite{siringo1} to obtain the gluon and ghost two-point functions through a massive perturbative expansion. Said solution has been obtained by expanding the theory around a massless, rather than massive, longitudinal gluon and ghost vacuum. Both these approaches lead essentially to the same conclusions with respect to the issues of mass generation and deconfinement. However, it turns out that only the one followed by ref.\cite{comitini2} gives rise to a perturbative series which at the one-loop order reproduces the non-perturbative results found on the lattice (see eg. \cite{bogolubsky}); on the downside, it provides an entropy which, in the GEP approximation, has negative values in a narrow range of temperatures below the predicted deconfinement temperature $T_{c}\approx 250$ MeV. Coversely, by working with massive ghosts, an entropy can be derived which stays positive for every value of the temperature. Therefore we believe that the present approach is still of interest from a physical point of view.\\
Since the time of writing this thesis, the author's point of view on the status of the coupling constant $\alpha_{s}$ which appears in the GEP equations has changed substantially. While both here and in ref.\cite{comitini} $\alpha_{s}$ is regarded as the standard running coupling constant of pure Yang-Mills theory (albeit subject to a non-standard running due to the influence of non-perturbative infrared effects) which must be fixed by the phenomenology, in ref.\cite{comitini2} $\alpha_{s}$ is treated as a bare coupling fixed by the principle of minimal sensitivity. For details on the renormalization of the Gaussian Effective Potential we refer the reader to ref.\cite{comitini2}.
\clearpage

\thispagestyle{empty}
\
\clearpage
\chapter*{Preface\index{Preface}}

Since its introduction in the `70s, quantum chromodynamics has been an active field of research both on the experimental and on the theoretical side. The discovery of asymptotic freedom in 1973 allowed to put on a firm theoretical ground the standard perturbation theory of QCD in the high energy limit, while at the same time showing that its low energy limit cannot be reached by a standard perturbative expansion. In order to gain information on the infrared behaviour of QCD, non-perturbative methods had to be devised such as discretization on the lattice, methods based on Schwinger-Dyson equations and variational methods. Despite the progress made in our understanding of the theory, till today a fully analytical description of QCD in the infrared is still missing.\\
In the last decade, improvements in lattice computations allowed us to gain essential insight into the behaviour of QCD at low energies. Unexpectedly, lattice data \cite{bogolubsky}-\cite{burgio15} showed that in the limit of vanishing momenta the Landau gauge gluon propagator develops an effective dynamical mass and remains finite, a phenomenon that in the literature is known as \textit{(dynamical) mass generation} \cite{cornwall}. What is unusual about mass generation is that gauge invariance should forbid the shift of the gluon mass pole. This is what would be found, for example, in the standard perturbation theory, where in the limit $p^{2}\to 0$ the gluon propagator is constrained to be singular. While different asymptotic behaviours based on analytic (or semi-analytic) methods have been proposed -- such as the Gribov-Zwanziger scenario \cite{dudal08}-\cite{dudal15} --, today it is generally believed that the correct infrared limit of the propagators is the one given by the lattice.\\
A second key topic in the understanding of QCD is confinement. The experimental lack of observation of free quarks or gluons leads us to postulate that such particles are dinamically constrained to exist as colorless bound states, although it must be noted that a rigorous theoretical proof of confinement in QCD is yet to be found. If an analytical description of the infrared behaviour of the theory were available, one could go on and discuss the issue of \textit{deconfinement} from first principles. A deconfinement phase transition occurs at the critical temperature beyond which colorful excitations of the quark and/or gluon fields cease to be confined and are allowed to propagate freely (or better almost freely, since they are still subject to interactions). Again, most of our knowledge on deconfinement comes from lattice data, which shows \cite{lucini}-\cite{boyd96} that at least in pure Yang-Mills theory a weakly first-order deconfinement phase transition occurs at a critical temperature $T_{c}$ of approximately 270 MeV and with a latent heat of $1.3$ - $1.5\ T_{c}^{4}$.\\
\\
Recent works of Pel\'{a}ez, Reinosa, Serreau, Tissier and Wschebor  \cite{tissier10}-\cite{tissier16} have established that the introduction by hand of a mass term for the gluons in the Lagrangian of pure Yang-Mills theory leads to a phenomenological model which at the one loop approximation reproduces very well the lattice data in the Landau gauge. The fact that the authors were able to obtain their results in a perturbative setting raises the question of whether the breakdown of perturbation theory at low energies could just be a consequence of a bad choice of the expansion point for the perturbative series. This is also suggested by the aforementioned fact that a massive dressed gluon propagator cannot be obtained in the standard perturbation theory. In order to address these issues, a non-standard perturbative expansion was proposed by F. Siringo in ref.~\cite{siringo1}-\cite{siringo3}. In the papers it is shown that by expanding the theory around massive rather than a massless vacuum, while not spoiling the original symmetries of the theory, dressed propagators can be derived which are in very good agreement with lattice data. Moreover, a running coupling constant can be defined and computed which stays finite and relatively small at all momenta, an indication of the fact that it should be possible to RG-improve the expansion without encountering Landau poles in the infrared. The massive expansion was initially defined \cite{siringo1} in the framework of pure Yang-Mills theory and recently extended to chiral QCD in ref.~\cite{siringo2}.\\
\\
While in \cite{siringo1}-\cite{siringo3} the validity of the massive expansion was justified only a posteriori by its agreement with the lattice data, in this thesis we show that a variational argument based on Gaussian Effective Potential (GEP) methods \cite{stevenson}-\cite{tedesco} naturally leads to the conclusion that an expansion around a massive rather than a massless vacuum is more suitable for the computations in pure Yang-Mills theory. Moreover, we show that an extension of the GEP formalism to finite temperature allows to predict the occurrence of a weakly first-order phase transition in the gluonic matter with the same thermodynamical properties as the deconfinement transition found in lattice computations.\\
\\

\clearpage


\tableofcontents{}

\clearpage{}
\thispagestyle{empty}
\
\\
\\
\clearpage
\newpage

\pagenumbering{arabic}

\addcontentsline{toc}{chapter}{Introduction}  \markboth{Introduction}{Introduction}

\chapter*{Introduction\index{Introduction}}

The objective of this thesis is to show how a variational argument based on the Gaussian Effective Potential (GEP) can be used to give evidence for mass generation and deconfinement in pure Yang-Mills SU(N) theory. Through a GEP analysis \cite{stevenson}-\cite{stevenson86} we will show that the standard massless vacuum of Yang-Mills theory (YMT) is perturbatively unstable towards a massive vacuum, indicating that the massive expansion introduced in \cite{siringo1}-\cite{siringo3} is the natural choice for computations in YMT. At finite temperature, the GEP approach will allow us to find an optimal temperature-dependent mass parameter $m(T)$ for the massive expansion. $m(T)$ will be shown to be discontinuous at a critical temperature $T_{c}\approx 0.35\ m_{0}$, where $m_{0}=m(0)$ is the optimal mass parameter at zero temperature. The entropy density $s(T)$ will also be shown to be slightly discontinuous at $T_{c}$, a feature that signals the occurrence of a weakly first-order phase transition. If one takes $m_{0}$ to be equal to the optimal mass parameter found in ref.\cite{siringo1}, $m_{0}=0.73$ GeV, a critical temperature of approximately 255 MeV is recovered. The latent heat of transition can also be estimated and is found to be equal to approximately 1.8 $T_{c}^{4}$. Both these predictions are in good agreement with the lattice results \cite{giusti} of $T_{c}\approx 270$ MeV and $\Delta h_{0}=1.3$ - $1.5\ T_{c}^{4}$.\\
\\
This thesis is organized as follows. In Chapter 1 we review the definition and the standard perturbative set-up of pure Yang-Mills SU(N) theory in the vacuum. Known results are collected with the purpose of illustrating the breakdown of the standard perturbation theory at low energies, and lattice data from ref.\cite{bogolubsky} is presented which attests the occurrence of mass generation in Yang-Mills theory. In order to show that the infrared limit of YMT can be described by a massive perturbative expansion, in Chapter 2 we review the contents and results of ref.\cite{siringo1}. Explicit expressions for the ghost and gluon dressed propagators in the Landau gauge are given and confronted with the lattice data of ref.\cite{bogolubsky}. Finally, in Chapter 3, the subjects of mass generation and deconfinement are addressed from the variational perspective of the GEP. The GEP is defined, motivated and applied to the study of pure Yang-Mills theory. Since most of the material presented in Chapter 3 is original, accurate derivations of the equations are presented.\\
This thesis is complemented by an Appendix. In Appendix A and B the formalism of quantum field theory at finite temperature is reviewed and applied to pure Yang-Mills theory. In Appendix C we prove the Jensen-Feynman inequality -- the variational statement which motivates the GEP approach. In Appendix D we derive explicit expressions for the thermal integrals involved in the computation of the GEP; these can be analytically evaluated up to a one-dimensional integration, which must be carried out numerically. In Appendix E we collect some of the numerical data needed for the GEP analysis at finite temperature.\\

\clearpage{}
\thispagestyle{empty}
\
\clearpage

\stepcounter{count}
\addcontentsline{toc}{chapter}{1 Pure Yang-Mills SU(N) vacuum theory: standard formulation and results}  \markboth{1 Pure Yang-Mills SU(N) vacuum theory: standard formulation and results}{1 Pure Yang-Mills SU(N) vacuum theory: standard formulation and results}
\chapter*{1\protect \\
\medskip{}
Pure Yang-Mills SU(N) vacuum theory: standard formulation and results\index{Pure Yang-Mills SU(N) vacuum theory: standard formulation and results}}

\clearpage{}


\addcontentsline{toc}{section}{1.1 Continuum field theory}  \markboth{1.1 Continuum field theory}{1.1 Continuum field theory}
\section*{1.1 Continuum field theory\index{Continuum field theory}}

In this section we will briefly review the definition of pure Yang-Mills SU(N) theory in the continuum and go through some of the results that are obtained from its standard formulation. In the standard formulation, the quantities of physical interest are computed by means of a perturbative expansion around a massless vacuum. As is well known, due to the presence of a Landau pole in the running of the coupling constant, these results lose their validity at energies lower than some mass scale -- the infrared or IR regime. For SU(3) in a physical setting, comparisons with both experiments and lattice data show that said regime lies at energies lower than the QCD scale, $\Lambda_{\text{QCD}}\approx 200$ MeV. The contents of this section may be found in standard textbook sources such as \cite{peskin}-\cite{itzykson}.\\
\\
This section is organized as follows. In section 1.1.1 we start by defining the theory both at the classical and at the quantum level. We then move on to describe in section 1.1.2 the standard set-up for the computation of the quantities of physical interest, namely, the standard perturbation theory. We will call such a set-up a massless perturbative expansion (MLPE) in order to distinguish it from the massive perturbative expansion (MSPE) that will be introduced in the following chapters. In section 1.1.3 we give a simple argument to show that mass generation is forbidden at any finite order in the MLPE. Finally, in section 1.1.4, we illustrate the break down of the MLPE at low energies by examining the behaviour of the running coupling constant in the infrared.\\
\\
\\
\\
\addcontentsline{toc}{subsection}{1.1.1 Yang-Mills action and the vacuum partition function}  \markboth{1.1.1 Yang-Mills action and the vacuum partition function}{1.1.1 Yang-Mills action and the vacuum partition function}
\subsection*{1.1.1 Yang-Mills action and the vacuum partition function\index{1.1.1 Yang-Mills action and the vacuum partition function}}

Pure Yang-Mills SU(N) theory in $d$-dimensions (YMT) is defined at the classical level by the action\\
\\
\begin{equation}\label{YMaction}
\mathcal{S}_{YM}[A]=-\frac{1}{2}\, \int\ \text{Tr}\Big(F[A]\wedge\star \,F[A]\Big)
\end{equation}\\
\\
The dynamical variable of the theory is the gauge potential $A$, a 1-form on Minkowski spacetime taking values in the adjoint representation of the Lie algebra $\mathfrak{su}$(N):\\
\\
\begin{equation}
A(x)=A_{\mu}^{a}(x)\ T_{a}\ dx^{\mu}
\end{equation}
\\
The $T_{a}$'s can be taken to be $N_{A}=N^{2}-1$ linearly independent $N\times N$ complex matrices forming a basis for the representation. The commutators of the $T_{a}$'s define the structure constants $f^{a}_{bc}$ of the representation through the position\\
\\
\begin{equation}\label{1}
[T_{a},T_{b}]=if^{c}_{ab}\ T_{c}
\end{equation}\\
\\
Since SU(N) is a compact simple Lie group, a specific choice of the set of matrices can always be made \cite{peskin} so that, with $\text{Tr}$ the trace operator on $N\times N$ matrices,\\
\\
\begin{equation}\label{2}
\text{Tr}\ \big(T_{a}\,T_{b}\big)=\frac{1}{2}\,\delta_{ab}
\end{equation}
\\
Such a choice implies the antisymmetry relations (with $f_{abc}=f^{a}_{bc}$)\\
\\
\begin{equation}
f_{abc}=-f_{bac}=f_{bca}
\end{equation}\\
\\
The gauge potential $A$ enters the action of the theory through its curvature 2-form $F[A]=F^{a}_{\mu\nu}[A]\ T_{a}\ dx^{\mu}\otimes dx^{\nu}$. With
\\
\begin{equation}
D_{\mu}=\partial_{\mu}-ig\ A_{\mu}^{a}\,T_{a}
\end{equation}
\\
the covariant derivative associated to $A$ (with coupling constant $g$) acting on fields in the fundamental representation of SU(N), $F$ is defined by the operatorial equation\\
\\
\begin{equation}
F_{\mu\nu}=\frac{i}{g}\ [D_{\mu},D_{\nu}]
\end{equation}
\\
It then follows from the definition \eqref{1} that the components of $F$ can be expressed as\\
\\
\begin{equation}
F_{\mu\nu}^{a}=\partial_{\mu}A_{\nu}^{a}-\partial_{\nu}A_{\mu}^{a}+g\,f^{a}_{bc}\ A^{b}_{\mu}\,A^{c}_{\nu}
\end{equation}\\
\\
By adopting the convention of eq.~\eqref{2}, the Lagrangian of pure Yang-Mills theory can be written in terms of the gauge potential $A$ as
\begin{align}\label{10}
\\
\notag\mathcal{L}_{YM}=-\frac{1}{2}\ \partial_{\mu}A_{\nu}^{a}\,(\partial^{\mu}A^{a\,\nu}-\partial^{\nu}A^{a\,\mu})-g\,f_{abc}\ \partial_{\mu}A_{\nu}^{a}\,A^{b\,\mu}\,A^{c\,\nu}-\frac{g^{2}}{4}\ f_{abc}\,f_{ade}\ A_{\mu}^{b}\,A_{\nu}^{c}\,A^{d\,\mu}\,A^{e\,\nu}
\end{align}
\\
where summation over the $a$ index in the first and last term is implied and the spacetime indices $\mu,\nu$ are raised and lowered through the Minkowski metric $\eta=\text{diag}(+1,-1,\dots,-1)$.\\
\\
\\
Amongst the defining properties of YMT is invariance under SU(N) gauge transformations. With $U(x)$ an arbitrary matrix field taking values in SU(N), such transformations act on the space of the gauge potentials $A$ as\\
\\
\begin{equation}\label{3}
A_{\mu}(x)\ \to U(x)\ \Big(A_{\mu}(x)+\frac{i}{g}\ \partial_{\mu}\Big)\ U^{\dagger}(x)
\end{equation}
The corresponding transformation on the space of curvature 2-forms simply reads\\
\\
\begin{equation}
F_{\mu\nu}(x)\to U(x)\ F_{\mu\nu}(x)\ U^{\dagger}(x)
\end{equation}
\\
Invariance of $S_{YM}$ under gauge transformations follows from the cyclic property of the trace operator. As a direct consequence of the existence of a local symmetry for the action, the Euler-Lagrange equations of the theory, namely the Yang-Mills equations,\\
\\
\begin{equation}\label{4}
\partial^{\mu}F_{\mu\nu}^{a}+g\,f^{a}_{bc}\ A^{b\,\mu}\,F_{\mu\nu}^{c}=0
\end{equation}
\\
form a set of underdetermined PDE's. The non-uniqueness of the solutions to the Dirichlet problem associated to the Yang-Mills equations points to the fact that some of the degrees of freedom (d.o.f.) of Yang-Mills theory are redundant. Such a redundancy must be dealt with when quantizing the theory.\\
\\
At the quantum level, the physical content of the theory \cite{weinberg} can be reconstructed from the vacuum partition function $Z$,\\
\\
\begin{equation}\label{6}
Z=\int\D A_{\mu}^{a}\ \ e^{i\mathcal{S}_{YM}[A]}
\end{equation}\\
\\
In order to make sense out of the RHS of eq.~\eqref{6}, one must first integrate over the gauge-equivalent configurations of the gauge potential. This is done by means of a Faddeev-Popov (FP) transformation, which allows to express $Z$ in the form\footnote{\ We restrict ourselves to FP transformations in a linear covariant gauge, since only these will be relevant to our study.}\\
\\
\begin{equation}\label{9}
Z=\mathcal{C}\, \int\D A_{\mu}^{a}\D\cbar^{a}\D c^{a}\ \ e^{i\mathcal{S}[A,\cbar,c]}=\mathcal{C}\, \int\D A_{\mu}^{a}\D\cbar^{a}\D c^{a}\ \ e^{i\BS_{YM}[A]+i\BS_{fix}[A]+i\BS_{FP}[A,\cbar,c]}
\end{equation}\\
\\
Here $\mathcal{C}$ is an inessential infinite constant factor (which we will drop in what follows), $c^{a}$ and $\cbar^{a}$ are a set of anticommuting ghost fields in the adjoint representation of SU(N), $\BS_{FP}$ is the action for the ghost fields and $\BS_{fix}$ is a gauge fixing term,\\
\\
\begin{equation}\label{sss}
\BS_{FP}[A,\cbar,c]=\int d^{d}x\ \partial^{\mu}\cbar^{a}\,D_{\mu}^{ab}c^{b}\quad;\qquad\quad \BS_{fix}[A]=-\int d^{d}x\ \ \frac{1}{2\xi}\ (\partial^{\mu}A_{\mu}^{a})^{2}
\end{equation}\\
\\
In eq.~\eqref{sss} $\xi$ is an arbitrary non-negative real number and the covariant derivative $D_{\mu}$ acts on fields in the adjoint representation as
\\
\begin{align}
D_{\mu}^{ab}c^{b}=\partial_{\mu}c^{a}+g\,f^{a}_{bc}\,A_{\mu}^{b}c^{c}
\end{align}
In the presence of a source term of the form\\
\\
\begin{equation}
\BS_{source}[A,J]=\int d^{d}x\ \ J^{\mu}_{a}(x)\,A_{\mu}^{a}(x)
\end{equation}
\\
where $J^{\mu}_{a}$ is an external vector field in the adjoint representation, the partition function $Z[J]$,\\
\\
\begin{equation}
Z[J]=\int\D A_{\mu}^{a}\D\cbar^{a}\D c^{a}\ \ \exp\Big(i\BS[A,\cbar,c]+i\BS_{source}[A,J]\Big)
\end{equation}\\
\\
generates the $n$-point correlation functions of the theory upon functional differentiation with respect to the external current $J$: with $\mathcal{A}_{\mu}^{a}(x)$ the Heisenberg-picture operator associated to the gauge potential $A_{\mu}^{a}(x)$,\\
\\
\begin{align}\label{8}
\\
\notag\bigg\{\frac{(-i)^{n}}{Z[J]}\frac{\delta^{(n)}Z[J]}{\delta J_{a_{1}}^{\mu_{1}}(x_{1})\cdots \delta J_{a_{n}}^{\mu_{n}}(x_{n})}\bigg\}\bigg|_{J=0}&=\frac{\int\D A_{\mu}^{a}\,\D\cbar^{a}\,\D c^{a}\ \ e^{i\BS}\ A_{\mu_{1}}^{a_{1}}(x_{1})\dots A_{\mu_{n}}^{a_{n}}(x_{n})}{\int\D A_{\mu}^{a}\,\D\cbar^{a}\,\D c^{a}\ \ e^{i\BS}}=\\
\notag\\
\notag&=\big\langle \,T\big\{\mathcal{A}_{\mu_{1}}^{a_{1}}(x_{1})\dots\mathcal{A}_{\mu_{n}}^{a_{n}}(x_{n})\big\}\big\rangle
\end{align}\\
\\
Correlation functions which include the ghost fields are generated by adding to the action analogous source terms for the ghosts.\\
\\
\\

\addcontentsline{toc}{subsection}{1.1.2 Standard perturbative expansion (MLPE)}  \markboth{1.1.2 Standard perturbative expansion (MLPE)}{1.1.2 Standard perturbative expansion (MLPE)}
\subsection*{1.1.2 Standard perturbative expansion (MLPE)\index{Standard perturbative expansion (MLPE)}}

Due to the presence of the non-linear interaction terms in $\BS$, the partition function \eqref{9} cannot be evaluated exactly and one has to resort to perturbative methods for its computation. As long as the coupling constant $g$ is assumed to be small, one can aim to obtain a perturbative expansion of $Z$ in powers of $g$.\\
\\
The $g$-dependence of the partition function comes from the exponential of the interaction terms $\BS_{int}:=\BS_{YM,int}+\BS_{FP,int}$,\\
\begin{align}
\\
\notag\BS_{int}=\int d^{d}x\ \ \Big\{-g\,f_{abc}\ \partial_{\mu}A_{\nu}^{a}\,A^{b\,\mu}\,A^{c\,\nu}-\frac{g^{2}}{4}\ f_{abc}\,f_{ade}\ A_{\mu}^{b}\,A_{\nu}^{c}\,A^{d\,\mu}\,A^{e\,\nu}+g\,f_{abc}\,\partial^{\mu}\cbar^{a}\ A_{\mu}^{b}c^{c}\Big\}
\end{align}\\
\\
By Taylor-expanding the interaction exponential $e^{i\BS_{int}}$, the partition function can be rewritten as\\
\\
\begin{equation}\label{12}
Z=\bigg(\int\D A_{\mu}^{a}\,\D\cbar^{a}\,\D c^{a}\ \ e^{i\BS_{0}}\bigg)\ \ \bigg(\sum_{n=0}^{+\infty}\ \frac{1}{n!}\ \frac{\int\D A_{\mu}^{a}\D\cbar^{a}\D c^{a}\ \ e^{i\BS_{0}}\ (i\BS_{int})^{n}}{\int\D A_{\mu}^{a}\,\D\cbar^{a}\,\D c^{a}\ \ e^{i\BS_{0}}}\bigg)
\end{equation}\\
\\
Here $\BS_{0}=\BS-\BS_{int}$,\\
\\
\begin{equation}\label{11}
\BS_{0}=\int d^{d}x\ \ \bigg\{-\frac{1}{2}\ \partial_{\mu}A_{\nu}^{a}\,(\partial^{\mu}A^{a\,\nu}-\partial^{\nu}A^{a\,\mu})-\frac{1}{2\xi}\ (\partial^{\mu}A_{\mu}^{a})^{2}+\partial^{\mu}\cbar^{a}\partial_{\mu}c^{a}\bigg\}
\end{equation}\\
\\
is the action for a free gauge vector field and a free ghost field. Eq.~\eqref{11} can be expressed in momentum space as\\
\begin{align}\label{13}
\\
\notag \BS_{0}=i\ \int \frac{d^{d}k}{(2\pi)^{d}}\ \bigg\{\frac{1}{2}\ A_{\mu}^{a}(k)\ \Big[\Delta^{\mu\nu}_{0\perp\,ab}(k)^{-1}+\Delta^{\mu\nu}_{0\parallel\, ab}(k)^{-1}\Big]\ A_{\nu}^{b}(k)^{*}+\cbar^{a}(k)\,\mathcal{G}_{0\,ab}(k)^{-1}\,c^{b}(k)\bigg\}
\end{align}\\
\\
where\\
\begin{equation}
\Delta^{\mu\nu}_{0\perp\,ab}(k)=\delta_{ab}\ \frac{-i\ t^{\mu\nu}(k)}{k^{2}}\quad;\quad\quad \Delta^{\mu\nu}_{0\parallel\,ab}(k)=\xi\ \delta_{ab}\ \frac{-i\ \ell^{\mu\nu}(k)}{k^{2}}
\end{equation}\\
\begin{equation}
\mathcal{G}_{0\,ab}(k)=\delta_{ab}\ \frac{i}{k^{2}}
\end{equation}\\
\\
are massless free particle propagators, with the transverse and longitudinal projectors $t^{\mu\nu},\,\ell^{\mu\nu}$ defined as\\
\\
\begin{equation}
t^{\mu\nu}(k)=\eta^{\mu\nu}-\frac{k^{\mu}k^{\nu}}{k^{2}}\quad;\quad\quad \ell^{\mu\nu}(k)=\frac{k^{\mu}k^{\nu}}{k^{2}}
\end{equation}\\
\\
For future reference, we recall that the action integral eq.~\eqref{13} can be continued from Minkowski space to Euclidean space, yielding (with $k^{0}_{E}=ik^{0}$, $k^{i}_{E}=k^{i}$, $k_{E}^{2}=\delta_{\mu\nu}\,k_{E}^{\mu}k_{E}^{\nu}=-k^{2}$, $\delta$ the Euclidean metric)\\
\begin{align}\label{14}
\\
\notag \BS_{0}&=i\ \int \frac{d^{d}k_{E}}{(2\pi)^{d}}\ \ \bigg\{\frac{1}{2}\ A_{\mu}^{a}(k_{E})\ \Big[D^{\mu\nu}_{0\perp\,ab}(k_{E})^{-1}+D^{\mu\nu}_{0\parallel\, ab}(k_{E})^{-1}\Big]\ A_{\nu}^{b}(k_{E})^{*}+\\
\notag&\qquad\qquad\qquad\qquad +\cbar^{a}(k_{E})\,G_{0\,ab}(k_{E})^{-1}\,c^{b}(k_{E})\bigg\}
\end{align}
Here\\
\begin{equation}
D^{\mu\nu}_{0\perp\,ab}(k_{E})=\delta_{ab}\ \frac{t^{\mu\nu}_{E}(k_{E})}{k_{E}^{2}}=\delta_{ab}\ \frac{\delta^{\mu\nu}-k_{E}^{\mu}k_{E}^{\nu}/k_{E}^{2}}{k_{E}^{2}}
\end{equation}\\
\begin{equation}
D^{\mu\nu}_{0\parallel\,ab}(k_{E})=\xi\ \delta_{ab}\ \frac{\ell^{\mu\nu}_{E}(k_{E})}{k_{E}^{2}}=\xi\ \delta_{ab}\ \frac{k_{E}^{\mu}k_{E}^{\nu}/k_{E}^{2}}{k_{E}^{2}}
\end{equation}\\
\begin{equation}
G_{0\,ab}(k_{E})=\delta_{ab}\ \frac{1}{k_{E}^{2}}
\end{equation}\\
\\
are massless free particle propagators in Euclidean space. We remark that in eq.~\eqref{14} the $0$-component of $A_{\mu}^{a}(k_{E})$ is related to the $0$-component of $A_{\mu}^{a}(k)$ by $A_{0}^{a}(k_{E})=-iA_{0}^{a}(k)$. If we define a Euclidean free action $\BS_{0}^{E}$ as\\
\\
\begin{equation}
\BS_{0}^{E}=-i\BS_{0}
\end{equation}\\
\\
the exponential $e^{i\BS_{0}}=e^{-\BS_{0}^{E}}$ takes the form of a Gaussian in the Euclidean Fourier components of the gauge and ghost fields.\\
\\
Going back to the evaluation of $Z$ in Minkowski space, since the functional integrands under the summation sign in eq.~\eqref{12} are equal to polynomials in the Fourier components of the fields times a Gaussian in the same components, each term of the sum can be explicitly evaluated as a Gaussian integral\footnote{\ To be more precise, one should first evaluate the integrals in Euclidean space and then switch back to Minkowski space.}. Moreover, any such term will be proportional to some power of the coupling constant $g$, so that the expression as a whole is given as a formal series in $g$. We will call such a series a massless perturbative expansion (MLPE), in order to distinguish it from the massive perturbative expansion (MSPE) that will be introduced in the following chapters. As is well known, expansions such as that in eq.~\eqref{12} organize themselves into the exponential of sums of connected Feynman diagrams,\\
\\
\begin{equation}
\ln\,Z=\ln\,Z_{0}+\sum\ \text{connected diagrams}
\end{equation}
\\
where\\
\begin{equation}
Z_{0}=\int\D A_{\mu}^{a}\,\D\cbar^{a}\,\D c^{a}\ \ e^{i\BS_{0}}
\end{equation}\\
\\
is the zeroth order approximation to $Z$ given by $\BS_{0}$. The vertices for a diagrammatic representation of the MLPE can be easily read out from the action: in Minkowski space\\
\\
\\
\begin{figure}[H]
\centering
\includegraphics[width=0.72\textwidth]{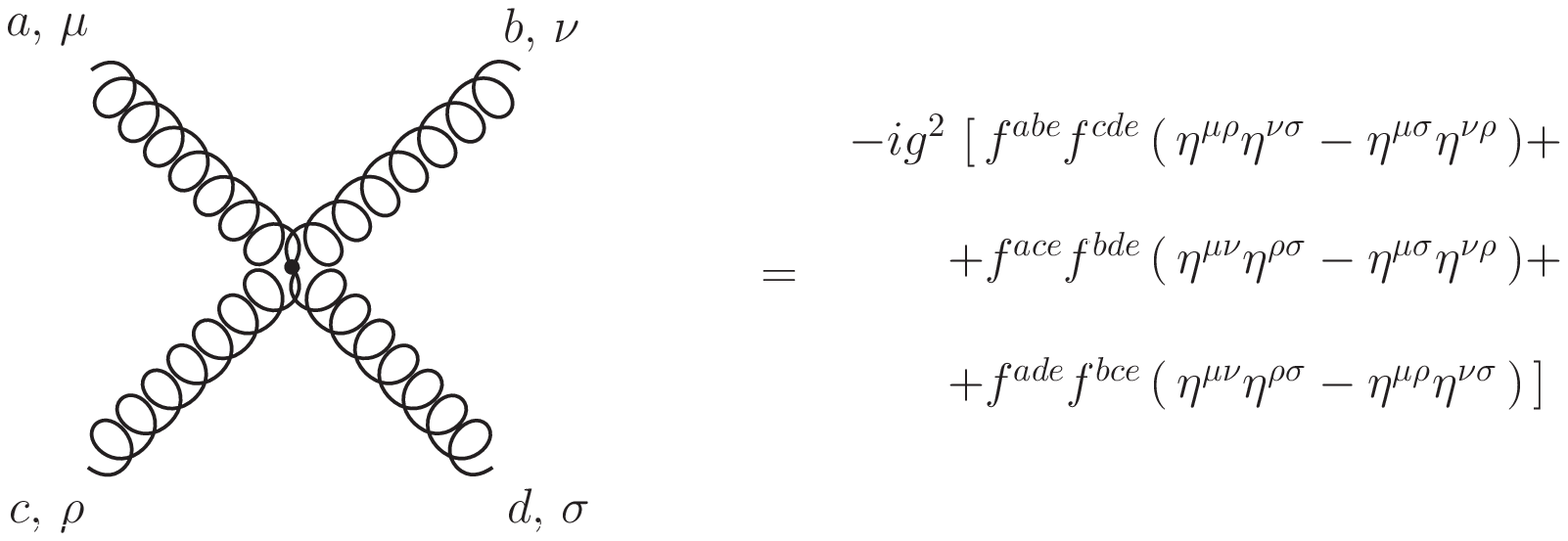}
\end{figure}\
\begin{figure}[H]
\centering
\includegraphics[width=0.82\textwidth]{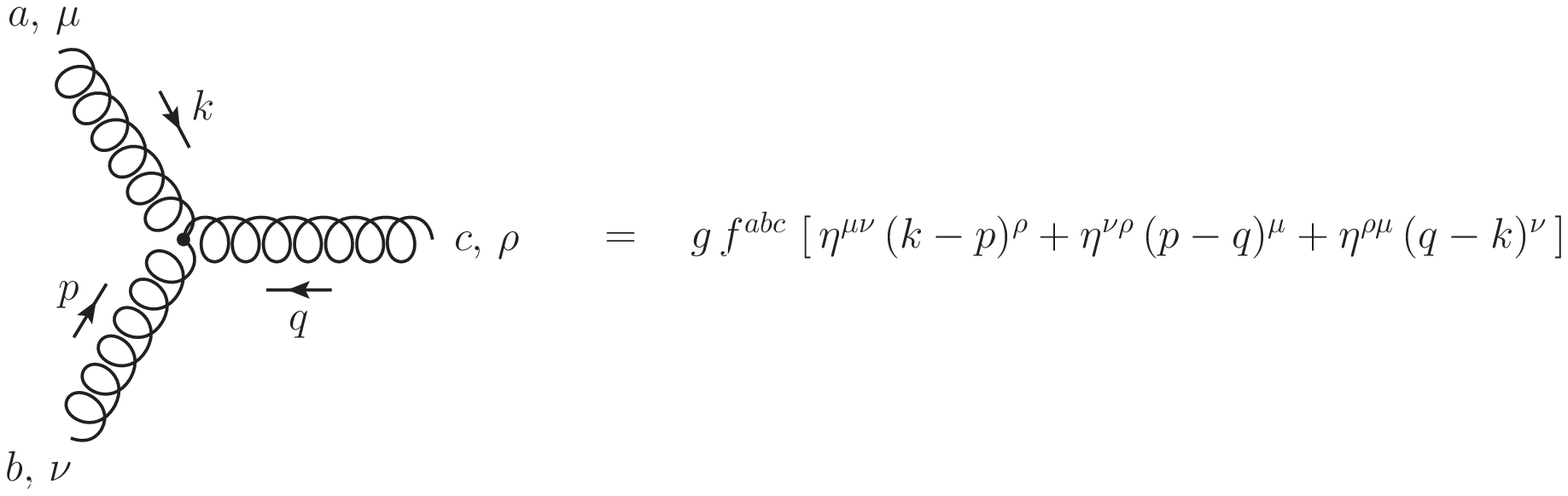}
\end{figure}\
\begin{figure}[H]
\centering
\includegraphics[width=0.50\textwidth]{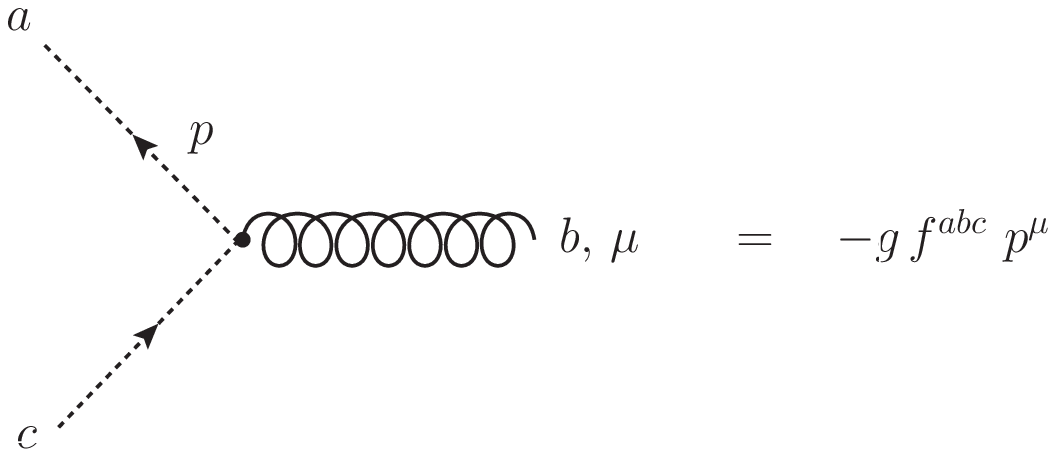}
\end{figure}\
\\
\\
\\
The propagators to be used in the MLPE diagrams are contained in the free massless action $\BS_{0}$:\\
\\
\\
\begin{figure}[H]
\centering
\includegraphics[width=0.80\textwidth]{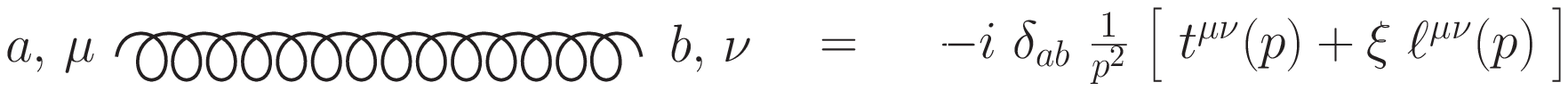}
\end{figure}\
\\
\begin{figure}[H]
\centering
\includegraphics[width=0.46\textwidth]{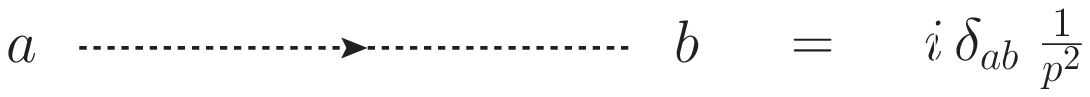}
\end{figure}

\addcontentsline{toc}{subsection}{1.1.3 Dressed gluon propagator in the MLPE}  \markboth{1.1.3 Dressed gluon propagator in the MLPE}{1.1.3 Dressed  gluon propagator in the MLPE}
\subsection*{1.1.3 Dressed  gluon propagator in the MLPE\index{Dressed  gluon propagator in the MLPE}}

The computation of the $n$-point correlation functions is carried out with the same method used to evaluate the partition function: one simply factors out the product $\mathcal{C}Z_{0}$ from \eqref{12} and integrates the connected diagrams together with an appropriate monomial in the fields. Of special interest to our study is the gluon 2-point function, i.e. the dressed gluon propagator, $\widetilde{\Delta}_{\mu\nu}^{ab}$,\\
\\
\begin{equation}
\widetilde{\Delta}_{\mu\nu}^{ab}(x,y)=\big\langle T\{\mathcal{A}_{\mu}^{a}(x)\,\mathcal{A}_{\nu}^{b}(y)\}\big\rangle
\end{equation}\\
\\
or equivalently, in momentum space,\\
\\
\begin{equation}
\widetilde{\Delta}_{\mu\nu}^{ab}(p)=\big\langle \mathcal{A}_{\mu}^{a}(p)\,\mathcal{A}_{\nu}^{b}(-p)\big\rangle
\end{equation}
\\
\\
In diagrammatic form, the dressed gluon propagator can be represented as\\
\\
\begin{figure}[H]
\centering
\includegraphics[width=\textwidth]{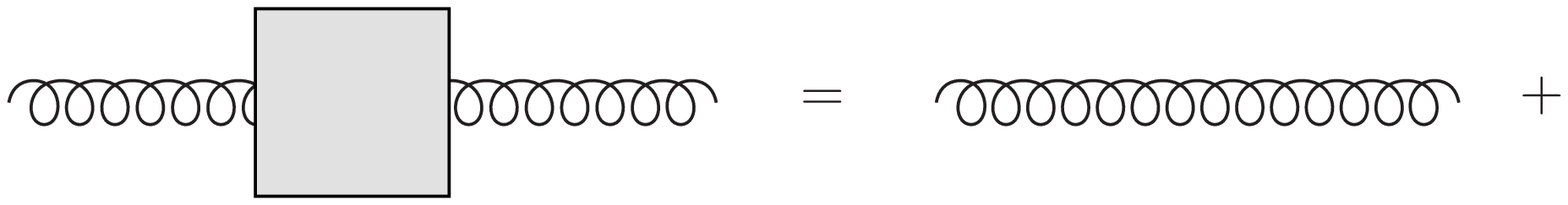}
\end{figure}\
\\
where the square denotes a sum over all possible diagrams and the blobs denote the sum over one-particle irreducible (1PI) diagrams. If we define the polarization tensor $\Pi_{\mu\nu}^{ab}(p)$ as\\
\\
\begin{figure}[H]
\centering
\includegraphics[width=0.35\textwidth]{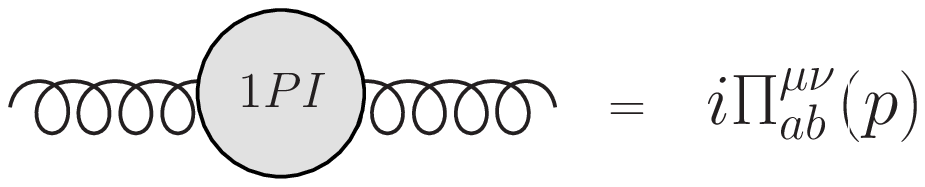}
\end{figure}\
\\
upon summation of the 1PI diagrams we obtain the dressed gluon propagator in the form\\
\\
\begin{equation}
\widetilde{\Delta}_{\mu\nu}^{ab}(p)=\delta^{ab}\ \bigg[\ \frac{-it_{\mu\nu}(p)}{p^{2}\big(1-\Pi_{\perp}(p)\big)}+\xi\ \frac{-i\ell_{\mu\nu}(p)}{p^{2}\big(1-\xi\,\Pi_{\parallel}(p)\big)}\bigg]
\end{equation}
\\
where\\
\begin{equation}
\Pi_{\perp}(p)=\frac{1}{(d-1)N_{A}}\ \frac{\delta_{ab}\,t^{\mu\nu}(p)\,\Pi_{\mu\nu}^{ab}(p)}{p^{2}}\quad;\quad\qquad \Pi_{\parallel}(p)=\frac{1}{N_{A}}\ \frac{\delta_{ab}\,\ell^{\mu\nu}(p)\,\Pi_{\mu\nu}^{ab}(p)}{p^{2}}
\end{equation}\\
\\
The fact that $\widetilde{\Delta}^{ab}_{\mu\nu}$ is proportional to a Kronecker delta in the algebra indices can be easily proven and is a consequence of gauge invariance. Another consequence of gauge invariance (see e.g. \cite{itzykson}) is that in any gauge -- both non-perturbatively and perturbatively at any given order in the MLPE -- the longitudinal polarization function $\Pi_{\parallel}(p)$ vanishes. In what follows we will use simple arguments based on dimensional analysis to show that at any finite order in the MLPE the gluon propagator is singular in the limit $p^{2}\to 0$. While our proof cannot be generalized to the more interesting case of Yang-Mills theory with massive quarks\footnote{\ Although in this case the same result holds due to gauge invariance.}, it clearly illustrates the fact that mass generation is forbidden in the MLPE.\\
\\
In principle, the transverse polarization function $\Pi_{\perp}(p)$ can be computed at any given order in perturbation theory by using standard diagrammatic techniques. These allow us to express $\Pi_{\perp}(p)$ in terms of loop integrals, some of which are divergent and need to be regularized. The functional form of the transverse polarization function can be guessed by noticing that, in our definition, $\Pi_{\perp}(p)$ is an adimensional function of the dimensionful variable $p$. This, together with Lorentz invariance, implies that $\Pi_{\perp}(p)$ can depend on $p$ only through a combination of the form $p^{2}/\mu^{2}$, where $\mu$ is some mass scale. Since pure Yang-Mills theory is scale-free, the scale $\mu$ can only arise from the regularization of the divergent loop integrals. If we assume, as is customary in the treatment of Yang-Mills theory, that loop integrals are defined in dimensional regularization, then a mass scale $\mu$ enters the computations through the dimensionful coupling constant $g\,\mu^{\epsilon}$, where $\epsilon=4-d$. Now, each diagram in the expansion of $\Pi_{\perp}(p)$ is proportional to some integer power $k$ of $\mu^{\epsilon}$, which in turn can be expressed as\\
\\
\begin{equation}\label{mass}
\mu^{k\epsilon}=\sum_{n=0}^{+\infty}\ \frac{1}{n!}\ \bigg(\,\frac{k\epsilon}{2}\ \ln\,\mu^{2}\,\bigg)^{n}
\end{equation}\\
\\
At finite order in perturbation theory, eq.~\eqref{mass} always multiplies a finite number of divergences. In dimensional regularization, these take the form of non-negative powers of $2/\epsilon$. In the limit $\epsilon\to 0$, only a finite number of logarithms survives. Since in dimensional regularization, as long as we use $g\mu^{\epsilon}$ as the coupling constant, all the quantities are dimensionally well-defined, every non-vanishing logarithm with argument $\mu^{2}$ must be accompanied by an equal and opposite logarithm with argument $-p^{2}$. Thus at any finite order in the MLPE we must have\\
\\
\begin{equation}\label{PI}
\Pi_{\perp}(p)=\sum_{n=0}^{h}\ c_{n}\ \bigg( \ln\,\frac{-p^{2}}{\mu^{2}}\,\bigg)^{n}
\end{equation}\\
\\
where $h$ is some integer and the $c_{n}$'s are numerical coefficients. At this stage, some of the coefficients in \eqref{PI} are still divergent in the limit $\epsilon\to 0$. These divergences can always be reabsorbed by using appropriate renormalization counterterms. Since such a procedure can be carried out diagrammatically, it must give as a result an expression which still has the functional form given by eq.~\eqref{PI}. Moreover, the renormalization procedure allows one to redefine the actual value of $\mu^{2}$, so that the mass scale can be given a physical significance by fixing appropriate renormalization conditions.\\
\\
From the argument given above it follows that at finite order in the MLPE the transverse part $\widetilde{\Delta}_{\perp}(p)$ of the renormalized dressed gluon propagator has the form\\
\\
\begin{equation}\label{pertprop}
\widetilde{\Delta}_{\perp}(p)^{-1}=ip^{2}\ \sum_{n=0}^{h}\ c_{n}\ \bigg( \ln\,\frac{-p^{2}}{\mu^{2}}\,\bigg)^{n}
\end{equation}\\
\\
In the limit $p^{2}\to0$, each of the terms in \eqref{pertprop} goes to zero. Hence we find that in the MLPE the gluon propagator is always singular:\\
\\
\begin{equation}
\lim_{p^{2}\to 0}\ \widetilde{\Delta}(p)\ =\ \infty 
\end{equation}\\
\\
We remark that the same conclusion obviously holds true for the continuation of $\widetilde{\Delta}(p)$ to Euclidean space, $\widetilde{D}(p_{E})$.\\
\\
For future reference, we recall that together with the dressed gluon propagator, a dressed ghost propagator $\widetilde{\mathcal{G}}^{ab}(p)$ is defined as\\
\\
\begin{equation}
\widetilde{\mathcal{G}}^{ab}(p)=\langle \,\mathcal{C}^{a}(p)\ \overline{\mathcal{C}}^{b}(p)\,\rangle
\end{equation}\\
\\
where $\mathcal{C}^{a}(p)$ and $\overline{\mathcal{C}}^{b}(p)$ are the Fourier transforms of the Heisenberg picture ghost field operators $\mathcal{C}^{a}(x)$ and $\overline{\mathcal{C}}^{b}(x)$. The same reasoning as above tells us that in the MLPE the ghost propagator is singular in the limit $p^{2}\to 0$.\\
\\
\addcontentsline{toc}{subsection}{1.1.4 Running of the coupling constant in the MLPE}  \markboth{1.1.4 Running of the coupling constant in the MLPE}{1.1.4 Running of the coupling constant in the MLPE}
\subsection*{1.1.4 Running of the coupling constant in the MLPE\index{Running of the coupling constant in the MLPE}}

Renormalization group methods allow us to enhance the perturbative expansion of the quantities of physical interest through the introduction of a momentum-dependent running coupling constant $\bar{g}$, to be used in place of the fixed coupling $g$. $\bar{g}$ is defined \cite{peskin} as the solution to the differential equation\\
\\
\begin{align}\label{run}
\frac{d\,\bar{g}}{d\ln(Q/\mu)}=\beta(\bar{g})
\end{align}\\
\\
where $\mu$ is a renormalization scale, $Q^{2}=s$ -- the center of mass energy of the process in consideration -- and at one loop order in the MLPE the beta function $\beta(\bar{g})$ reads\\
\\
\begin{equation}
\beta(\bar{g})=-b_{0}\ \frac{\bar{g}^{3}}{16\pi^{2}}\ ;\qquad\qquad b_{0}=\frac{11N}{3}
\end{equation}
Eq.~\eqref{run} can be easily solved to yield\\
\\
\begin{equation}\label{run2}
\bar{\alpha}_{s}(Q)=\frac{\bar{\alpha}_{s}(\mu)}{1+\frac{b_{0}}{4\pi}\,\bar{\alpha}_{s}(\mu)\, \ln(Q^{2}/\mu^{2})}
\end{equation}\\
\\
where $\bar{\alpha}_{s}=\bar{g}^{2}/4\pi$ and $\bar{\alpha}(\mu)$ is a renormalized coupling constant to be deduced from the phenomenology. If we define a energy scale $\Lambda_{\text{SU(N)}}$ by setting\\
\\
\begin{equation}
1+\frac{b_{0}}{4\pi}\,\bar{\alpha}_{s}(\mu)\, \ln(\Lambda_{\text{SU(N)}}^{2}/\mu^{2})=0
\end{equation}\\
\\
then we can put eq.~\eqref{run2} in the form\\
\\
\begin{equation}\label{run3}
\bar{\alpha}_{s}(Q)=\frac{4\pi}{b_{0}\,\ln(Q^{2}/\Lambda_{\text{SU(N)}}^{2})}
\end{equation}\\
\\
It is easy to see that as $Q^{2}$ decreases to $\Lambda_{\text{SU(N)}}$, $\bar{\alpha}_{s}(Q)$ grows to infinity. Such a singularity is not suppressed by higher order corrections and is known in the literature as a Landau pole. The presence of an IR Landau pole tells us that only at energies sufficiently higher than $\Lambda_{\text{SU(N)}}$ the coupling constant is small enough to yield a sensible perturbative expansion; this in turn implies that the MLPE cannot be trusted at low energies. As for the actual value of $\Lambda_{\text{SU(N)}}$, this cannot be predicted from first principles and must come from phenomenology. For N $=3$ we can take it to be of the order of the analogous mass scale of QCD (i.e. of YMT with quarks), $\Lambda_{\text{QCD}}\approx 200$ MeV. Then eq.~\eqref{run3} tells us that the MLPE yields a sensible approximation to pure YMT only at energies higher than a few GeV. The low energy limit of the theory, on the other hand, cannot be reached by a massless perturbative expansion.\\
\\
We remark that the presence of the Landau pole in the MLPE coupling does not imply that any perturbative expansion of Yang-Mills theory would fail at low energies. For all we know, the pole may well be an artifact of the standard perturbation theory.\\
\clearpage


\addcontentsline{toc}{section}{1.2 Lattice field theory: $d=4$, N = 3 propagators in the deep IR in the Landau gauge}  \markboth{1.2 Lattice field theory: $d=4$, N = 3 propagators in the deep IR in the Landau gauge}{1.2 Lattice field theory: $d=4$, N = 3 propagators in the deep IR in the Landau gauge}
\section*{1.2 Lattice field theory: $\boldsymbol{d=4}$, N = 3 propagators in the deep IR in the Landau gauge\index{Lattice field theory: $d=4$, N = 3 propagators in the deep IR in the Landau gauge}}

The most popular alternative to perturbative methods in pure Yang-Mills theory is given by lattice field theory. In the lattice approach \cite{greiner}, a finite lattice of discrete spacetime points replaces the spacetime continuum and group variables are used in place of the algebra variables $A_{\mu}^{a}$. The quantities of physical interest are defined in terms of Euclidean path integrals, so that the computations can be carried out by numerically integrating over the configurations of a finite number of group variables. Provided that a suitable Euclidean action is defined, one expects to recover the standard continuum theory in the limit in which the lattice points are infinitely close and the lattice is of infinite extension.\\
In this section we present some of the results of lattice computations in the IR for the case $d=4$, N~$=3$. The data refers to the lattice analogue of the Landau gauge and is taken from ref.\cite{bogolubsky}. As we saw in sec. 1.1, the low energy limit of YMT cannot be reached by a MLPE; therefore, lattice data is essential to our knowledge of YMT in said regime. In the Euclidean formalism, the dressed ghost and gluon propagators $\widetilde{G}$ and $\widetilde{D}$ depend upon a Euclidean momentum variable which in this section we will denote with $p$. As we will soon see, contrarily to what would be expected from the MLPE approach, lattice data tells us that in the limit $p^{2}\to 0$ the gluon propagator reaches a finite value. This feature can be understood as the appearance of a dynamical effective mass due to the interactions amongst gluons. The question of whether such a feature can be reproduced by a perturbative expansion in the continuum formalism will be addressed in the next chapter.\\
\\
\\
\\
\begin{figure}[H]
\centering
\includegraphics[angle=270, origin=c,width=0.65\textwidth]{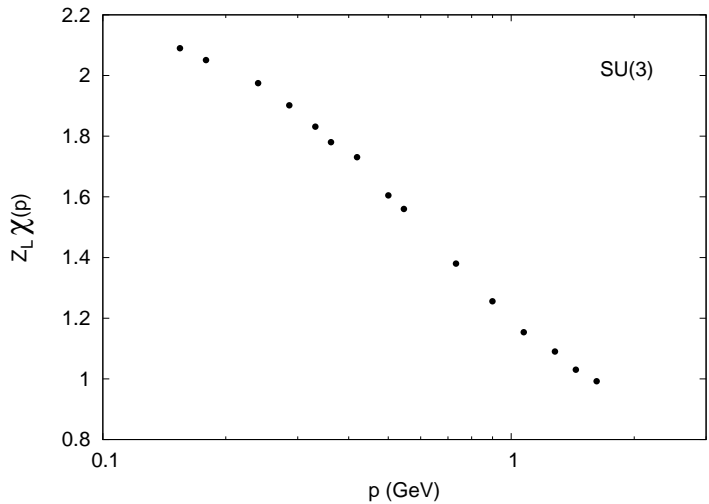}
\vspace{-8mm}
\caption{Lattice data points for the ghost dressing function $\chi(p)$ as a function of momentum (ref.\cite{bogolubsky}). $Z_{L}$ is a dimensionless normalization factor that in this section plays no role and is kept only for future convenience.}
\end{figure}\
\\
\\
In Fig. 1 lattice data points for the ghost dressing function $\chi(p)$,\\
\\
\begin{equation}
\chi(p)=p^{2}\ \widetilde{G}(p)
\end{equation}\\
\\
are plotted as a function of the euclidean momentum $p$. As $p$ goes to zero, $\chi(p)$ approaches a finite value, which in turn implies that in the limit $p^{2}\to 0$ the ghost propagator is singular. This is in agreement with the MLPE constraints.\\
In Fig. 2 lattice data points for the dressed gluon propagator $\widetilde{D}(p)$ are plotted as a function of the euclidean momentum $p$. In the limit $p\to 0$, $\widetilde{D}(p)$ saturates and reaches a finite value, in contrast with what happens in the MLPE. If one postulates for the $p^{2}\to 0$ limit of the propagator a functional dependence of the form\\
\\
\begin{equation}
\widetilde{D}(p)\ \to\ \frac{Z}{p^{2}+M^{2}}
\end{equation}\\
\\
where $M$ is some mass scale and $Z$ is a constant, then the data tells us that $M$ is different from zero. This is the phenomenon of mass generation in Yang-Mills theory.\\
\\
\\
\\
\\
\\
\begin{figure}[H]
\centering
\includegraphics[angle=270, origin=c,width=0.65\textwidth]{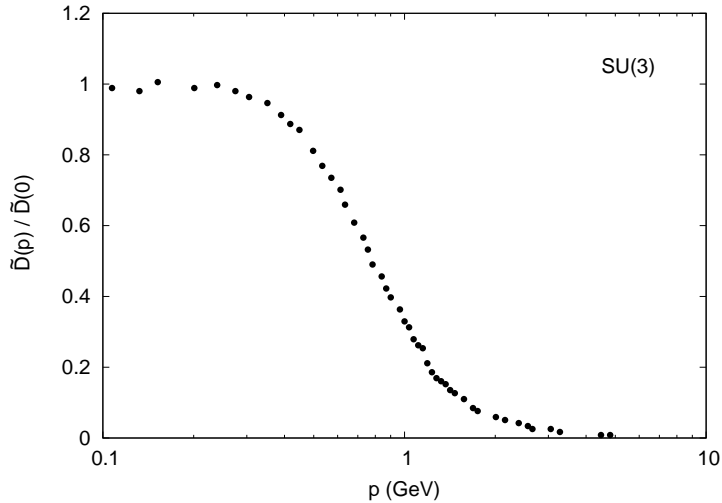}
\vspace{-8mm}
\caption{Lattice data points for the dressed gluon propagator $\widetilde{D}(p)$ as a function of momentum (ref.\cite{bogolubsky}). $\widetilde{D}(0)$ is a dimensionful normalization factor that in this section plays no role and is kept only for future convenience.}
\end{figure}\

\clearpage{}


\stepcounter{count}
\addcontentsline{toc}{chapter}{2 Pure Yang-Mills SU(N) vacuum theory: massive expansion and results}  \markboth{2 Pure Yang-Mills SU(N) theory: massive expansion and results}{2 Pure Yang-Mills SU(N) theory: massive expansion and results}
\chapter*{2\protect \\
\medskip{}
Pure Yang-Mills SU(N) theory: massive expansion and results\index{Pure Yang-Mills SU(N) theory: massive expansion and results}}

In Chapter 1 we saw that the running coupling constant of pure Yang-Mills theory, as given by the MLPE, develops a Landau pole in the IR, so that such an expansion cannot be trusted at low energies. In the absence of alternative analytical computational methods, most of the information we have on the behaviour of the theory in the IR comes from lattice computations in Euclidean space. Of special interest to our study is the low energy behaviour of the gluon propagator. As reviewed in sec. 1.2, lattice data shows that in $d=4$, N~$=3$ and in the Landau gauge the dressed gluon propagator develops a finite effective mass, a feature that, as seen in sec. 1.1.3, cannot be reproduced at any finite order in the MLPE perturbation theory. While in itself the difference between lattice and MLPE results could be traced back to the break down of the latter at low energies, the impossibility of obtaining a finite value of $\widetilde{\Delta}(0)$ in the MLPE points to the fact that non-pertubative effects, in general, are not negligible in Yang-Mills theory. It is then clear that some non-perturbative computational tool is needed in order to account for such effects. In this chapter we present one possible such tool in the form of a seemingly perturbative technique, namely, the massive perturbative expansion (MSPE). The MSPE is a straightforward generalization of the standard perturbative expansion of YMT whose motivation stems from the acknowledgement that, since the gluon propagator acquires a non-zero effective mass in the limit $p^{2}\to 0$, a perturbative expansion around a massless vacuum may not be at all suitable for computations in Yang-Mills theory. The contents of this chapter are taken from ref.\cite{siringo1}-\cite{siringo3}.\\
\\
This chapter is organized as follows. In sec. 2.1 we define the MSPE and discuss its significance in relation to the MLPE. In sec. 2.2 we give explicit expressions for the MSPE ghost and gluon dressed propagators at one loop order and examine their $p^{2}\to 0$ limit. In sec. 2.3 we compare our results with lattice data for $d=4$ and N~$=3$.\\

\clearpage{}

\addcontentsline{toc}{section}{2.1 Massive perturbative expansion (MSPE)}  \markboth{2.1 Massive perturbative expansion (MSPE)}{2.1 Massive perturbative expansion (MSPE)}
\section*{2.1 Massive perturbative expansion (MSPE)\index{Massive perturbative expansion (MSPE)}}

We start again from the definition of the partition function eq.~\eqref{9} and notice that an expansion in terms of Gaussian integrals is still obtained if we shift the free action $\BS_{0}$ by an amount $\delta \BS$ that is quadratic in the field variables. In order for the total action to remain unchanged, such a shift must be compensated by an opposite shift in the interaction action. We may thus define\\
\\
\begin{equation}\label{738}
\BS_{0}'=\BS_{0}+\delta \BS;\qquad\BS_{int}'=\BS_{int}-\delta \BS
\end{equation}
\\
so that
\\
\begin{equation}\label{SHIFT}
\BS_{0}'+\BS_{int}'=\BS_{0}+\BS_{int}
\end{equation}
\\
and expand perturbatively around the vacuum described by $\BS_{0}'$ rather than $\BS_{0}$, with a new interaction term $-\delta \BS$. As we saw in Chapter 1, in the Landau gauge the gluon propagator acquires a transverse mass due to non-perturbative effects. This suggests that an expansion around a massive -- rather than a massless -- transverse vacuum could be more suitable for the computations in Yang-Mills theory. We then fix $\delta\BS$ by the requirement that $\BS_{0}'$ be the kinetic action for a set of massive transverse gauge bosons (together with massless longitudinal gluons and ghosts), namely\\
\\
\begin{align}\label{737}
\BS_{0}'&=i\ \int \frac{d^{d}k}{(2\pi)^{d}}\ \frac{1}{2}\ A_{\mu}^{a}(k)\ \Big[\Delta^{\mu\nu}_{m\,\perp ab}(k)^{-1}\ +\Delta^{\mu\nu}_{0\,\parallel ab}(k)^{-1}\Big]\ A_{\nu}^{b}(k)^{*}+\text{ghost term}
\end{align}\\
\\
where\\
\\
\begin{equation}\label{740}
\Delta^{\mu\nu}_{m\perp\,ab}(k)=\delta_{ab}\ \frac{-i\ t^{\mu\nu}(k)}{k^{2}-m^{2}}
\end{equation}\\
\\
is a massive transverse boson propagator. In order to obtain \eqref{737} one must set\\
\begin{align}\label{739}
\delta\BS&=i\ \int \frac{d^{d}k}{(2\pi)^{d}}\ \frac{1}{2}\ A_{\mu}^{a}(k)\ \Big[\ \Delta^{\mu\nu}_{m\perp\,ab}(k)^{-1}-\Delta^{\mu\nu}_{0\perp\, ab}(k)^{-1}\ \Big]\ A_{\nu}^{b}(k)^{*}=\\
&\notag= \int \frac{d^{d}k}{(2\pi)^{d}}\ \frac{1}{2}\ A_{\mu}^{a}(k)\ m^{2}\ t^{\mu\nu}(k)\ A_{\nu}^{b}(k)^{*}
\end{align}\\
\\
We will call an expansion with $S_{0}'$ as the kinetic action a massive perturbative expansion (MSPE). The Feynman rules for the MSPE are the same as those given in Chapter 1 for the MLPE, except that now the gluon propagator is given by Fig. 3 and we have a new vertex, given by Fig. 4.\\
\\
\begin{figure}[H]
\centering
\includegraphics[width=0.75\textwidth]{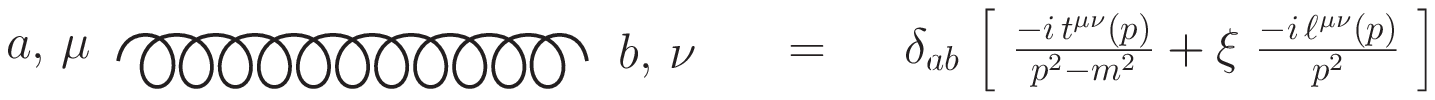}
\caption{MSPE gluon propagator.}
\end{figure}\
\\
\\
\begin{figure}[H]
\centering
\includegraphics[width=0.75\textwidth]{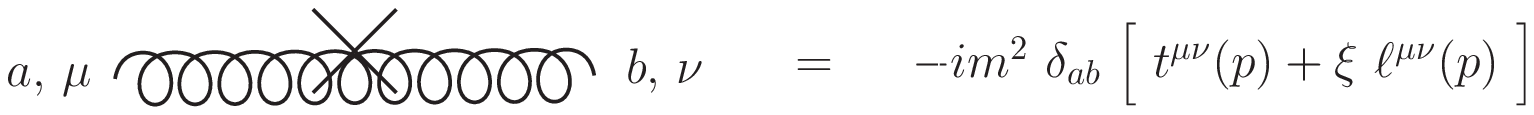}
\caption{MSPE mass counterterm.}
\end{figure}\
\\
\\
\\
\\
In principle, the $m^{2}$ in eqq.~\eqref{737}-\eqref{739} is an arbitrary positive parameter. In practice, we expect eq.~\eqref{737} to be a sensible kinetic action for the expansion of YMT provided that we choose a specific value of $m^{2}$. As the action with which we started was scale-free, such a value cannot be determined a priori and must come from the phenomenology.\\
\\
If we identify $m$ with the dynamical mass acquired by the gluons\footnote{\ Such an identification is only approximate due to radiative corrections, but this fact does not spoil our reasoning.}, then $m$ must be in some way related to the coupling constant of the theory. This in turn implies that the MSPE is not a perturbative expansion at all, or at least not in a rigorous sense. Consider the vertex that arises from the shift in the propagator, Fig. 4. We will call such a vertex the mass counterterm. When one computes the relevant quantities in terms of standard MLPE Feynman diagrams, one organizes the perturbative series in powers of the coupling constant and discards all the diagrams of order greater than some fixed value. By contrast, in the MSPE the diagrams that contain the mass counterterm are not explicitly proportional to any single power of $g$. Nevertheless, as we said, the actual value of $m$ must implicitly depend on the coupling; therefore the $m$-dependence of the mass counterterm -- and of the gluon propagator of course -- effectively introduces a mixing between MLPE diagrams of different order in the coupling constant, yielding a formally perturbative but actually non-perturbative series in $g$. Since such a series could be obtained only by adding up an infinite number of renormalized MLPE diagrams, this explains why the MSPE can incorporate non-pertubative effects that could not be taken into account through a simple MLPE.\\

\clearpage

\addcontentsline{toc}{section}{2.2 MSPE dressed propagators in the Landau gauge}  \markboth{2.2 MSPE dressed propagators in the Landau gauge}{2.2 MSPE dressed propagators in the Landau gauge}
\section*{2.2 MSPE dressed propagators in the Landau gauge\index{MSPE dressed propagators in the Landau gauge}}

In this section we will give explicit expressions for the dressed MSPE ghost and gluon Euclidean propagators to one loop order in the Landau gauge. Currently, we are not able to fix an objective criterion for the number of insertions of the mass counterterm to be added at any given order in perturbation theory. Our not so arbitrary prescription consists in adding as many counterterms as the number of powers of $\BS_{int}$ required to obtain the analogous MLPE diagrams at the given order. The reader is referred to \cite{siringo1} for the details of the computations.\\

\addcontentsline{toc}{subsection}{2.2.1 Ghost propagator}  \markboth{2.2.1 Ghost propagator}{2.2.1 Ghost propagator}
\subsection*{2.2.1 Ghost propagator\index{Ghost propagator}}

Let us define the ghost self-energy $\Sigma(p)$ as\\
\\
\begin{figure}[H]
\centering
\includegraphics[width=0.4\textwidth]{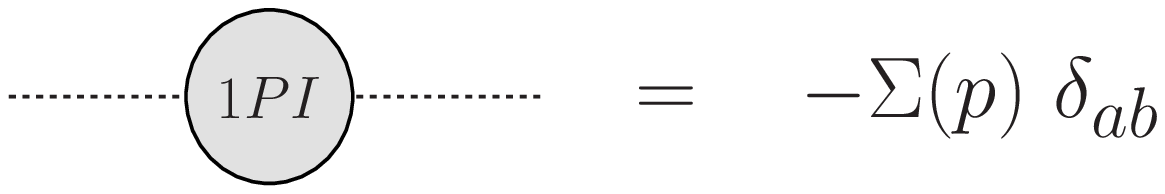}
\end{figure}
\
\\
where the blob on the LHS denotes the sum of the one-particle irreducible graphs. The diagrams that contribute to the one-loop order ghost self-energy are displayed in Fig. 5. An explicit computation shows that\\
\\
\begin{equation}
\Sigma(p)=\Sigma^{\epsilon}(p)+\Sigma^{f}(p)
\end{equation}\\
\\
where $\Sigma^{\epsilon}(p)$ is a divergent contribution,\\
\\
\begin{equation}
\Sigma^{\epsilon}(p)=-\frac{\alpha p^{2}}{9}\ \bigg(\ \frac{2}{\epsilon}+\ln \frac{\mu^{2}}{m^{2}}\ \bigg)
\end{equation}\\
\\
with $\epsilon=4-d$ and $\mu$ the renormalization scale, and $\Sigma^{f}(p)$ is a finite contribution,\\
\\
\\
\\
\begin{figure}[H]
\centering
\includegraphics[width=0.7\textwidth]{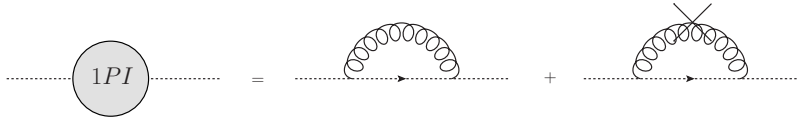}
\caption{1PI diagrams for the ghost self-energy at one-loop order.}
\end{figure}\
\\
\begin{equation}\label{sf}
\Sigma^{f}(p)=\frac{4\alpha p^{2}}{9}\ \bigg(L(s)-\frac{2}{3}\bigg)\quad;\qquad\qquad\qquad s=\frac{p^{2}}{m^{2}}
\end{equation}\\
\\
Here the effective coupling $\alpha$ is defined as\\
\\
\begin{equation}
\alpha=\frac{27N}{16\pi}\ \alpha_{s}\ ;\qquad\qquad \alpha_{s}=\frac{g^{2}}{4\pi}
\end{equation}
\\
and the function $L(s)$ reads\\
\\
\begin{equation}\label{L}
L(s)=\frac{1}{12}\ \bigg[\ \frac{(1+s)^{2}\,(2s-1)}{s^{2}}\ \ln(1+s)-2s\,\ln s+\frac{1}{s}+2\ \bigg]
\end{equation}\\
\\
The constant in parentheses in \eqref{sf} depends on the renormalization scheme and has thus no direct significance. In the $\overline{\text{MS}}$ scheme, the counterterm needed to eliminate the divergence in $\Sigma(p)$ is\\
\\
\begin{equation}\label{ctg}
\delta Z_{c}=-\frac{2\alpha}{9\epsilon}
\end{equation}\\
\\
If we define the ghost dressing function $\chi(p)$ as\\
\\
\begin{equation}
\chi(p)=p^{2}\ \widetilde{G}(p)
\end{equation}\\
\\
where $\widetilde{G}(p)$, apart from the tensorial factor $\delta_{ab}$, is the dressed ghost propagator, by summing the 1PI diagrams to one loop order together with the counterterm \eqref{ctg}  we find that\\
\\
\begin{equation}
\chi(s)^{-1}=1+\frac{4\alpha}{9}\ \bigg[\, L(s)-\frac{2}{3}-\frac{1}{4}\ \ln\,\frac{\mu^{2}}{m^{2}}\ \bigg]
\end{equation}\\
\\
modulo a constant proportional to $\alpha$. The constant is automatically included if we express the inverse dressing function in the general form\\
\\
\begin{equation}\label{chi}
\chi(s)^{-1}=\chi(s_{0})^{-1}+\frac{4\alpha}{9}\ \big(L(s)-L(s_{0})\big)
\end{equation}\\
\\
where $s_{0}\,m^{2}=p_{0}^{2}$ is some momentum scale. In the limit $s\to 0$, $L(s)$ is finite. This implies that in the MSPE the ghost pole is not shifted from its original position $p^{2}=0$.\\

\addcontentsline{toc}{subsection}{2.2.2 Gluon propagator}  \markboth{2.2.2 Gluon propagator}{2.2.2 Gluon propagator}
\subsection*{2.2.2 Gluon propagator\index{Gluon propagator}}

Let us define the gluon polarization tensor $\Pi^{\mu\nu}_{ab}(p)$ as\\
\\
\begin{figure}[H]
\centering
\includegraphics[width=0.52\textwidth]{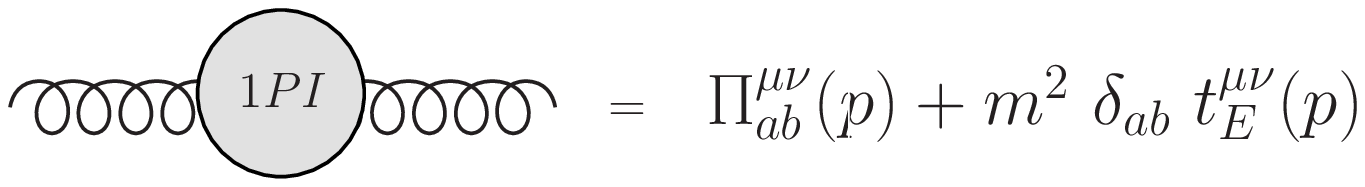}
\end{figure}
\
\\
The reason for this definition will become clear in a moment. The diagrams that contribute to the gluon propagator at one loop order are shown in Fig. 6. The counterterm depicted in the first diagram is equal to $m^{2}\,\delta_{ab}\,t^{\mu\nu}_{E}(p)$; our definition simply subtracts the contribution due to said diagram from the rest of the polarization. Since in the Landau gauge only the transverse degrees of freedom propagate, we may set\\
\\
\begin{equation}
\Pi^{\mu\nu}_{ab}(p)=\Pi(p)\ \delta_{ab}\ t^{\mu\nu}_{E}(p)
\end{equation}\\
\\
where $\Pi(p)$ is a polarization function,\\
\\
\begin{equation}
\Pi(p)=\frac{1}{N_{A}(d-1)}\ \delta^{ab}\ t_{E\,\mu\nu}(p)\ \Pi^{\mu\nu}_{ab}(p)
\end{equation}\\
\\
Notice that this definition of $\Pi(p)$ differs from that of $\Pi_{\perp}(p)$ in sec. 1.1.3 for a factor of $p^{2}$. An explicit computation shows that\\
\\
\begin{equation}
\Pi(p)=\Pi^{\epsilon}(p)+\Pi^{f}(p)
\end{equation}\\
\\
\\
\\
\begin{figure}[H]
\centering
\includegraphics[width=0.95\textwidth]{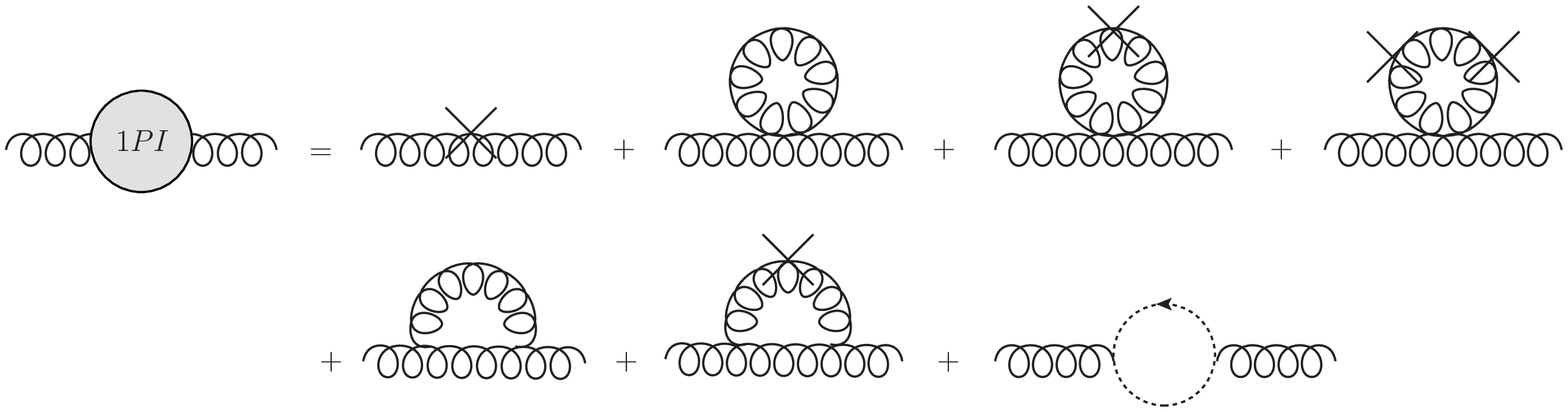}
\caption{1PI diagrams for the gluon polarization at one-loop order.}
\end{figure}\
\\
\\
where $\Pi^{\epsilon}(p)$ is a divergent contribution,\\
\\
\begin{equation}\label{glud}
\Pi^{\epsilon}(p)=\frac{26\alpha p^{2}}{81}\ \bigg(\ \frac{2}{\epsilon}+\ln\,\frac{\mu^{2}}{m^{2}}\ \bigg)
\end{equation}\\
\\
and $\Pi^{f}(p)$ is a finite contribution,\\
\\
\begin{equation}\label{gluf}
\Pi^{f}(p)=-\frac{4\alpha p^{2}}{9}\ \bigg(F(s)-\frac{197}{108}\ \bigg)
\end{equation}\\
\\
Here $\alpha$ is as in 2.2.1 and the function $F(s)$ reads\\
\\
\begin{align}
F(s)&=\frac{1}{72}\ \bigg\{\ \frac{3s^{3}-34s^{2}-28s-24}{s}\ \sqrt{\frac{4+s}{s}}\ \ln\, \bigg(\frac{\sqrt{4+s}-\sqrt{s}}{\sqrt{4+s}+\sqrt{s}}\bigg)\ +\\
\notag&\qquad\qquad+\frac{2(1+s)^{2}}{s^{3}}\ (3s^{3}-20s^{2}+11s-2)\ \ln(1+s)+\\
\notag&\qquad\qquad+(2-3s^2)\ \ln\,s-\frac{4+s}{s}\ (s^2-20s+12)+\\
\notag&\qquad\qquad+\frac{2(1+s)^{2}}{s^2}\ (s^2-10s+1)+\frac{2}{s^2}+2-s^2\ \bigg\}+\frac{5}{8s}
\end{align}\\
\\
Again, the constant in parentheses in \eqref{gluf} depends on the renormalization scheme. The counterterm needed to remove the divergence in \eqref{glud} is\\
\\
\begin{equation}\label{ctgl}
\delta Z_{A}=-\frac{52\alpha}{81\epsilon}
\end{equation}\\
\\
If we define the gluon dressing function $J(p)$ as\\
\\
\begin{align}
J(p)=p^{2}\ \widetilde{D}(p)
\end{align}\\
\\
where $\widetilde{D}(p)$, apart from the tensorial factor $\delta_{ab}\ t^{\mu\nu}(p)$, is the dressed gluon propagator in the Landau gauge, by summing the 1PI diagrams together with the counterterm \eqref{ctgl} we find that\\
\\
\begin{equation}
J(s)^{-1}=1+\frac{4\alpha}{9}\ \bigg[\, F(s)-\frac{197}{108}-\frac{13}{18}\ \ln\,\frac{\mu^{2}}{m^{2}}\ \bigg]
\end{equation}
modulo a constant proportional to $\alpha$. As in 2.2.1, the constant is automatically included if we express the inverse dressing function in the general form\\
\\
\begin{equation}\label{J}
J(s)^{-1}=J(s_{0})^{-1}+\frac{4\alpha}{9}\ \big(F(s)-F(s_{0})\big)
\end{equation}\\
\\
In the limit $s\to 0$, $F(s)$ goes to infinity as $s^{-1}$:\\
\\
\begin{equation}
\lim_{s\to 0}\ F(s)\ =\ \lim_{s\to 0}\ \frac{5}{8s}
\end{equation}\\
\\
It then follows that the MSPE dressed propagator is non-singular in the limit $p^{2}\to0$:\\
\\
\begin{equation}\label{gen}
\lim_{s\to 0}\ J(s)\ =\ \lim_{s\to 0}\ \frac{18s}{5\alpha}\qquad\Longrightarrow\qquad\widetilde{D}(0)^{-1}\ =\ \frac{5\alpha m^{2}}{18}
\end{equation}\\
\\
Eq.~\eqref{gen} shows that the MSPE is not subject to the singularity contraints of the MLPE and proves that the phenomenon of mass generation can in fact be described by a non-standard perturbative expansion of Yang-Mills theory.\\
\\

\clearpage

\addcontentsline{toc}{section}{2.3 Comparison with lattice data: $d=4$, N~$=3$}  \markboth{2.3 Comparison with lattice data: $d=4$, N~$=3$}{2.3 Comparison with lattice data: $d=4$, N~$=3$}
\section*{2.3 Comparison with lattice data: $\boldsymbol{d=4}$, N~$\boldsymbol{=3}$\index{Comparison with lattice data: $d=4$, N~$=3$}}

The propagators of the last section can be compared with the lattice data of sec. 1.2 by noticing that eqq.~\eqref{chi} and \eqref{J} predict a functional relation of the form\\
\\
\begin{align}
[Z_{L}\chi(s)]^{-1}=L(s)+L_{0}\\
\notag\\
[Z_{F}J(s)]^{-1}=F(s)+F_{0}
\end{align}\\
\\
between the functions $L(s)$ and $F(s)$ and the respective dressing functions, where $Z_{L}$, $Z_{F}$, $L_{0}$ and $F_{0}$ are dimensionless constants. In the expressions given above, $Z_{L}$ and $Z_{F}$ play the role of field strength renormalization factors and are thus as much dependent on the subtraction point chosen for the renormalization as the constants $L_{0}$ and $F_{0}$. Since we have not imposed any explicit renormalization condition on our expressions, in order to meaningfully relate our propagators to the lattice data we must regard $Z_{L}$ and $Z_{F}$ as independent constants. This approach has the additional advantage of hiding the dependence on the coupling, so that no specific value needs to be chosen for $\alpha$.\\
In Figg. 7 and 8 lattice data points for the ghost dressing function and the dressed gluon propagator are plotted as a function of momentum together with their MSPE counterpart. The values of $m$, $Z_{L}$, $Z_{F}$, $L_{0}$ and $F_{0}$ that best reproduce the lattice data are shown in Table 1. The MSPE propagators are found in very good agreement with the lattice data of ref.\cite{bogolubsky} if one fixes the momentum scale by setting $m$=0.73 MeV.\\
\\
\\
\\
\begin{figure}[H]
\centering
\includegraphics[angle=270, origin=c,width=0.65\textwidth]{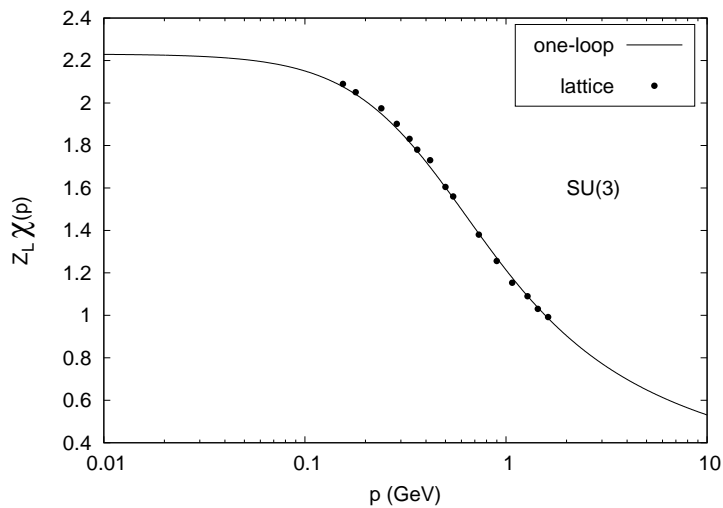}
\vspace{-8mm}
\caption{Lattice data points for the ghost dressing function $\chi(p)$ and its MSPE counterpart as a function of momentum normalized by the constant $Z_{L}$ (N~$=3$).}
\end{figure}\
\\
\begin{figure}[H]
\centering
\includegraphics[angle=270, origin=c,width=0.65\textwidth]{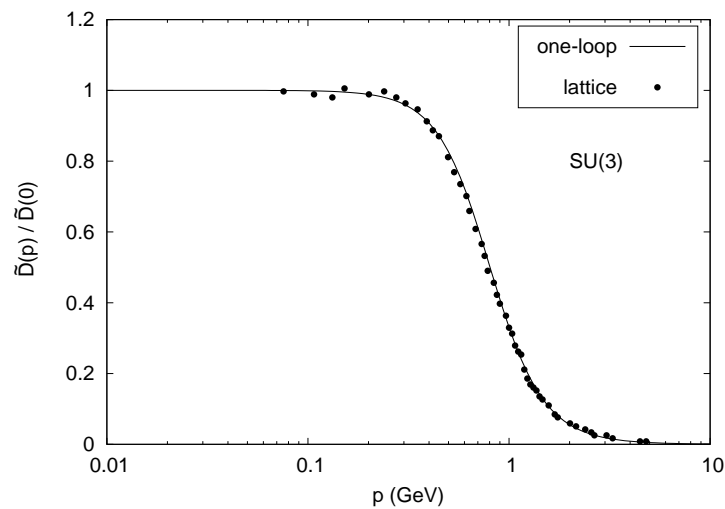}
\vspace{-8mm}
\caption{Lattice data points for the dressed gluon propagator $\widetilde{D}(p)$ and its MSPE counterpart as a function of momentum normalized by the MSPE limiting value $\widetilde{D}(0)$ (N~$=3$).}
\end{figure}\
\\
\\
\\
\\
\begin{table}[H]
\centering
\begin{tabular}{ccccc}
\hline\hline
$m$&$Z_{L}$ & $L_{0}$ & $Z_{F}$ & $F_{0}$\\ \hline
\\ \hline
0.73 GeV&0.637&0.24&0.30&-1.05\\ \hline
\end{tabular}
\vspace{3mm}
\caption{Values of the renormalization constants $Z_{L}$, $Z_{F}$, $L_{0}$ and $F_{0}$.}
\end{table}\
\\
\\
\\

\clearpage


\stepcounter{count}
\addcontentsline{toc}{chapter}{3 The Gaussian Effective Potential: mass generation and deconfinement in $\boldsymbol{d=4}$ Yang-Mills theory}  \markboth{3 The Gaussian Effective Potential: mass generation and deconfinement in d=4 Yang-Mills theory}{3 The Gaussian Effective Potential: mass generation and deconfinement in d=4 Yang-Mills theory}
\chapter*{3\protect \\
\medskip{}
The Gaussian Effective Potential: mass generation and deconfinement in $\boldsymbol{d=4}$ Yang-Mills theory\index{The Gaussian Effective Potential: mass generation and deconfinement in d=4 Yang-Mills theory}}

In Chapter 2 we saw that a massive perturbative expansion of Yang-Mills theory correctly reproduces the infrared behaviour of the vacuum gluon and ghost propagators. There the introduction of a mass parameter $m$ was justified only a posteriori by the suitability of the expansion, and its actual value could not be predicted from first principles, since in the vacuum YMT is free of any mass-scale. In this chapter we will present evidence for mass generation and deconfinement in pure Yang-Mills theory, as obtained from a GEP analysis in the thermal formalism. The Gaussian Effective Potential (GEP) \cite{stevenson}-\cite{stevenson86} is a simple variational tool that allows one to find the best zeroth order approximation to the free energy density of a system. Once such an approximation is found, one can go on and compute the successive orders in the resulting perturbative series. In the thermal formalism, we are free to choose a temperature-dependent mass parameter $m(T)$ for the MSPE. In principle, the mass parameter can be chosen arbitrarily; in practice, one should choose for $m(T)$ a value that minimizes the contribution of the higher order terms, so as to optimize the finite order truncation of the perturbative series. In this respect, the GEP approach offers us a criterion for such a choice.\\
In the limit $T\to 0$, the value $m_{0}=m(0)$ that optimizes the perturbative series will be shown to depend on the coupling constant $\alpha_{s}=g^{2}/4\pi$ and on an unknown mass scale $\Lambda$ that arises from the renormalization of the divergent one-loop integrals. If we take $\Lambda$ to be different from zero, it follows that $m_{0}$ too is different from zero. This may be interpreted as evidence for mass generation in pure Yang-Mills theory. At finite temperature, one may trade the dependence of the GEP on the parameters $(\Lambda,\alpha_{s})$, with a dependence on the parameters $(m_{0},\alpha_{s})$. Both $m_{0}$ and $\alpha_{s}$ have been estimated in \cite{siringo1}; $m_{0}$ was found to be equal to 0.73 GeV, while in the IR $\alpha_{s}$ was found to lie in the range $[\,0.4,\,1.2\,]$. As we will see, the optimal mass parameter $m(T)$ turns out to be discontinuous at a temperature $T_{c}\approx 0.35\, m_{0}$. This leads to a modest discontinuity in the entropy of the system, which in turn signals the presence of a (weakly) first-order phase transition at $T=T_{c}$. If we take $m_{0}$ to be equal to 730 MeV and $\alpha_{s}$ to lie in the physical range $[\,0.4,\,1.2\,]$, a critical temperature of approximately 255 MeV is recovered. The latent heat of the transition can also be estimated, and is found to be approximately equal to 1.8 $T_{c}^{4}$. These results are in good agreement with lattice computations \cite{lucini}-\cite{boyd96}, which show that a weakly first-order deconfinement transition is found in pure YMT at a critical temperature of approximately 270 MeV and with a latent heat of 1.3 - 1.5 $T_{c}^{4}$. Part of the contents of this chapter has been presented in \cite{comitini}.\\
\\
This chapter is organized as follows. The definition of the GEP and its relationship with the Jensen-Feynman inequality (see Appendix C) are reviewed in section 3.1. In section 3.2 we define and compute in the Landau gauge the GEP of YMT in $d=4$. Due to issues between the Jensen-Feynman inequality and the presence of anticommuting fields in the action of YMT, we will employ a different expansion than that which was used in Chapter 2 to derive the vacuum propagators. All the details are given in sec. 3.2.1. While in the vacuum the GEP analysis can be carried out analytically, at finite temperature the potential involves quantities that can be evaluated only up to a one-dimensional integration, which must be carried out numerically. Expressions for such integrals are derived in Appendix D, while numerical tables are presented in Appendix E. In section 3.3 we carry out the GEP analysis and discuss the issue of mass generation and deconfinement in YMT. A brief review of the thermal formalism and of the computational tools needed to derive the GEP are given in Appendix A and B.\\

\clearpage

\addcontentsline{toc}{section}{3.1 The Gaussian Effective Potential}  \markboth{3.1 The Gaussian Effective Potential}{3.1 The Gaussian Effective Potential}
\section*{3.1 The Gaussian Effective Potential\index{The Gaussian Effective Potential}}

In a thermodynamical setting, the Gaussian Effective Potential (GEP) \cite{stevenson}-\cite{stevenson86} is loosely defined as the free energy density of a physical system, computed to first order in its interactions. The motivation that underlies this definition is given by the Jensen-Feynman inequality, of which we give formal proof in Appendix C. In this section will briefly review the significance of the Jensen-Feynman inequality and its connection to the GEP formalism. The latter will be generalized in the next section to YMT.\\
\\
\\
Let $\mathcal{I}=\mathcal{I}_{0}+\mathcal{I}_{int}$ be the thermal action of a field theory whose degrees of freedom are described in terms of a set of commuting fields $\mathscr{F}^{I}$. Supposing that $\mathcal{I}_{0}$ can be expressed as a positive-definite functional on the space of field variables, the Jensen-Feynman inequality states that\\
\\
\begin{equation}\label{487}
\F\leq\F_{0}+\frac{T}{V}\ \int\D\mathscr{F}^{I}\ e^{-\mathcal{I}_{0}}\ \mathcal{I}_{int}
\end{equation}\\
\\
where $T$ and $V$ are the temperature and spatial volume of the system,\\
\\
\begin{equation}\label{479}
\F=-\frac{T}{V}\ \ln\ \int\D\mathscr{F}^{I}\ e^{-\mathcal{I}}
\end{equation}\\
\\
is the free energy density of the system and\\
 \\
\begin{equation}
\F_{0}=-\frac{T}{V}\ \ln\ \int\D\mathscr{F}^{I}\ e^{-\mathcal{I}_{0}}
\end{equation}\\
\\
is the zeroth-order approximation to $\F$ given by the kinetic action $\mathcal{I}_{0}$. Let us restrict ourselves to $\mathcal{I}_{0}$'s which are Gaussian in the field variables. It may happen, e.g. as a consequence of a specific choice of the split $\mathcal{I}=\mathcal{I}_{0}+\mathcal{I}_{int}$ or due to a change of variables of integration in the exact free energy density \eqref{479}, that the approximation given by the RHS of eq.~\eqref{487} depends on some set of free parameters $\{\lambda\}$. In this case we may define a $\lambda$-dependent Gaussian Effective Potential $\F_{G}(\lambda,T)$ as\\
\\
\begin{equation}
\F_{G}(\lambda,T)=\F_{0}(\lambda,T)+\frac{T}{V}\ \int\D\mathscr{F}^{I}\ e^{-\mathcal{I}_{0}(\lambda,T)}\ \mathcal{I}_{int}(\lambda,T)
\end{equation}\\
\\
The Jensen-Feynman inequality then simply states that, for any value of $\lambda$ and $T$,\\
\\
\begin{equation}
\F(T)\leq\F_{G}(\lambda,T)
\end{equation}
If for fixed $T$ the GEP has a global minimum for some $\lambda=\overline{\lambda}(T)$, the Jensen-Feynman inequality tells us that\\
\\
\[
\F(T)\leq\F_{G}(\overline{\lambda}(T),T)\leq \F_{G}(\lambda,T)
\]\\
\\
meaning that the best approximation to $\mathcal{F}(T)$ is obtained if one evaluates $\F_{G}(\lambda,T)$ for $\lambda=\overline{\lambda}(T)$. We then expect a finite-order perturbative expansion of $\F(T)$ with $\mathcal{F}_{0}(\overline{\lambda}(T))$ as its zeroth order to yield a better approximation than that obtained for general $\lambda$'s.\\
\\
The free parameters of the GEP are usually taken to be a mass parameter $m$, introduced through a shift of the kinetic action similar to that operated in Chapter 2, and the thermal average $\langle\mathscr{F}^{I}\rangle$ of the fields $\mathscr{F}^{I}$, introduced through a change of variables of integration $\mathscr{F}^{I}\to \langle\mathscr{F}^{I}\rangle+\mathscr{F}^{I}$ in the path integral \eqref{479}. In the case of Yang-Mills theory, such an average takes the form $\langle A_{\mu}^{a}\,\rangle$. Since Lorentz and gauge-invariance are most likely to constrain the latter to be equal to zero, we may focus our attention on a GEP that depends only on the mass parameter. By minimizing the GEP with respect to $m$ (or, equivalently, with respect to $m^{2}$) for each value of the temperature, we will find a temperature-dependent optimized $m(T)$ which can then be used in the shifted kinetic action of a thermal MSPE.\\

\clearpage{}

\addcontentsline{toc}{section}{3.2 The GEP in Yang-Mills theory}  \markboth{3.2 The GEP in Yang-Mills theory}{3.2 The GEP in Yang-Mills theory}
\section*{3.2 The GEP in Yang-Mills theory\index{The GEP in Yang-Mills theory}}

\addcontentsline{toc}{subsection}{3.2.1 Definition and discussion}  \markboth{3.2.1 Definition and discussion}{3.2.1 Definition and discussion}
\subsection*{3.2.1 Definition and discussion\index{Definition and discussion}}

As reviewed in Appendix A, the free energy density of Yang-Mills theory at temperature $T=\beta^{-1}$ can be expressed as\\
\\
\begin{equation}\label{409}
\F=-\frac{T}{V}\ \ln\, \mathcal{Z}
\end{equation}
\\
where $\Z$ is the thermal partition function\footnote{\ Notice that in our convention the exponential of the action is defined with a minus sign.}\\
\\
\begin{equation}\label{408}
\Z=\int_{\text{per.}}\mathcal{D}A_{\mu}^{a}\,\D\cbar^{a}\,\D c^{a}\ e^{-\mathcal{S}^{th}}
\end{equation}\\
\\
In eq.~\eqref{408}, $\BS^{th}$ is the thermal action of YMT,\\
\\
\begin{equation}
\mathcal{S}^{th}=\int_{0}^{\beta}d\tau\int d^{d-1}x\ \ \ \frac{1}{4}\ \delta^{\mu\sigma}\delta^{\nu\lambda}\ F_{\mu\nu}^{a}\,F_{\sigma\lambda}^{a}+\frac{1}{2\xi}\ \big(\delta^{\mu\nu}\partial_{\mu}A_{\nu}^{a}\big)^{2}+\delta^{\mu\nu}\partial_{\mu}\cbar^{a}\,D_{\nu}^{ab}c^{b}
\end{equation}\\
\\
($\partial/\partial x^{0}=\partial/\partial \tau$) and the subscript ``per.'' reminds us that we are to functionally integrate over field configurations which are periodic in the variable $\tau$. The thermal GEP $\F_{G}$ of YMT is then defined as\\
\\
\begin{equation}\label{407}
\F_{G}=-\frac{T}{V}\ \ln\, \mathcal{Z}_{0}'+\frac{T}{V}\ \int_{\text{per.}}\mathcal{D}A_{\mu}^{a}\,\D\cbar^{a}\,\D c^{a}\ e^{-\mathcal{S}_{0}^{th\,'}}\ \BS_{int}^{th\,'}
\end{equation}\\
\\
where\\
\begin{equation}
\mathcal{Z}_{0}'=\int_{\text{per.}}\mathcal{D}A_{\mu}^{a}\,\D\cbar^{a}\,\D c^{a}\ e^{-\mathcal{S}_{0}^{th\,'}}
\end{equation}\\
\\
gives the zeroth order approximation to $\F$ and\\
\\
\begin{equation}
\BS_{0}^{th\,'}=\BS_{0}^{th}+\delta\BS^{th}\quad;\qquad\qquad \BS_{int}^{th\,'}=\BS_{int}^{th}-\delta\BS^{th}
\end{equation}\\
\\
are shifted kinetic and interaction terms. As anticipated in the introduction to this chapter, for our GEP analysis we will choose a different shift than that which was used in Chapter 2 to derive the vacuum propagators: given the thermal Fourier expansions \eqref{349}-\eqref{350} of Appendix A, we define a new shifted kinetic action $\BS_{0}^{th\,'}$ as
\\
\begin{align}
\\
\notag \BS_{0}^{th\,'}&=V\ \sum_{n}\int \frac{d^{d-1}K}{(2\pi)^{d-1}}\ \ \bigg\{\frac{1}{2}\ A_{\mu}^{a}(K)\ \beta^{2}\, \Big[\ D^{\mu\nu}_{m\perp\,ab}(K)^{-1}+D^{\mu\nu}_{m\parallel\, ab}(K)^{-1}\ \Big]\ A_{\nu}^{b}(K)^{*}+\\
\notag&\qquad\qquad\qquad\qquad +\cbar^{a}(K)\,\beta^{2}\, G_{m\,ab}(K)^{-1}\,c^{b}(K)\bigg\}
\end{align}\\
\\
where\\
\\
\begin{equation}
D^{\mu\nu}_{m\perp\,ab}(K)=\delta_{ab}\ \frac{t^{\mu\nu}_{E}(K)}{K^{2}+m^{2}}=\delta_{ab}\ \frac{\delta^{\mu\nu}-K^{\mu}K^{\nu}/K^{2}}{K^{2}+m^{2}}
\end{equation}\\
\begin{equation}
D^{\mu\nu}_{m\parallel\,ab}(K)=\xi\ \delta_{ab}\ \frac{\ell^{\mu\nu}_{E}(K)}{K^{2}+m^{2}}=\xi\ \delta_{ab}\ \frac{K^{\mu}K^{\nu}/K^{2}}{K^{2}+m^{2}}
\end{equation}\\
\begin{equation}
G_{m\,ab}(K)=\delta_{ab}\ \frac{1}{K^{2}+m^{2}}
\end{equation}\\
\\
are massive euclidean free particle propagators. The counterterm $\delta\BS^{th}$ that produces the shift is
\\
\begin{align}
\\
\notag \delta\BS^{th}&=V\ \sum_{n}\int \frac{d^{d-1}K}{(2\pi)^{d-1}}\ \ \bigg\{\frac{1}{2}\ A_{\mu}^{a}(K)\ \beta^{2}\, \Big[\ m^{2}\ t^{\mu\nu}_{E}(K)+m^{2}\ \xi^{-1}\ \ell^{\mu\nu}_{E}(K)\ \Big]\ A_{\nu}^{a}(K)^{*}+\\
\notag&\qquad\qquad\qquad\qquad +\cbar^{a}(K)\,\beta^{2}\, m^{2}\,c^{a}(K)\bigg\}
\end{align}\\
\\
In the previous equations, $m^{2}$ is the variational parameter with respect to which we will minimize the GEP.\\
\\
The expressions given above treat the ghosts and longitudinal modes of the gluons, as well as their transverse modes, as massive fields. The choice we made is suggested by physical considerations. First of all, we have to deal with the well known fact that the Jensen-Feynman inequality may not hold in the presence of anticommuting fields \cite{ibanez}. In order to address this matter, we notice that the ghosts of any gauge theory are not ordinary anticommuting fields, in that they serve to cancel the unphysical d.o.f. contained in the gauge-fixed action. As the Jensen-Feynman inequality certainly holds in any ghost-free gauge, we expect it to hold as well in some form in the linear covariant gauges. In this respect, we believe that at the GEP level of approximation ghosts with the same mass as the transverse modes may be more effective in cancelling the contributions due to the unphysical degrees of freedom, thus leading to a more suitable potential. If we assume this to be the case, we must also give the longitudinal modes of the gluons a mass $m$, for otherwise there would still be a mismatch between the mass of one gluonic d.o.f. and that of a ghost d.o.f.. For example, it is easy to see that only in the MSPE scheme presented above the cancellation mechanism works at the zeroth order approximation to $\F$, namely, at the ideal gas level: the transverse modes contribute with $N_{A}(d-1)$ massive ideal gas terms while due to their anticommuting nature the ghosts contribute with $-2N_{A}$ such terms; only by giving mass to the longitudinal modes one can obtain the correct d.o.f. count of $N_{A}(d-2)$ massive gluon modes. The introduction of a longitudinal mass may of course seem somewhat unnatural, as gauge invariance \cite{itzykson} forbids both non-perturbatively and perturbatively in the MLPE the dressing of the longitudinal propagator and the shift of its mass pole. However, it is important to realize that the same need not be true with regard to a non-perturbative approximation like the one given by the GEP, which still needs to be improved by adding successive corrections.\\
The second reason for our choice is given by the fact that a MSPE that treats the ghosts and the longitudinal gluons as massless would lead in the GEP approximation to a negative entropy at low temperatures\footnote{\ This has been verified through numerical computations which we won't present for reasons of brevity. Negative entropy in temperature intervals below the critical temperature seems to be a common trait of massive perturbative expansions in the Landau gauge, see however ref.~\cite{tissier16}.}. While this is not by itself a sufficient reason to discard such a potential -- again, the GEP needs higher order contributions in order to well approximate the exact results --, we expect the GEP approach to yield better results if unphysical features such as a negative entropy do not make comparison in our computations.\\
\\
At first order in the MSPE interaction, the GEP receives contributions from the diagrams shown in Fig. 9, where the loops with no vertices represent the zeroth order logarithm and the crosses represent the mass counterterms. The vertices needed for its evaluation are  shown in Fig. 10.\\
\\
\\
\begin{figure}[H]
\centering
\includegraphics[width=0.8\textwidth]{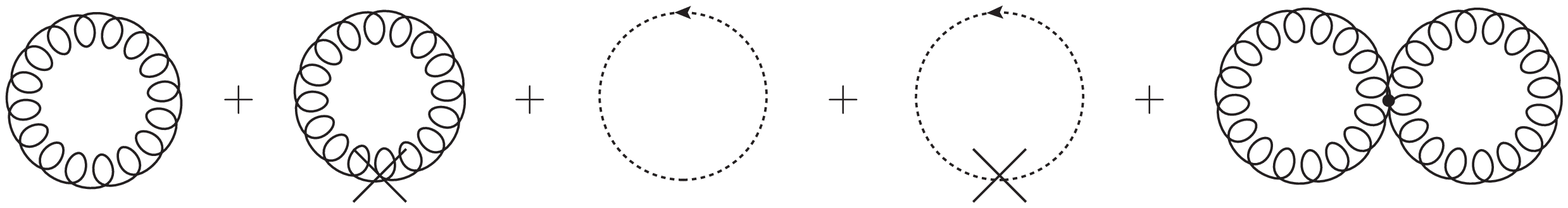}
\vspace{5mm}
\caption{Diagrams that contribute to the GEP. The crosses represent the mass counterterms.}
\end{figure}\
\begin{figure}[H]
\centering
\includegraphics[width=0.78\textwidth]{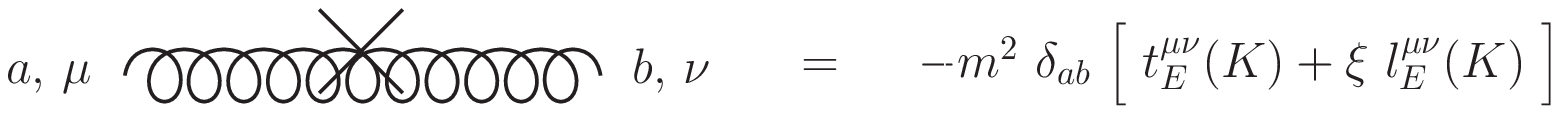}
\vspace{3mm}
\caption{Gluon mass counterterm.}
\end{figure}
\newpage
\
\begin{figure}[H]
\centering
\includegraphics[width=0.48\textwidth]{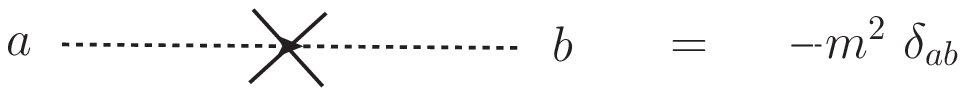}
\vspace{3mm}
\caption{Ghost mass counterterm.}
\end{figure}\
\\
\\
\\
\begin{figure}[H]
\centering
\includegraphics[width=0.78\textwidth]{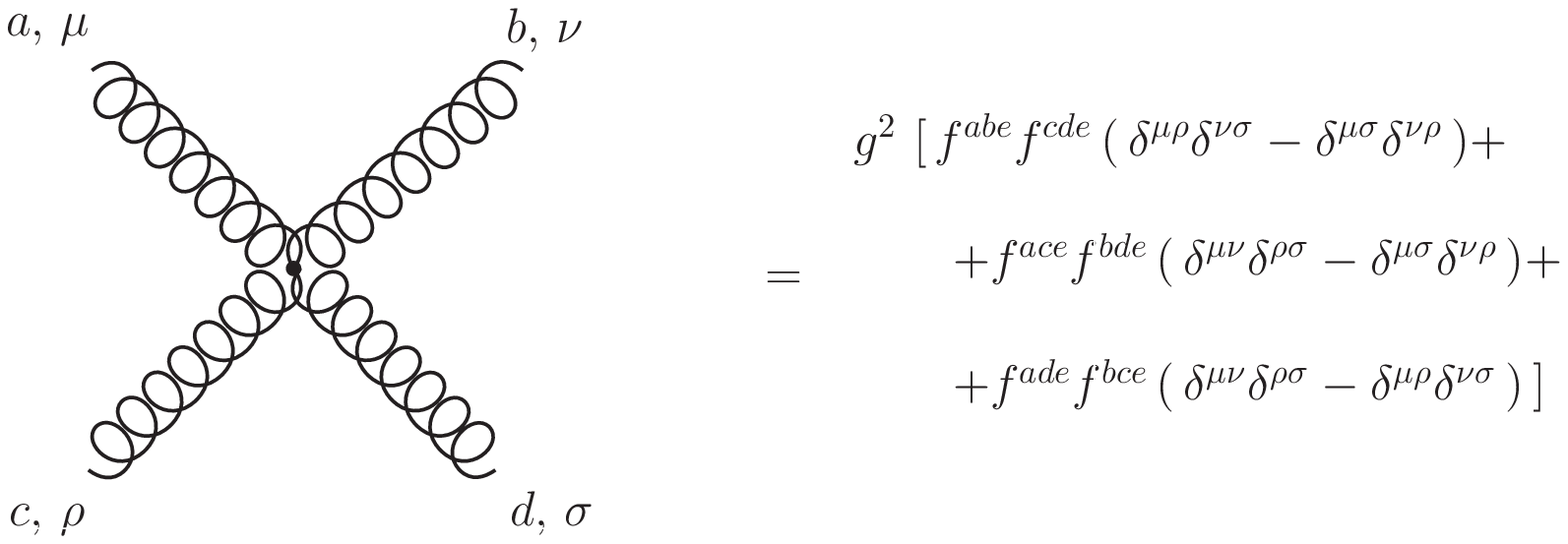}
\vspace{5mm}
\caption{4-gluon vertex.}
\end{figure}\
\\
\\
\\
\\
\\
\addcontentsline{toc}{subsection}{3.2.2 Computation of the GEP in the Landau gauge}  \markboth{3.2.2 Computation of the GEP in the Landau gauge}{3.2.2 Computation of the GEP in the Landau gauge}
\subsection*{3.2.2 Computation of the GEP in the Landau gauge\index{Computation of the GEP in the Landau gauge}}

Let us write\\
\\
\begin{equation}
\mathcal{F}_{G}=\mathcal{F}_{0}+\mathcal{F}_{11}+\mathcal{F}_{12}
\end{equation}
where
\\
\begin{equation}
\mathcal{F}_{0}=-\frac{T}{V}\ \ln\, \mathcal{Z}_{0}
\end{equation}\\
\begin{equation}
\mathcal{F}_{11}=\frac{T}{V}\ \big\langle-\delta\BS^{th}\,\big\rangle
\end{equation}\\
\begin{equation}
\mathcal{F}_{12}=\frac{T}{V}\ \big\langle\,\BS^{th}_{A^{4}}\, \big\rangle
\end{equation}
With reference to Fig. 9, $\F_{0}$, $\F_{11}$ and $\F_{12}$ are represented respectively by the loops with no vertices, the loops with mass counterterms and the double loop.\\
\\
The evaluation of $\mathcal{F}_{0}$ and $\mathcal{F}_{11}$ is straightforward. In a general linear covariant gauge we have, modulo an inessential constant,
\\
\begin{align}
\\
\notag\ln\,\mathcal{Z}_{0}= -\frac{1}{2}\ \ln\ \det\,\Big[\ \beta^{2}\, \Big(\ D^{\mu\nu}_{m\perp\,ab}(K)^{-1}+D^{\mu\nu}_{m\parallel\, ab}(K)^{-1}\Big)\Big]+\ln\ \det\ \big(\beta^{2}\, G_{m\,ab}(K)^{-1}\big)
\end{align}\\
\\
By using the matrix identity $\ln\, \det\, M=\text{Tr}\, \ln\, M$, the ghost determinant can be put in the form
\\
\begin{align*}
&\ln\ \det\,\Big(\ \beta^{2}\, G_{m\,ab}(K)^{-1}\ \Big)=\text{Tr}\ \ln\ \Big(\ \beta^{2}\, G_{m\,ab}(K)^{-1}\ \Big)=\\
&=N_{A}V\ \sum_{n}\ \int\frac{d^{d-1}K}{(2\pi)^{d-1}}\ \ln\,\beta^{2}(K^{2}+m^{2})
\end{align*}\\
\\
The gluonic determinant splits into the product of determinants on the transverse and longitudinal subspaces; using again $\ln\, \det\, M=\text{Tr}\, \ln\, M$, modulo a $\xi$-dependent inessential constant,\\
\\
\begin{align*}
&\ln\ \det\,\Big[\ \beta^{2}\, \Big(\ D^{\mu\nu}_{m\perp\,ab}(K)^{-1}+D^{\mu\nu}_{m\parallel\, ab}(K)^{-1}\Big)\Big]=\\
&=\ln\ \det\,\Big[\ \beta^{2}\ D^{\mu\nu}_{m\perp\,ab}(K)^{-1}\ \Big]+\ln\ \det\ \Big[\ \beta^{2}\ D^{\mu\nu}_{m\parallel\, ab}(K)^{-1}\ \Big]=\\
&=\text{Tr}\ \ln\ \Big[\ \beta^{2}\ D^{\mu\nu}_{m\perp\,ab}(K)^{-1}\Big]+\text{Tr}\ \ln\ \Big[\ \beta^{2}\ D^{\mu\nu}_{m\parallel\, ab}(K)^{-1}\ \Big]=\\
&=dN_{A}V\ \sum_{n}\ \int\frac{d^{d-1}K}{(2\pi)^{d-1}}\ \ln\,\beta^{2}(K^{2}+m^{2})
\end{align*}\\
\\
Thus we find
\\
\begin{equation}\label{4929}
\F_{0}=(d-2)N_{A}\, K_{m} 
\end{equation}
\\
where, setting\\
\begin{equation}
\int_{K}\ =\ \ T\ \sum_{n}\int\frac{d^{d-1}K}{(2\pi)^{d-1}}
\end{equation}\\
\\
we have defined\\
\\
\begin{equation}
K_{m}=\frac{1}{2}\ \int_{K}\ln\,(K^{2}+m^{2})
\end{equation}\\
\\
In a general linear covariant gauge, given the vertices in Figg. 10 and 11, the mass counterterm contribution $\F_{11}$ reads\\
\\
\begin{align*}
\F_{11}&=-\frac{1}{2}\ m^{2}\ \int_{K}\ \Big(\ D^{\mu\nu}_{m\perp\,ab}(K)+D^{\mu\nu}_{m\parallel\, ab}(K)\Big)\ \delta^{ab}\ \Big(\ t^{\mu\nu}_{E}(K)+ \xi^{-1}\ \ell^{\mu\nu}_{E}(K)\ \Big)+\\
&\qquad+m^{2}\  \int_{K}\ G_{ab}(K)\ \delta^{ab}=\\
&=-\frac{d N_{A}}{2}\ m^{2}\ \int_{K}\frac{1}{K^{2}+m^{2}}+N_{A}\ m^{2}\ \int_{K}\frac{1}{K^{2}+m^{2}}
\end{align*}\\
\\
We see that, since the longitudinal mass counterterm is proportional to $\xi^{-1}$, even in the Landau gauge the longitudinal modes contribute to $\F_{11}$. If we define\\
\\
\begin{equation}
J_{m}=\int_{K}\frac{1}{K^{2}+m^{2}}
\end{equation}\\
\\
we obtain for $\F_{11}$ the expression\\
\\
\begin{equation}
\F_{11}=-\frac{N_{A}}{2}\, (d-2)\ J_{m}
\end{equation}\\
\\
As for the $\F_{12}$ term, this can be computed by integrating the tadpole diagram Fig. 13 in Matsubara/momentum space together with a gluon propagator. This can be done directly in the Landau gauge, where the tadpole reads\\
\\
\begin{equation}
\text{tadpole}=g^{2}N\ \delta_{ab}\ \Big((d-2)\ J_{m}\ \delta^{\mu\nu}+I_{m}^{\mu\nu}\Big)
\end{equation}
\\
\begin{figure}[H]
\centering
\includegraphics[width=0.25\textwidth]{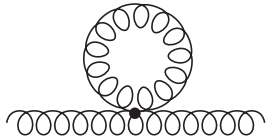}
\vspace{3mm}
\caption{Tadpole diagram.}
\end{figure}\
\\
Here we have defined\\
\\
\begin{equation}
I^{\mu\nu}_{m}=\int_{K}\frac{K^{\mu}K^{\nu}}{K^2(K^{2}+m^{2})}
\end{equation}\\
\\
The tadpole is a constant tensor, thus its contraction with the gluon propagator simply equals the product between the former and the integral of the propagator. The latter is given by\\
\\
\begin{equation}
\int_{K}D_{m\,\perp\,\mu\nu}^{ab}(K)=\delta^{ab}\ \Big(J_{m}\ \delta_{\mu\nu}-I_{m\,\mu\nu}\Big)
\end{equation}\\
\\
Taking into account a symmetry factor of $1/4$ and multiplying by appropriate factors of $\beta$ and $V$, we find that\\
\\
\begin{align}
\F_{12}&=\frac{g^{2}N N_{A}}{4}\ \Big((d-2)\ J_{m}\ \delta^{\mu\nu}+I_{m}^{\mu\nu}\Big)\Big(J_{m}\ \delta_{\mu\nu}-I_{m\,\mu\nu}\Big)=\\
&\notag=\frac{g^{2}N N_{A}}{4}\ \Big([(d-1)(d-2)+1]\ J_{m}^{2}-I^{\mu\nu}_{m}\,I_{m\,\mu\nu}\Big)
\end{align}\\
\\
In Appendix D it is shown that by a suitable definition of the integral $H_{m}$ (see ahead) the product $I^{\mu\nu}_{m}I_{m\,\mu\nu}$ can be put in the form\\
\\
\begin{align}
I^{\mu\nu}_{m}I_{m\,\mu\nu}=\frac{1}{d}\,J_{m}^{2}+\frac{d}{d-1}\,\bigg(\frac{1}{d}\ J_{m}-H_{m}\bigg)^{2}
\end{align}
Then, using the equality\\
\\
\[
(d-1)(d-2)+1-\frac{1}{d}=\frac{(d-1)^{3}}{d}
\]\\
\\
we obtain the following expression for $\F_{12}$:\\
\\
\begin{equation}\label{4928}
\F_{12}=\frac{g^{2}NN_{A}}{4}\ \bigg[\ \frac{(d-1)^{3}}{d}\ J_{m}^{2}-\frac{d}{d-1}\ \bigg(\frac{1}{d}\ J_{m}-H_{m}\bigg)^{2}\ \bigg]
\end{equation}\\
\\
Finally, if we define an effective coupling constant $\alpha$ as\\
\\
\begin{equation}
\alpha\ =\ \frac{(d-1)^{3}N}{2\pi d(d-2)}\ \ \alpha_{s}\ =\ \frac{N(d-1)^{3}}{2\pi d(d-2)}\ \ \frac{g^{2}}{4\pi}
\end{equation}\\
\\
we obtain our final expression for the GEP in the form\\
\\
\begin{equation}\label{GEP}
\mathcal{F}_{G}(m,T)=(d-2)\,N_{A}\ \bigg\{K_{m}-\frac{1}{2}\,m^{2}J_{m}+2\pi^{2} \alpha\ \bigg[\ J_{m}^{2}-\frac{d^{2}}{(d-1)^{4}}\ \bigg(\frac{1}{d}\ J_{m}-H_{m}\bigg)^{2}\ \bigg]\bigg\}
\end{equation}\\
\\
The integrals $K_{m}$, $J_{m}$ and $H_{m}$ are explicitly computed in Appendix D up to a one-dimensional integration which must be carried out numerically for each value of the parameters $m^{2}$ and $T$. It is shown in Appendix B that every sum over the Matsubara frequencies can be decomposed into a vacuum part and a thermal part, the latter being defined by its vanishing for $T\to0$. The thermal contributions to $K_{m}$, $J_{m}$ and $H_{m}$ are given by\\
\\
\begin{equation}\label{INT1}
K_{m}^{th}=-\frac{1}{6\pi^{2}}\ \int_{0}^{+\infty}dk\ \ k^{4}\ \frac{n_{\beta}\big(\varepsilon_{m}(k)\big)}{\varepsilon_{m}(k)}
\end{equation}\\
\\
\begin{equation}
J_{m}^{th}=\frac{1}{2\pi^{2}}\ \int_{0}^{+\infty}dk\ \ k^{2}\ \frac{n_{\beta}\big(\varepsilon_{m}(k)\big)}{\varepsilon_{m}(k)}
\end{equation}\\
\\
\begin{equation}\label{INT3}
H_{m}^{th}=-\frac{3}{m^{2}}\ K_{m}^{th}+ J_{m}^{th}-\frac{\pi^{2}T^{4}}{30m^{2}}
\end{equation}\\
\\
The vacuum contributions to $K_{m}$, $J_{m}$ and $H_{m}$ need to be renormalized. Since the results that we will present in the next section strongly depend on the choice of a renormalization scheme, it is worth to spend some words on this subject. It is easy to see from its definition that the integral $K_{m}$ satisfies the differential equation\\
\\
\begin{equation}\label{DIFF}
\frac{\partial K_{m}}{\partial m^{2}}=\frac{1}{2}\ J_{m}
\end{equation}\\
\\
The latter can be inverted to give a renormalized $K_{m}^{v}=K_{m}\big|_{T=0}$ once a renormalization for $J_{m}^{v}=J_{m}\big|_{T=0}$ is chosen. Moreover, in Appendix D it is shown that $H_{m}^{v}=H_{m}\big|_{T=0}=J_{m}^{v}/d$. Thus it is enough to discuss the renormalization of the integral $J_{m}^{v}$.\\
\\
As shown in Appendix D, $J_{m}^{v}$ is equal to the divergent integral\\
\\
\begin{equation}
J_{m}^{v}=\int\frac{d^{4}K}{(2\pi)^{4}}\ \frac{1}{K^{2}+m^{2}}
\end{equation}\\
\\
As is customary in the treatment of Yang-Mills theory, the integral can be defined in dimensional regularization to yield\\
\\
\begin{align}
J_{m}^{v}&=\lim_{d\to 4}\ \frac{m^{2}}{16\pi^{2}}\ \Gamma(1-d/2)\ \bigg(\frac{m^{2}}{4\pi}\bigg)^{d/2-2}=\\
\notag&=\lim_{d\to 4}\ -\frac{m^{2}}{16\pi^{2}}\ \bigg(\frac{2}{\epsilon}-\gamma_{E}+\ln\,4\pi+1-\ln\, m^{2}\bigg)
\end{align}\\
\\
where $\epsilon=4-d$ and $\gamma_{E}$ is the Euler-Mascheroni constant. Just by dimensional analysis, it is easy to see that any arbitrary renormalization scheme\footnote{\ Here we are implicitly assuming that such a scheme does not modify the dependence of $J_{m}^{v}$ on the mass parameter.} would bring $J_{m}^{v}$ to the form\\
\\
\[
J_{m}^{v}=\frac{m^{2}}{16\pi^{2}}\bigg(\ln\, \frac{m^{2}}{Q^{2}}+A\,\bigg)
\]\\
\\
where $Q$ is a mass scale and $A$ is an adimensional constant. As $Q$ is unknown, we can define a second mass scale $\Lambda$ by setting\\
\\
\[
\ln\,\Lambda^{2}=\ln\,Q^{2}-A
\]
and rewrite $J_{m}^{v}$ as
\\
\begin{align}\label{JMV}
J_{m}^{v}=\frac{m^{2}}{16\pi^{2}}\ \ln\, \frac{m^{2}}{\Lambda^{2}}
\end{align}\\
\\
This is the expression that we will use in our GEP analysis. From eq.~\eqref{DIFF} it follows that, modulo an inessential constant, $K_{m}^{v}$ is renormalized as\\
\\
\begin{align}\label{KMV}
K_{m}^{v}=\frac{m^{4}}{64\pi^{2}}\ \bigg(\ln\, \frac{m^{2}}{\Lambda^{2}}-\frac{1}{2}\bigg)
\end{align}\\
\\
Together with
\\
\begin{equation}
H_{m}^{v}=\frac{1}{4}\ J_{m}^{v}
\end{equation}\\
\\
this completes our list of expressions for the integrals involved in the computation of the thermal GEP.\\

\clearpage{}

\addcontentsline{toc}{section}{3.3 Mass generation and deconfinement in $d=4$ Yang-Mills theory}  \markboth{3.3 Mass generation and deconfinement in $d=4$ Yang-Mills theory}{3.3 Mass generation and deconfinement in $d=4$ Yang-Mills theory}
\section*{3.3 Mass generation and deconfinement in $\boldsymbol{d=4}$ Yang-Mills theory\index{Mass generation and deconfinement in $d=4$ Yang-Mills theory}}

In this section we will carry out the GEP analysis of pure Yang-Mills theory in $d=4$. In 3.3.1 we will take the limit $T\to 0$ and study the extrema of the vacuum GEP $\F_{G}^{v}=\F_{G}(m,T=0)$. As we will see, for any finite value of the coupling constant, $\F_{G}^{v}$ possesses a global minimum $m_{0}$ which is given by the so called mass gap equation. The mass gap equation only tells us that $m_{0}$ is non vanishing: its actual value must come from the phenomenology and can be taken to be of the order of the mass parameter used in Chapter 2 to recover the ghost and gluon propagators, namely 730 MeV. In 3.3.2 we will express the thermal GEP in terms of $m_{0}$ to obtain a full determination of the temperature-dependent mass parameter $m(T)$. Such a determination is carried out numerically for $T$ in the range from zero to $m_{0}$. Our results will be shown to depend on the coupling constant $\alpha_{s}$. In ref.\cite{siringo1} it was found that in the MSPE the IR value of $\alpha_{s}$ lies in the range $[\,0.4,\,1.2\,]$. As we will see, if we limit ourselves to this range the physical predictions of the GEP approach are found not to be very sensitive to the actual value of the coupling.\\

\addcontentsline{toc}{subsection}{3.3.1 The GEP in the vacuum: mass generation}  \markboth{3.3.1 The GEP in the vacuum: mass generation}{3.3.1 The GEP in the vacuum: mass generation}
\subsection*{3.3.1 The GEP in the vacuum: mass generation\index{The GEP in the vacuum: mass generation}}

By taking the limit $d\to 4$, $T\to 0$ of eq.~\eqref{GEP} and recalling that $H_{m}^{v}=J_{m}^{v}/4$, we find the following expression for the vacuum GEP of YMT in four dimensions:\\
\\
\begin{equation}\label{GEPV}
\mathcal{F}_{G}^{v}(m)=\mathcal{F}_{G}(m,T)\Big|_{T=0}=2\,N_{A}\ \bigg\{K_{m}^{v}-\frac{1}{2}\,m^{2}\, J_{m}^{v}+2\pi^{2} \alpha\ J_{m}^{v\,2}\bigg\}
\end{equation}\\
\\
where in $d=4$
\\
\begin{equation}
\alpha=\frac{27N\alpha_{s}}{16\pi}
\end{equation}\\
\\
From the renormalized expressions given in eqq.~\eqref{JMV}-\eqref{KMV}, it is easily seen that each term in eq.~\eqref{GEPV} vanishes in the limit $m\to 0$, so that\\
\\
\begin{equation}
\lim_{m\to 0}\ \mathcal{F}_{G}^{v}(m)=0
\end{equation}\\
\\
Let us study the extrema of $\mathcal{F}_{G}^{v}(m)$. As was noted in the last section, the integral $K_{m}$ satisfies the differential equation\\
\\
\[
\frac{\partial K_{m}}{\partial m^{2}}=\frac{1}{2}\ J_{m}
\]\\
\\
The latter is still verified in the limit $T\to 0$ and can be used  to express the derivative of $\mathcal{F}_{G}^{v}$ in the form\\
\\
\begin{equation}\label{GEPVD}
\frac{\partial \mathcal{F}_{G}^{v}}{\partial m^{2}}=-N_{A}\ \Big(m^{2}-8\pi^{2}\alpha\ J_{m}^{v}\Big)\ \frac{\partial J_{m}^{v}}{\partial m^{2}}
\end{equation}\\
\\
The extremum equation\\
\\
\begin{equation}
\frac{\partial \mathcal{F}_{G}^{v}}{\partial m^{2}}\bigg|_{m^{2}=m^{2}_{0}}=0
\end{equation}\\
\\
has then two solutions. The first one is given by the vanishing of the derivative of $J_{m}^{v}$, namely\\
\\
\begin{equation}
0=\frac{\partial J_{m}^{v}}{\partial m^{2}}=\frac{1}{16\pi^{2}}\ \bigg(\ln\,\frac{m^{2}}{\Lambda^{2}}+1\bigg)\qquad\Longleftrightarrow\qquad m^{2}=\Lambda^{2}/e
\end{equation}\\
\\
for which we find\\
\\
\[
\bigg[\, K_{m}^{v}-\frac{1}{2}\,m^{2}\,J_{m}^{v}\, \bigg]_{m^{2}=\Lambda^{2}/e}=\quad \frac{m^{4}}{128\pi^{2}}\bigg|_{m^{2}=\Lambda^{2}/e}>0
\]\\
\\
Since the $\alpha$-dependent term in eq.~\eqref{GEPV} is non-negative and $\mathcal{F}_{G}^{v}\big|_{m^{2}=0}=0$, the inequality\\
\\
\begin{equation}\label{maxineq}
\mathcal{F}_{G}^{v}\big|_{m^{2}=\Lambda^{2}/e}\ >\ \mathcal{F}_{G}^{v}\big|_{m^{2}=0}
\end{equation}\\
\\
holds. The second extremum, which we will denote $m_{0}$, is found by solving the so called mass gap equation,\\
\\
\begin{equation}
m^{2}_{0}=8\pi^{2}\alpha\,J_{m_{0}}^{v}
\end{equation}\\
\\
In our renormalization scheme, the latter is equivalent to\\
\\
\begin{equation}
1=\frac{\alpha}{2}\ \ln\,\frac{m^{2}_{0}}{\Lambda^{2}}\qquad\Longleftrightarrow\qquad m_{0}=\Lambda\ e^{1/\alpha}
\end{equation}\\
\\
By plugging $m_{0}=\Lambda\ e^{1/\alpha}$ back into eq.~\eqref{GEPV} evaluated at $m=m_{0}$ we find that\\
\\
\begin{equation}\label{minineq}
\mathcal{F}_{G}^{v}(m_{0})=-\frac{N_{A}\,m_{0}^{4}}{64\pi^{2}}\quad <\quad 0=\mathcal{F}_{G}^{v}(0)
\end{equation}\\
\\
For any value of $\alpha$, $m_{0}$ falls to the right of the extremum $m=\Lambda/\sqrt{e}$:\\
\\
\begin{equation}
m_{0}>\Lambda/\sqrt{e}
\end{equation}\\
\\
The inequality \eqref{maxineq} then tells us that $m=\Lambda/\sqrt{e}$ is a local maximum; it follows that $m=m_{0}$ is a local minimum, which is also global due to the inequality \eqref{minineq}. This behaviour is illustrated in Fig. 14, where an adimensionalized version of $\F_{G}^{v}$ is plotted as a function of $m/m_{0}$ for those values of $\alpha$ which correspond to $\alpha_{s}$ in the physical range $[\,0.4,\, 1.2\,]$.\\
\\
\\
\\
\\
\begin{figure}[H]
\centering
\includegraphics[angle=270, origin=c,width=0.7\textwidth]{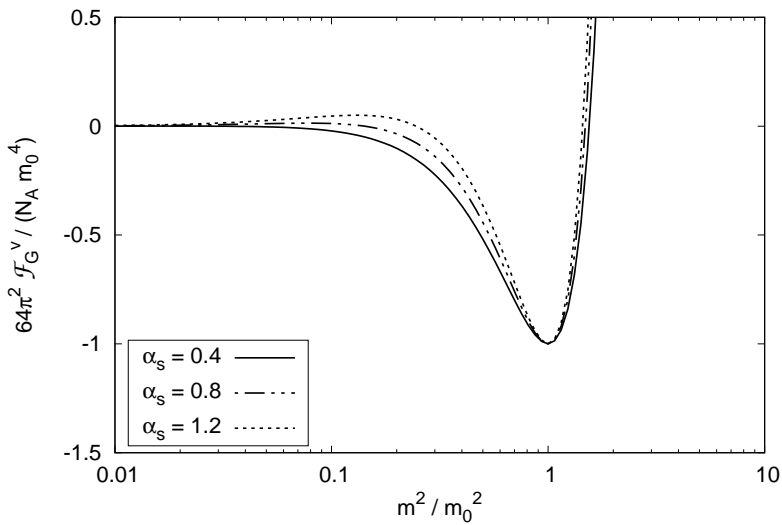}
\vspace{-12mm}
\caption{The vacuum GEP as a function of the mass parameter for $\alpha_{s}\in[\,0.4,\,1.2\,]$.}
\end{figure}\
\\
\\
\\
As was anticipated in the introduction to this chapter, finding a minimum for the vacuum GEP is not enough to give us an explicit value for $m_{0}$: since the mass-scale $\Lambda$ comes from renormalization, its value is unknown and cannot be used to determine $m_{0}$. In the vacuum, the mass gap equation only tells us that for any reasonable (i.e. non-zero) value of $\Lambda$ the optimal mass parameter $m_{0}$ is different from zero. We interpret this result as evidence for mass generation in $d=4$ pure Yang-Mills theory. On the other hand, the mass gap equation itself may be used to rewrite the vacuum contribution to the GEP as a function of $m_{0}$ and $\alpha$, rather than of $\Lambda$ and $\alpha$. By doing so, we obtain the following expression for $\F_{G}^{v}(m)$:\\
\\
\begin{equation}\label{GEPVM0}
\F_{G}^{v}(m)=\frac{N_{A}\,m^{4}}{64\pi^{2}}\ \bigg\{\,\alpha\, \bigg(\ln\,\frac{m^{2}}{m_{0}^{2}}\,\bigg)^{2}+2\,\ln\,\frac{m^{2}}{m_{0}^{2}}-1\,\bigg\}
\end{equation}\\
\\
Eq.~\eqref{GEPVM0} of course still holds at finite temperature. In the next section it will be used to obtain a full determination of the temperature-dependent mass parameter $m(T)$ in terms of the optimal vacuum mass parameter $m_{0}$.\\

\addcontentsline{toc}{subsection}{3.3.2 The GEP at finite temperature: deconfinement}  \markboth{3.3.2 The GEP at finite temperature: deconfinement}{3.3.2 The GEP at finite temperature: deconfinement}
\subsection*{3.3.2 The GEP at finite temperature: deconfinement\index{The GEP at finite temperature: deconfinement}}

The mass gap equation allows us to drop the dependence on the unknown mass scale $\Lambda$ and express the thermal GEP in terms of the vacuum mass parameter $m_{0}$. A straightforward calculation shows that at finite temperature $\F_{G}$ can be put in the form\\
\\
\begin{align}\label{GEPT}
&\quad\ \F_{G}(m,T)=\\
&\notag=\ \F_{G}^{v}(m)+2N_{A}\ \bigg\{\ K_{m}^{th}+\frac{\alpha m^{2}}{4}\ J_{m}^{th}\ \ln\,\frac{m^{2}}{m_{0}^{2}}+2\pi^{2}\alpha\ \bigg[\ J_{m}^{th\,2}-\frac{16}{81}\ \bigg(\frac{1}{4}\,J_{m}^{th}-H_{m}^{th}\bigg)^{2}\ \bigg]\bigg\}
\end{align}\\
\\
where $\F_{G}^{v}$ is given by eq.~\eqref{GEPVM0} and $K_{m}^{th}$, $J_{m}^{th}$ and $H_{m}^{th}$ are given by eqq.~\eqref{INT1}-\eqref{INT3}.\\
\\
\\
\\
\begin{figure}[H]
\centering
\begin{subfigure}[H]{0.45\textwidth}
\includegraphics[trim={0 85 0 100}, clip, angle=270, origin=c,width=1\textwidth]{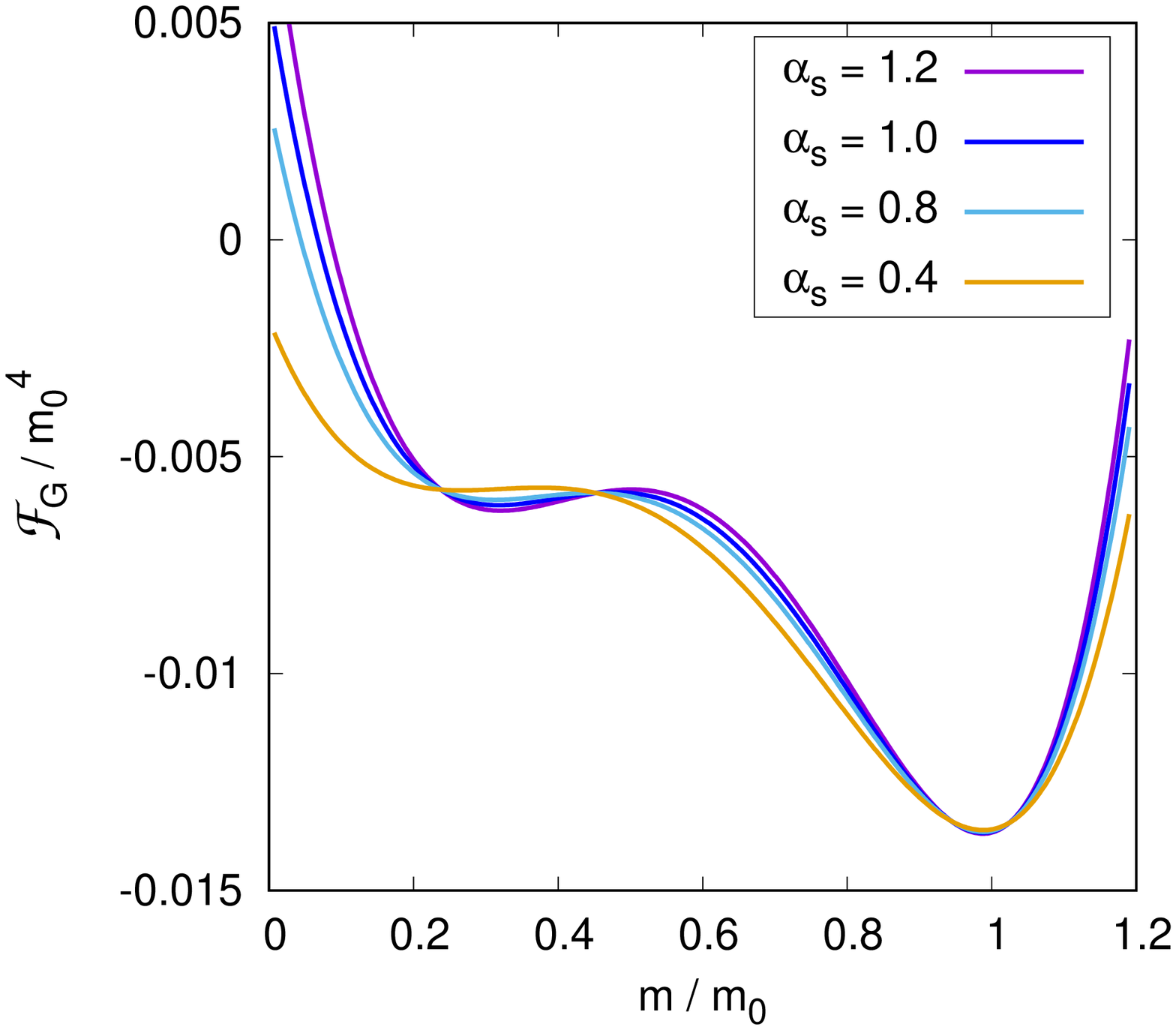}
\end{subfigure}\quad\quad
\begin{subfigure}[H]{0.45\textwidth}
\includegraphics[trim={0 85 0 100}, clip, angle=270, origin=c,width=1\textwidth]{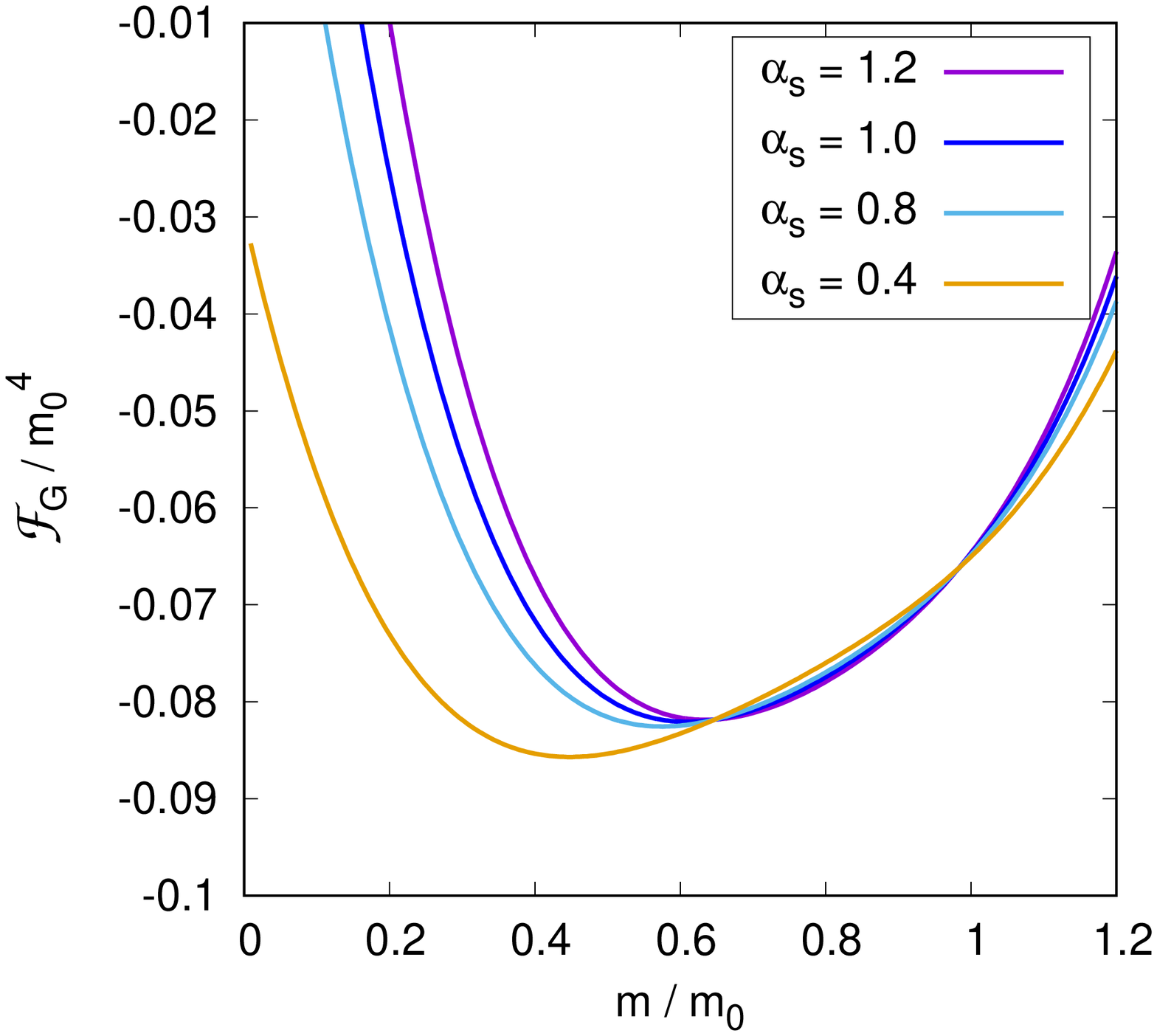}
\end{subfigure}
\caption{The GEP at $T=0.25\ m_{0}$ (left), $T=0.5\ m_{0}$ (right) as a function of the mass parameter for different values of $\alpha_{s}=1$.}
\end{figure}\
\\
In Fig. 15 the GEP is shown as a function of the mass parameter for different values of the coupling constant and two sample values of the temperature. It is clear from the figure that the GEP does not depend very much on $\alpha_{s}$ in the physical range [\,0.4, 1.2\,], especially at low temperatures and around its minima. In what follows, we limit our discussion to the case $\alpha_{s}=1$. Later on we will show that the physical predictions of the GEP approach are not very sensitive to the actual value of $\alpha_{s}$ in said range.\\
\\
\\
\\
\\
\begin{figure}[H]
\centering
\includegraphics[angle=270, origin=c,width=0.8\textwidth]{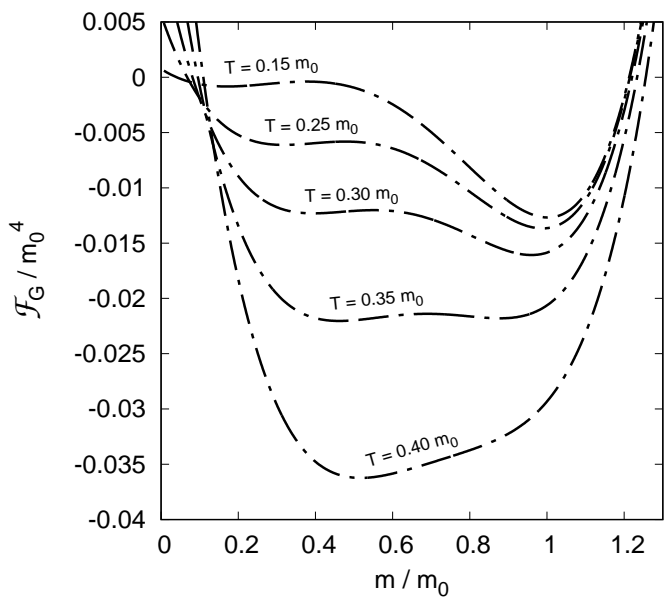}
\vspace{-16mm}
\caption{The GEP as a function of the mass parameter for $\alpha_{s}=1$ and different values of the temperature.}
\end{figure}\
\\
\\
\\
\\
In Fig. 16 we give a plot of the GEP as a function of the mass parameter for $\alpha_{s}=1$ and different values of the temperature. At low temperatures, the GEP possesses two minima. The rightmost one can be recognized as the global minimum already found in the vacuum; we will refer to it as the $m=m_{0}$ minimum. The leftmost one has evolved from $m=0$; we will refer to it as the $m=0$ minimum. As the temperature increases, the $m=0$ minimum sinks deeper and increases in value, while at the same time the $m=m_{0}$ minimum decreases in value and deepens more slowly than the $m=0$ minimum. Around $T = 0.35\, m_{0}$, the two minima align. At higher temperatures, the $m=0$ minimum grows deeper than the $m=m_{0}$ minimum until the latter completely disappears. At the critical temperature $T_{c}\approx 0.35\ m_{0}$ the two minima are separated by a distance of approximately 0.4 $m_{0}$: the optimal mass parameter $m(T)$ is discontinuous across $T_{c}$. This is made explicit in Fig. 17, where the optimal mass parameter is plotted as a function of the temperature. In Fig. 18 we show the GEP evaluated on the two minima near the critical temperature. It is clear that at the intersection point the derivatives of the two curves are different. At the GEP level of approximation, the derivative of the GEP is equal to minus the entropy\\
\begin{figure}[H]
\centering
\vspace{2mm}
\includegraphics[angle=270, origin=c,width=0.55\textwidth]{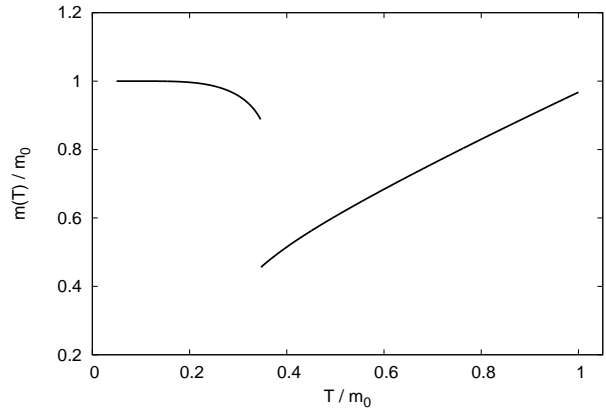}
\vspace{-12mm}
\caption{The optimal mass parameter as a function of the temperature for $\alpha_{s}=1$.}
\end{figure}\
\begin{figure}[H]
\centering
\includegraphics[angle=270, origin=c,width=0.65\textwidth]{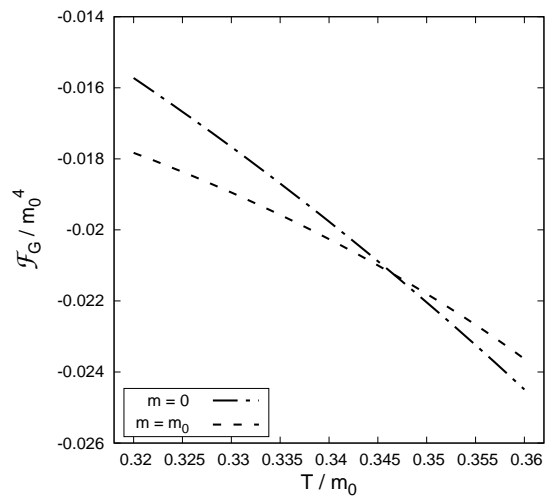}
\vspace{-12mm}
\caption{The GEP evaluated on the two minima as a function of the temperature for $\alpha_{s}=1$.}
\end{figure}\
\\
\\
density of the system (Fig. 19). The discontinuity of the derivative then signals the presence of a first-order phase transition at $T_{c}\approx 0.35\ m_{0}$. It is interesting to notice that the entropy nearly vanishes in a wide range of temperatures below $T_{c}$. This feature may be interpreted as evidence for the fact that below the critical temperature the gluonic degrees of freedom, due to confinement, are frozen.\\
\\
\begin{figure}[H]
\centering
\includegraphics[angle=270, origin=c,width=0.55\textwidth]{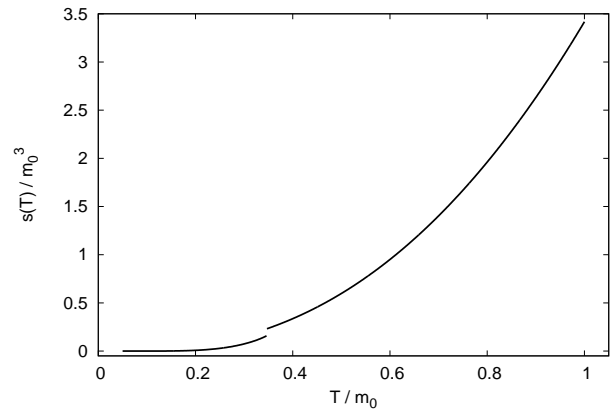}
\vspace{-12mm}
\caption{Entropy density in the GEP approximation as a function of the temperature for $\alpha_{s}=1$}
\end{figure}\
\\
In Fig. 20 we show the critical temperature and latent heat of transition as functions of the coupling constant $\alpha_{s}$ in the physical range $[\,0.4,\ 1.2\,]$. In said range, the critical temperature is seen to vary by less than 1\% from its middle value of $0.347\ m_{0}$. As for the latent heat $\Delta h_{0}\approx 1.8\,T_{c}^{4}$, this is found within 30\% from its lattice value of 1.3 - 1.5 $T_{c}^{4}$. Since the GEP needs higher order corrections in order to well approximate the free energy density of YMT, such an error should not concern us; on the contrary, it is sufficiently small to suggest that the GEP approach is meaningful in the study of the thermodynamics of YMT. This suggestion is enforced by the good match between the GEP and lattice critical temperatures: taking $m_{0}$ = 730 MeV, a critical temperature of $T_{c}\approx$ 250-255 MeV can be read out from Fig. 20, a value that lies within a 6-8 \% from its lattice value of 270 MeV.\\
\\
\begin{figure}[H]
\centering
\begin{subfigure}[H]{0.45\textwidth}
\includegraphics[trim={0 85 0 100}, clip, angle=270, origin=c,width=0.98\textwidth]{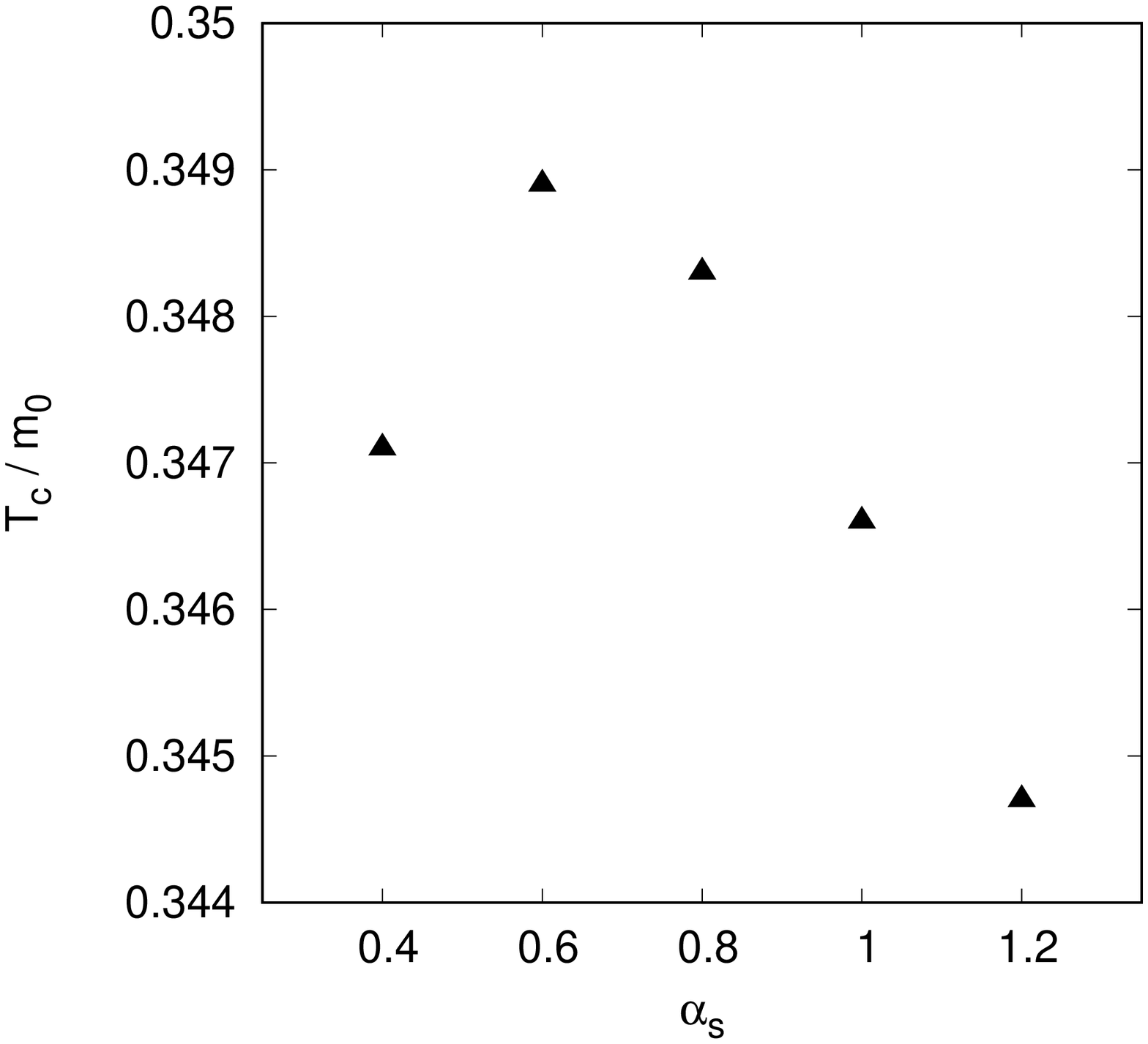}
\end{subfigure}
\quad\quad
\begin{subfigure}[H]{0.45\textwidth}
\includegraphics[trim={0 85 0 100}, clip, angle=270, origin=c,width=0.98\textwidth]{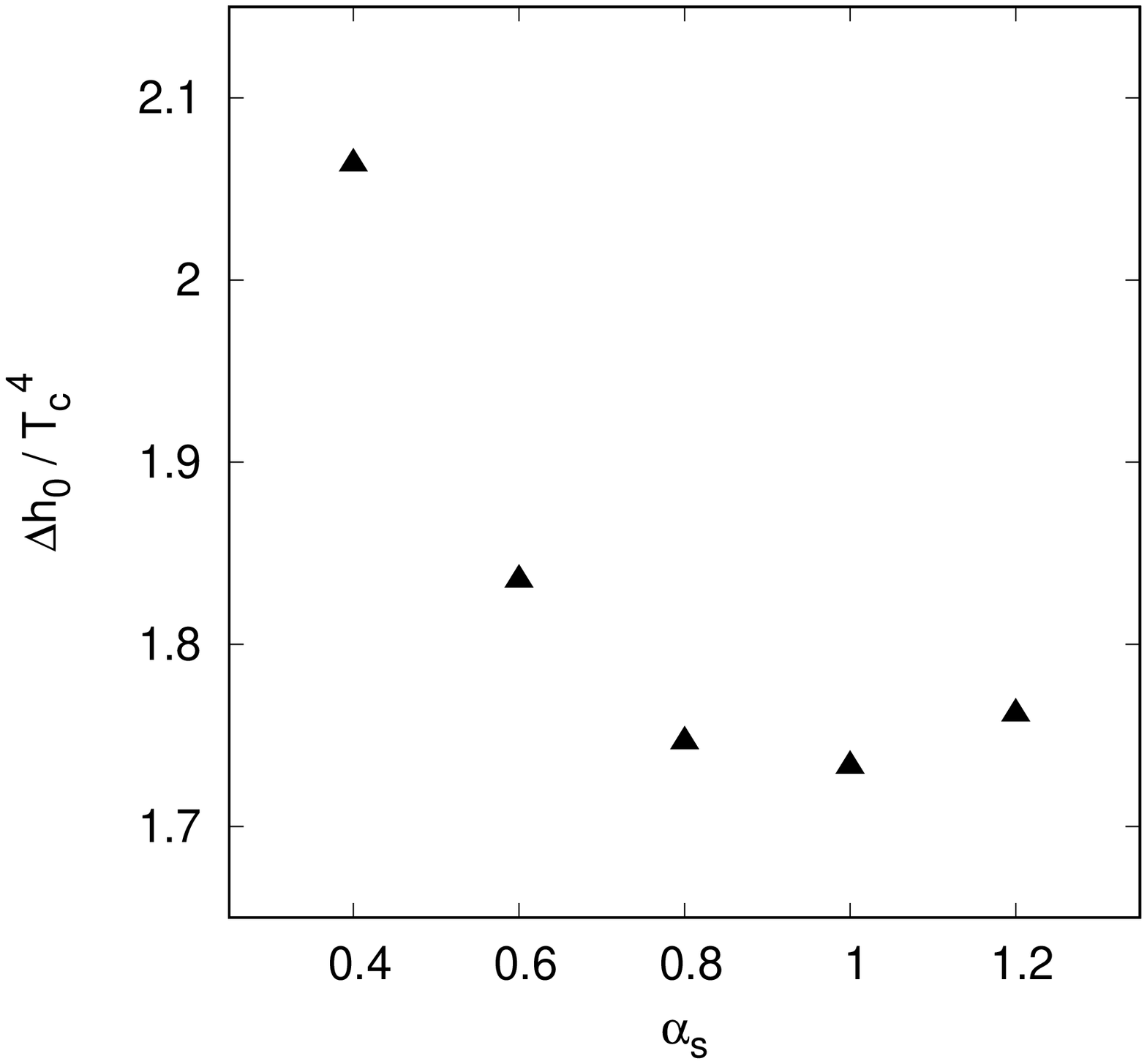}
\end{subfigure}
\caption{Critical temperature (left) and latent heat of transition (right) as functions of the coupling $\alpha_{s}$}
\end{figure}
\clearpage{}
\thispagestyle{empty}
\
\clearpage


\addcontentsline{toc}{chapter}{Conclusions}  \markboth{Conclusions}{Conclusions}

\chapter*{Conclusions\index{Conclusions}}

The main findings of this thesis can be summarised as follows. In the vacuum, the existence of a sharp global minimum for the GEP at a non-zero value of the mass parameter indicates that the standard (i.e. massless) vacuum of Yang-Mills theory is unstable towards the massive vacuum of ref.\cite{siringo1}. This result (\textit{i}) can be interpreted as evidence for the phenomenon of dynamical mass generation in YMT and (\textit{ii}) enforces from first principles the validity of the massive perturbative expansion of ref.\cite{siringo1}. At finite temperature, the GEP provides us with an optimal temperature-dependent mass parameter $m(T)$, which is found to be discontinuous at a critical temperature $T_{c}\approx\, 0.35\ m_{0}$. The GEP approximation to the entropy density $s(T)$ too is found to be discontinuous at $T_{c}$, signaling the occurrence of a (weakly) first-order phase transition. If for $m_{0}$ we use the value given in ref.\cite{siringo1}, $m_{0}=730$ MeV, we find that the transition has the following thermodynamical properties: (\textit{i}) it occurs at a critical temperature $T_{c}\approx 255$ MeV; (\textit{ii}) it has a latent heat $\Delta h_{0}\approx 1.8\, T_{c}^{4}$. These values are in good agreement with lattice data, which exhibit a weakly first-order deconfinement transition at a critical temperature of approximately 270 MeV and with a latent heat of 1.3 - 1.5 $T_{c}^{4}$.\\
\\
\\
\\
\clearpage{}
\thispagestyle{empty}
\
\clearpage


\newcounter{count2}
\setcounter{count2}{1}
\setcounter{equation}{0}
\renewcommand{\theequation}{\Alph{count2}.\arabic{equation}}
\addcontentsline{toc}{chapter}{Appendix}  \markboth{Appendix}{Appendix}

\chapter*{Appendix\index{Appendix}}

\addcontentsline{toc}{section}{A. Thermal field theory and applications to Yang-Mills theory}  \markboth{A. Thermal field theory and applications to Yang-Mills theory}{A. Thermal field theory and applications to Yang-Mills theory}
\section*{A. Thermal field theory and applications to Yang-Mills theory\index{Thermal field theory and applications to Yang-Mills theory}}

The statistical behaviour of a quantum system in thermodynamic equilibrium is described \cite{kapusta} by its density matrix $\rho_{\beta}$,\\
\\
\begin{equation}
\rho_{\beta}=\frac{e^{-\beta H}}{\text{Tr}\,\{e^{-\beta H}\}}
\end{equation}\\
\\
Here $H$ is the Hamiltonian of the system, $\beta=1/T$ is the inverse temperature and $\text{Tr}$ denotes the trace operator over Hilbert space endomorphisms. Through the density matrix one can express the thermal average of any operator $\mathcal{O}$ at temperature $T$ as\\
\\
\begin{equation}
\langle\mathcal{O}\rangle_{\beta}=\text{Tr}\big\{\rho_{\beta}\,\mathcal{O}\big\}
\end{equation}
\\
If we define the thermal partition function $\mathcal{Z}$ of the system as\\
\\
\begin{equation}
\mathcal{Z}=\text{Tr}\,\{e^{-\beta H}\}
\end{equation}
\\
then the mean energy $E=\text{Tr}\{\rho_{\beta}\, H\}$ and entropy $S=-\text{Tr}\{\rho_{\beta}\,\ln\rho_{\beta}\}$ of the system can be obtained by differentiating its logarithm,\\
\\
\begin{equation}\label{344}
E=-\frac{\partial}{\partial \beta}\ \ln \text{Tr}\,\{e^{-\beta H}\}
\end{equation}\\
\begin{equation}\label{345}
S=\frac{\partial}{\partial T}\ \ T\, \ln \text{Tr}\,\{e^{-\beta H}\}
\end{equation}\\
\\
while the free energy density $\F$ of the system is given by\\
\\
\begin{equation}
\F=-\frac{T}{V}\ \ln \Z
\end{equation}
with $V$ the $(d-1)$-dimensional spatial volume. When the degrees of freedom of the system are described in terms of bosonic fields, the partition function can be expressed in functional form by noticing that in this case the trace of a generic operator $\mathcal{O}$ can be computed as\\
\\
\[
\text{Tr}\{\mathcal{O}\}=\int d\mathscr{F}^{I}(\vec{x})\ \ \bra{\mathscr{F}^{I}(\vec{x})}\, \mathcal{O}\, \ket{\mathscr{F}^{I}(\vec{x})}
\]
\\
where $\ket{\mathscr{F}^{I}(\vec{x})}$ is an eigenstate of the Schr\"{o}dinger picture field operators $\hat{\mathscr{F}}^{I}(\vec{x})$ with eigenvalue $\mathscr{F}^{I}(\vec{x})$. With this in mind, we can write\\
\\
\begin{equation}\label{346}
\mathcal{Z}=\int d\mathscr{F}^{I}(\vec{x})\ \ \bra{\mathscr{F}^{I}(\vec{x})}\ e^{-\beta H}\, \ket{\mathscr{F}^{I}(\vec{x})}
\end{equation}\\
\\
Eq.~\eqref{346} is formally identical to the integral of a transition amplitude from a state with field configuration $\mathscr{F}^{I}(\vec{x})$ at time $t=0$ to a state with the same configuration at time $t=-i\beta$. In complete analogy with the case of dynamical evolution in real time, we can then express the partition function as the path integral\footnote{\ Here we are implicitly assuming that the Hamiltonian $H$ is quadratic in the momenta conjugate to the field variables. This is of course the case for YMT.}\\
\\
\begin{equation}
\mathcal{Z}=\int d\mathscr{F}^{I}(\vec{x})\ \int_{_{\tilde{\mathscr{F}}(\vec{x},0)=\tilde{\mathscr{F}}(\vec{x},-i\beta)=\mathscr{F}(\vec{x})}}\D\tilde{\mathscr{F}}^{I}(\vec{x})\ \exp\bigg(i\int_{0}^{-i\beta}dt\int d^{d-1}x\ \ \mathcal{I}[\tilde{\mathscr{F}},\dot{\tilde{\mathscr{F}}}]\bigg)
\end{equation}\\
\\
where $\mathcal{I}$ is the Lagrangian of the theory. Equivalently,\\
\\
\begin{equation}\label{347}
\mathcal{Z}=\int_{\text{per.}}\D\mathscr{F}^{I}(\vec{x})\ \exp\bigg(-\int_{0}^{\beta}d\tau\int d^{d-1}x\ \ \mathcal{I}^{E}[\mathscr{F},\partial\mathscr{F}/\partial \tau]\bigg)
\end{equation}\\
\\
where\\
\\
\begin{equation}
\mathcal{I}^{E}[\mathscr{F},\partial\mathscr{F}/\partial \tau]=-\mathcal{I}[\mathscr{F},i\partial\mathscr{F}/\partial \tau]
\end{equation}\\
\\
and the subscript ``per.'' indicates that we are to integrate over periodic $\tau$-boundary configurations.\\
\\
\\
In the case of Yang-Mills theory, eq.~\eqref{347} translates into\footnote{\ It can be shown \cite{kapusta} that the presence of the ghost fields does not spoil the validity of the equation. This is done by defining $\mathcal{Z}$ to be the partition function of Yang-Mills theory in a ghost-free gauge such as the axial gauge. After restoring the gauge invariance of $\mathcal{S}_{YM}$ and applying the FP procedure to the result, one obtains $\mathcal{Z}$ in the form given in eq.~\eqref{348}.}
\\
\begin{equation}\label{348}
\mathcal{Z}=\int_{\text{per.}}\mathcal{D}A_{\mu}^{a}\,\D\cbar^{a}\,\D c^{a}\ e^{-\mathcal{S}^{th}}
\end{equation}\\
\\
where the thermal action $\mathcal{S}^{th}$ is defined as\\
\\
\begin{equation}\label{3480}
\mathcal{S}^{th}=\int_{0}^{\beta}d\tau\int d^{d-1}x\ \ \ \frac{1}{4}\ \delta^{\mu\sigma}\delta^{\nu\lambda}\ F_{\mu\nu}^{a}\,F_{\sigma\lambda}^{a}+\frac{1}{2\xi}\ \big(\delta^{\mu\nu}\partial_{\mu}A_{\nu}^{a}\big)^{2}+\delta^{\mu\nu}\partial_{\mu}\cbar^{a}\,D_{\nu}^{ab}c^{b}
\end{equation}\\
\\
($\partial/\partial x^{0}=\partial/\partial \tau$). In order to evaluate the thermal partition function, we notice that eq.~\eqref{348} is very similar in form to eq.~\eqref{9} continued to Euclidean space, the only difference being that in the thermal action \eqref{3480} the imaginary-time domain of integration is bounded, $\tau\in[0,\beta]$, and that the integration is carried over $\tau$-periodic field configurations. Periodic boundary conditions on a bounded domain can be implemented in the computations by expressing the fields as sums over discrete Fourier components: we set\\
\\
\begin{equation}\label{349}
A_{\mu}^{a}(\vec{x},\tau)=\sqrt{\frac{\beta}{V}}\ \ \sum_{K}\ \ e^{iK\cdot X}\ A_{\mu}^{a}(K)
\end{equation}
\begin{equation}
c^{a}(\vec{x},\tau)=\sqrt{\frac{\beta}{V}}\ \ \sum_{K}\ \ e^{iK\cdot X}\ c^{a}(K)
\end{equation}
\begin{equation}\label{350}
\cbar^{a}(\vec{x},\tau)=\sqrt{\frac{\beta}{V}}\ \ \sum_{K}\ \ e^{-iK\cdot X}\ \cbar^{a}(K)
\end{equation}\\
\\
where $K=(K^{0},\vec{K})=(2\pi nT,\vec{K})$, $K\cdot X=(K^{0}\tau+\vec{K}\cdot\vec{x})$ and the factor of $\sqrt{\beta/V}$ is introduced to adimensionalize the Fourier components. In the spatial continuum limit the summation sign is to be interpreted as\\
\\
\begin{equation}
\sum_{K}\ \longrightarrow\ V\ \sum_{n}\ \int\frac{d^{d-1}K}{(2\pi)^{d-1}}
\end{equation}\\
\\
The frequencies $\omega_{n}=K^{0}_{n}=2\pi n T$, $n\in\Bbb{Z}$, are known as Matsubara frequencies. Inserting eqq.~\eqref{349}-\eqref{350} back into \eqref{3480}, the thermal partition function is brought to the same form of the vacuum partition function continued to Euclidean space, with integrals over the continuous parameter $k^{0}_{E}$ everywhere replaced by sums over discrete Matsubara frequencies. Therefore $\Z$ can be perturbatively evaluated with the same methods used to compute $Z$ in the vacuum formalism, namely the MLPE or MSPE of Chapters 1-2, with Matsubara sums everywhere replacing $k^{0}_{E}$ integrals. Such sums are to be handled with a general method that we will describe in the next appendix.\\
\stepcounter{count2}
\setcounter{equation}{0}
\addcontentsline{toc}{section}{B. Matsubara sums}  \markboth{B. Matsubara sums}{B. Matsubara sums}
\section*{B. Matsubara sums\index{Matsubara sums}}

Consider a Matsubara sum of the form\\
\\
\begin{equation}\label{B1}
I=T\ \sum_{n}\ f(2\pi nT)
\end{equation}\\
\\
with $f$ a complex-analytic function with no poles on the real axis such that the infinite sum on the RHS is finite. Keeping in mind that the function $\coth(iz)$ has poles of the first order for $z\in\pi\Bbb{Z}$, a straightforward calculation \cite{kapusta} shows that the sum can be expressed as\\
\\
\begin{equation}
T\ \sum_{n}\ f(2\pi nT)=\frac{1}{4\pi}\ \ \oint_{\Gamma}\ dz\ \ f(z)\,\coth\bigg(\frac{iz}{2T}\bigg)
\end{equation}\\
\\
where $\Gamma$ is the contour shown in Fig. 21 for $\varepsilon\to0$ ($\Gamma$ is assumed to close at infinity). If we split $\Gamma$ into the upper and lower curves $\Gamma_{+}$ and $\Gamma_{-}$, we may rewrite the integral as the sum of two terms,\\
\\
\[
I=\frac{1}{4\pi}\ \int_{-\infty-i\varepsilon}^{+\infty-i\varepsilon}dz\ \ f(z)\,\coth\bigg(\frac{iz}{2T}\bigg)-\frac{1}{4\pi}\ \int_{-\infty+i\varepsilon}^{+\infty+i\varepsilon}dz\ \ f(z)\,\coth\bigg(\frac{iz}{2T}\bigg)
\]\\
\\
\\
\\
\\
\begin{figure}[H]
\centering
\frame{\includegraphics[width=0.6\textwidth]{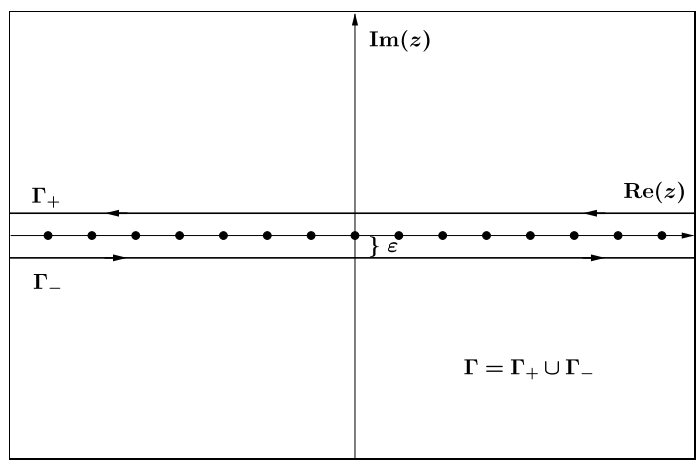}}
\vspace{5mm}
\caption{Contour of integration for a typical Matsubara sum.}
\end{figure}\
\\
\\
Changing variable of integration from $z$ to $-z$ in the second integral, since $\coth(-z)=-\coth(z)$, we obtain\\
\\
\begin{equation}\label{378}
I=\frac{1}{4\pi}\ \ \int_{-\infty-i\varepsilon}^{+\infty-i\varepsilon}dz\ \ \big(f(z)+f(-z)\big)\,\coth\bigg(\frac{iz}{2T}\bigg)
\end{equation}\\
\\
Eq.~\eqref{378} can be further simplified by expressing the hyperbolic cotangent as\\
\\
\[
\coth\bigg(\frac{iz}{2T}\bigg)=\frac{e^{iz/T}+1}{e^{iz/T}-1}=2\,\bigg[\frac{1}{2}+\frac{1}{e^{i\beta z}-1}\bigg]
\]
\\
so that
\\
\begin{equation}\label{387}
I=\frac{1}{2\pi}\ \ \int_{-\infty-i\varepsilon}^{+\infty-i\varepsilon}dz\ \ \big(f(z)+f(-z)\big)\,\bigg[\frac{1}{2}+\frac{1}{e^{i\beta z}-1}\bigg]
\end{equation}\\
\\
The sum can be subdivided into two terms,\\
\\
\begin{equation}\label{379}
I=I^{vac}+I^{th}
\end{equation}
\\
where
\\
\begin{equation}\label{380}
I^{vac}=\int_{-\infty}^{+\infty}\frac{dz}{2\pi}\ \frac{f(z)+f(-z)}{2}
\end{equation}\\
\\
\begin{equation}\label{381}
I^{th}=\frac{1}{2\pi}\ \ \int_{-\infty-i\varepsilon}^{+\infty-i\varepsilon}dz\ \ \big(f(z)+f(-z)\big)\ \frac{1}{e^{i\beta z}-1}
\end{equation}\\
\\
In eq.~\eqref{380} we have dropped the $i\varepsilon$-dependence of the domain of integration since we have assumed that $f$ has no poles on the real axis. In eq.~\eqref{381}, as $T$ goes to zero $\beta$ goes to infinity; taking into account the $\varepsilon$-dependence of the domain of integration, we see that the denominator also goes to infinity, telling us that $I^{th}(T\to 0)=0$. Thus $I^{th}$ can be interpreted as the thermal contribution to $I$. On the other hand, $I^{vac}$ does not depend on the temperature. Thus it can be interpreted as the vacuum contribution to $I$.\\
\\
\\
Matsubara sums arise in thermal loop diagrams of perturbative expansions in the form\\
\\
\begin{equation}\label{382}
I=T\ \sum_{n}\ \int\frac{d^{d-1}K}{(2\pi)^{d-1}}\ f(2\pi nT,\,\vec{K})
\end{equation}
Any such quantity can again be computed as\\
\begin{equation}\label{383}
I=I^{vac}+I^{th}
\end{equation}
where\\
\begin{equation}\label{384}
I^{vac}=\int\frac{d^{d}K}{(2\pi)^{d}}\ \frac{f(K^{0},\vec{K})+f(-K^{0},\vec{K})}{2}
\end{equation}\\
\\
\begin{equation}\label{385}
I^{th}=\int_{K^{0}=-\infty-i\varepsilon}^{K^{0}=+\infty-i\varepsilon} \frac{d^{d}K}{(2\pi)^{d}}\ \ \big(f(K^{0},\vec{K})+f(-K^{0},\vec{K})\big)\ n_{\beta}\big(iK^{0}\big)
\end{equation}\\
\\
and we have defined the Bose distribution $n_{\beta}(x)$ as\\
\\
\begin{equation}\label{386}
n_{\beta}(x)=\frac{1}{e^{\beta x}-1}
\end{equation}\\
\\
In general, the vacuum contribution to any such integral may be divergent. Divergences can be eliminated through renormalization procedures analogous to those employed in the vacuum formalism. The Bose distribution, on the other hand, ensures that the thermal contributions are finite.\\

\stepcounter{count2}
\setcounter{equation}{0}
\addcontentsline{toc}{section}{C. The Jensen-Feynman inequality}  \markboth{C. The Jensen-Feynman inequality}{C. The Jensen-Feynman inequality}
\section*{C. The Jensen-Feynman inequality\index{The Jensen-Feynman inequality}}

Let $d\mu$ be a probability measure on some measure space $X$. Define the average of a real function $f:X\to \Bbb{R}$ to be\\
\\
\[
\langle \,f\,\rangle=\int_{X} d\mu\ \ f
\]\\
\\
Rewriting the average of the exponential of $f$ as\\
\\
\begin{align*}
\langle e^{f}\rangle=e^{\langle f \rangle}\ \langle{e^{f-\langle f\rangle}}\rangle
\end{align*}
\\
and using the inequality $e^{x}\geq 1+x$, from the positive-definiteness of the measure $d\mu$ it follows that\\
\\
\[
\langle e^{f}\rangle\geq e^{\langle f \rangle}\ \big\langle 1+f-\langle f\rangle \big\rangle=e^{\langle f \rangle}
\]
i.e. that\\
\begin{equation}
\langle e^{f}\rangle\geq e^{\langle f \rangle}
\end{equation}
\\
The former is known as the Jensen inequality for the exponential function. In the context of thermal field theory, the Jensen inequality can be used to prove an inequality for the free energy density of a system of commuting fields. Consider the free energy density in the functional formulation of the theory,\\
\\
\begin{equation}\label{FEabc}
\F=\F_{0}-\frac{T}{V}\ \ln\ \bigg(\frac{\int\D\mathscr{F}\ e^{-\mathcal{I}_{0}}\ e^{-\mathcal{I}_{int}}}{\int\D\mathscr{F}\ e^{-\mathcal{I}_{0}}}\bigg)
\end{equation}\\
\\
where $\mathcal{I}_{0}\geq 0$ is the thermal action of the system, $\mathcal{I}_{int}$ is the thermal interaction action and
\\
\[
\F_{0}=-\frac{T}{V}\ \ln\ \int\D\mathscr{F}\ e^{-\mathcal{I}_{0}}
\]\\
\\
is the ideal gas contribution to the free energy density. The exponential $e^{-\mathcal{I}_{0}}$ defines a probability measure on the space of the functionals of the fields, namely\\
\\
\[
\langle\, f\,\rangle=\frac{\int\D\mathscr{F}\ e^{-\mathcal{I}_{0}}\ f[\mathscr{F}]}{\int\D\mathscr{F}\ e^{-\mathcal{I}_{0}}}
\]\\
\\
From the Jensen inequality $\langle e^{f}\rangle\geq e^{\langle f \rangle}$ it then follows that\\
\\
\[
\frac{\int\D\mathscr{F}\ e^{-\mathcal{I}_{0}}\ e^{-\mathcal{I}_{int}}}{\int\D\mathscr{F}\ e^{-\mathcal{I}_{0}}}\geq \exp\bigg(\frac{\int\D\mathscr{F}\ e^{-\mathcal{I}_{0}}\ \big(-\mathcal{I}_{int}\big)}{\int\D\mathscr{F}\ e^{-\mathcal{I}_{0}}}\bigg)
\]\\
\\
which in turn implies\\
\\
\begin{equation}\label{ineqabc}
\ln\ \bigg(\frac{\int\D\mathscr{F}\ e^{-\mathcal{I}_{0}}\ e^{-\mathcal{I}_{int}}}{\int\D\mathscr{F}\ e^{-\mathcal{I}_{0}}}\bigg)\geq -\ \frac{\int\D\mathscr{F}\ e^{-\mathcal{I}_{0}}\ \mathcal{I}_{int}}{\int\D\mathscr{F}\ e^{-\mathcal{I}_{0}}}=-\langle\ \mathcal{I}_{int}\,\rangle
\end{equation}\\
\\
By plugging eq.~\eqref{ineqabc} back into eq.~\eqref{FEabc}, we find that\\
\\
\begin{equation}
\F\leq \F_{0}+\frac{T}{V}\ \langle\, \mathcal{I}_{int}\,\rangle=\F_{1}
\end{equation}
\\
$\F_{1}$ is none other than the free energy density of the system computed to first order in $\mathcal{I}_{int}$.\\

\stepcounter{count2}
\setcounter{equation}{0}
\addcontentsline{toc}{section}{D. Thermal integrals in $d=4$}  \markboth{D.Thermal integrals in $d=4$}{D. Thermal integrals in $d=4$}
\section*{D. Thermal integrals in $\boldsymbol{d=4}$\index{Thermal integrals in $d=4$}}

In this appendix we will explicitly compute the integrals $K_{m}$, $J_{m}$ and $H_{m}$ that appear in the expression \eqref{GEP} for the GEP of YMT in $d=4$. The vacuum contributions will need to be regularized. A discussion of the renormalization scheme to be applied in our computations is given in sec. 3.2.2.\\
\\
\\
The integral $J_{m}$ is defined as\\
\\
\begin{equation}
J_{m}:=\int_{K}\ \frac{1}{K^{2}+m^{2}}
\end{equation}\\
\\
where the symbol $\int_{K}$ stands for integration over spatial momenta, summation over Matsubara frequencies and multiplication by the temperature $T$,\\
\\
\begin{equation}
\int_{K}\ \longrightarrow\quad T\ \sum_{n}\ \int\frac{d^{3}k}{(2\pi)^{3}}
\end{equation}\\
\\
As we did in Appendix B for a general Matsubara sum, we can identify a vacuum and a thermal contribution to $J_{m}$,
\\
\begin{equation}
J_{m}=J_{m}^{v}+J_{m}^{th}
\end{equation}
\\
where, being $J_{m}$'s integrand symmetric with respect to the exchange $K^{0}\to -K^{0}$,\\
\\
\begin{equation}
J_{m}^{v}=\int \frac{d^{4}K}{(2\pi)^{4}}\ \frac{1}{K^{2}+m^{2}}
\end{equation}\\
\\
\begin{equation}\label{D839}
J_{m}^{th}=2\ \int_{-\infty-i\varepsilon}^{+\infty-i\varepsilon}\frac{dK^{0}}{2\pi}\int \frac{d^{3}K}{(2\pi)^{3}}\ \frac{1}{K^{2}+m^{2}}\ n_{\beta}(iK^{0})
\end{equation}\\
\\
Recall that $n_{\beta}(x)$ is the Bose distribution, i.e.\\
\\
\[
n_{\beta}(x)=\frac{1}{e^{\beta x}-1}
\]\\
\\
Let us start from the thermal contribution $J_{m}^{th}$. The $K^{0}$ integral in eq.~\eqref{D839} reads\\
\\
\[
\int_{-\infty-i\varepsilon}^{+\infty-i\varepsilon} \frac{dK^{0}}{2\pi}\ \frac{1}{(K^{0})^{2}+|\vec{K}|^{2}+m^{2}}\ n_{\beta}(iK^{0})
\]
and can be evaluated with elementary methods. Closing the contour of integration as in Fig. 22, the integral is equal to $-2\pi i$ times the residue of the integrand at $K^{0}=-i\,(m^{2}+|\vec{K}|^{2})^{1/2}$. Setting\\
\\
\begin{equation}
\varepsilon_{m}(x)=\sqrt{m^{2}+x^{2}}
\end{equation}
\\
we find\\
\\
\[
\int_{-\infty-i\varepsilon}^{+\infty-i\varepsilon} \frac{dK^{0}}{2\pi}\ \frac{n_{\beta}(iK^{0})}{K^{2}+m^{2}}=\frac{1}{2}\ \frac{n_{\beta}\big(\varepsilon_{m}(|\vec{K}|)\big)}{\varepsilon_{m}(|\vec{K}|)}
\]\\
\\
so that a trivial integration over the angular variables gives us\\
\\
\begin{equation}\label{D90}
J_{m}^{th}=\frac{1}{2\pi^{2}}\ \int_{0}^{+\infty}dk\ \ k^{2}\ \frac{n_{\beta}\big(\varepsilon_{m}(k)\big)}{\varepsilon_{m}(k)}
\end{equation}\\
\\
The vacuum contribution to $J_{m}$ is divergent. In dimensional regularization, setting $\epsilon=4-d$, it reads\\
\\
\begin{equation}\label{D91}
J_{m}^{v}=-\frac{m^{2}}{16\pi^{2}}\ \bigg(\frac{2}{\epsilon}-\gamma+\ln\,4\pi-\ln\, m^{2}+1\bigg)
\end{equation}\\
\\
\\
\\
\begin{figure}[H]
\centering
\frame{\includegraphics[width=0.55\textwidth]{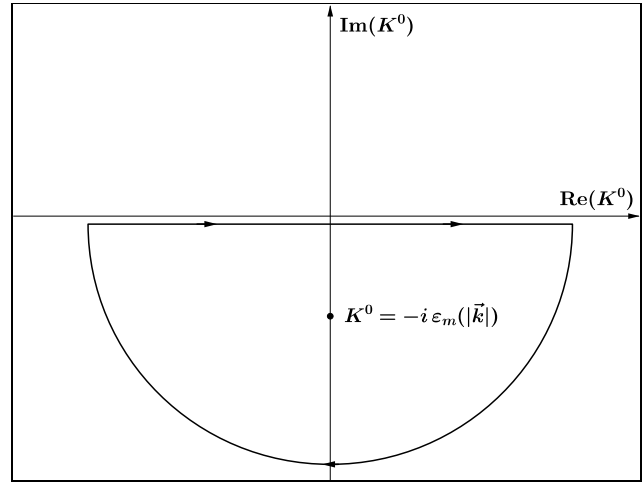}}
\vspace{5mm}
\caption{Contour of integration for the integral $J_{m}^{th}$.}
\end{figure}\
\\
The integral $K_{m}$ is defined as\\
\\
\begin{equation}
K_{m}:=\frac{1}{2}\ \int_{K}\ \ln\ \big[\beta^{2}(K^{2}+m^{2})\big]
\end{equation}\\
\\
$K_{m}$ satisfies the differential equation\\
\\
\begin{equation}\label{A1}
\frac{\partial K_{m}}{\partial m^{2}}=\frac{1}{2}\ J_{m}
\end{equation}\\
\\
and coincides (see e.g. ref.\cite{kapusta}) with the free energy density of an ideal gas of interacting massive bosons. In the limit $T\to 0$, upon inversion of eq.~\eqref{A1}, we see that $K_{m}^{v}$ can be evaluated as\\
\\
\begin{equation}
K_{m}^{v}=\frac{1}{2}\ \int dm^{2}\ J_{m}^{v}+\text{const.}
\end{equation}\\
\\
where the constant does not depend on $m^{2}$ and $T$ and is thus inessential. An elementary integration of eq.~\eqref{D91} gives us the value of $K_{m}^{v}$ in dimensional regularization:\\
\\
\\
\begin{equation}
K_{m}^{v}=-\frac{m^{4}}{64\pi^{2}}\ \bigg(\frac{2}{\epsilon}-\gamma+\ln\,4\pi-\ln\, m^{2}+\frac{1}{2}\bigg)
\end{equation}\\
\\
Subtracting from eq.~\eqref{A1} the vacuum contribution, we find an analogous integral equation for $K_{m}^{th}$,\\
\\
\begin{equation}\label{A2}
K_{m}^{th}=\frac{1}{2}\ \int_{0}^{m^{2}} d\bar{m}^{2}\ J_{\bar{m}}^{v}+K_{0}^{th}
\end{equation}\\
\\
where this time the integration constant $K_{0}^{th}$ depends on the temperature and cannot be discarded. Eq.~\eqref{A2} can be solved by using the following trick. Let us write $J_{m}^{th}$ as\\
\\
\[
J_{m}^{th}=\frac{1}{6\pi^{2}}\ \int_{0}^{+\infty}dk\ \ \frac{dk^{3}}{dk}\ \frac{n_{\beta}\big(\varepsilon_{m}(k)\big)}{\varepsilon_{m}(k)}
\]\\
\\
Integrating by parts, we obtain\\
\\
\[
J_{m}^{th}=-\frac{1}{6\pi^{2}}\ \int_{0}^{+\infty}dk\ \ k^{3}\ \frac{d}{dk}\, \frac{n_{\beta}\big(\varepsilon_{m}(k)\big)}{\varepsilon_{m}(k)}=-\frac{1}{3\pi^{2}}\ \int_{0}^{+\infty}dk\ \ k^{4}\ \frac{d}{dk^{2}}\, \frac{n_{\beta}\big(\varepsilon_{m}(k)\big)}{\varepsilon_{m}(k)}
\]\\
\\
The argument of the derivative depends on $k^{2}$ through the combination $m^{2}+k^{2}$. Hence\\
\\
\[
-\frac{1}{3\pi^{2}}\ \int_{0}^{+\infty}dk\ \ k^{4}\ \frac{d}{dk^{2}}\, \frac{n_{\beta}\big(\varepsilon_{m}(k)\big)}{\varepsilon_{m}(k)}=-\frac{1}{3\pi^{2}}\ \int_{0}^{+\infty}dk\ \ k^{4}\ \frac{\partial}{\partial m^{2}}\, \frac{n_{\beta}\big(\varepsilon_{m}(k)\big)}{\varepsilon_{m}(k)}
\]\\
\\
and we can bring the derivative out of the integral sign, yielding\\
\\
\[
J_{m}^{th}=\frac{\partial}{\partial m^{2}}\ \bigg[-\frac{1}{3\pi^{2}}\ \int_{0}^{+\infty}dk\ \ k^{4}\ \frac{n_{\beta}\big(\varepsilon_{m}(k)\big)}{\varepsilon_{m}(k)}\ \bigg]
\]\\
\\
Putting this back into eq.~\eqref{A2} we find\\
\\
\begin{equation}\label{D902}
K_{m}^{th}=-\frac{1}{6\pi^{2}}\ \int_{0}^{+\infty}dk\ \ k^{4}\ \frac{n_{\beta}\big(\varepsilon_{m}(k)\big)}{\varepsilon_{m}(k)}+\frac{1}{6\pi^{2}}\ \int_{0}^{+\infty}dk\ \ k^{3}\ n_{\beta}(k)+K_{0}^{th}
\end{equation}\\
\\
where the middle term can be evaluated explicitly and is equal to $\pi^{2}T^{4}/90$. Now, $K_{0}^{th}$ is defined as the thermal contribution to\\
\\
\[
K_{0}=\frac{1}{2}\ \int_{K}\ \ln\ \big(\beta^{2}K^{2}\big)
\]\\
\\
This is simply the free energy density of an ideal gas of massless bosons, whose thermal contribution \cite{kapusta} reads\\
\\
\[
K_{0}^{th}=-\frac{\pi^{2}T^{4}}{90}
\]\\
\\
Hence the second and third terms in eq.~\eqref{D902} cancel and we are left with\\
\\
\begin{equation}\label{D930}
K_{m}^{th}=-\frac{1}{6\pi^{2}}\ \int_{0}^{+\infty}dk\ \ k^{4}\ \frac{n_{\beta}\big(\varepsilon_{m}(k)\big)}{\varepsilon_{m}(k)}
\end{equation}\\
\\
$H_{m}$ is defined as follows. Consider the integral $I^{\mu\nu}_{m}$,\\
\\
\begin{equation}
I^{\mu\nu}_{m}=\int_{K}\frac{K^{\mu}K^{\nu}}{K^{2}\,(K^{2}+m^{2})}
\end{equation}\\
\\
Due to the antisymmetry of the integrand with respect to the reflections $K^{\mu}\to -K^{\mu}$ for each value of $\mu$ separately, $I^{\mu\nu}_{m}$ is diagonal as a matrix. Moreover, due to $(d-1)$-dimensional rotational invariance, every diagonal entry except from $I^{00}_{m}$ is equal to each other. Hence we may set\\
\\
\begin{equation}\label{D17}
I^{\mu\nu}_{m}=\delta^{\mu}_{0}\delta^{\nu}_{0}\ \int_{K}\frac{(K^{0})^{2}}{K^{2}\,(K^{2}+m^{2})}+\frac{(\delta^{\mu\nu}-\delta^{\mu}_{0}\delta^{\nu}_{0})}{(d-1)}\ \int_{K} \frac{|\vec{K}|^{2}}{K^{2}\,(K^{2}+m^{2})}
\end{equation}\\
\\
By subtracting from the first term and adding to the second term the quantity\\
\\
\[
\frac{(\delta^{\mu\nu}-\delta^{\mu}_{0}\delta^{\nu}_{0})}{(d-1)}\ \int_{K} \frac{(K^{0})^{2}}{K^{2}\,(K^{2}+m^{2})}
\]\\
\\
\eqref{D17} can be put in the form\\
\\
\begin{equation}
I^{\mu\nu}_{m}=\frac{(d\,\delta^{\mu}_{0}\delta^{\nu}_{0}-\delta^{\mu\nu})}{(d-1)}\ \int_{K}\frac{(K^{0})^{2}}{K^{2}\,(K^{2}+m^{2})}+\frac{(\delta^{\mu\nu}-\delta^{\mu}_{0}\delta^{\nu}_{0})}{(d-1)}\ \int_{K} \frac{1}{K^{2}+m^{2}}
\end{equation}\\
\\
We see that the second term involves the integral $J_{m}$. As for the first term, we define $H_{m}$ as\\
\\
\begin{equation}
H_{m}=\int_{K}\frac{(K^{0})^{2}}{K^{2}\,(K^{2}+m^{2})}
\end{equation}\\
\\
so that\\
\\
\begin{equation}\label{D903}
I^{\mu\nu}_{m}=\frac{(d\,\delta^{\mu}_{0}\delta^{\nu}_{0}-\delta^{\mu\nu})}{(d-1)}\ H_{m}+\frac{(\delta^{\mu\nu}-\delta^{\mu}_{0}\delta^{\nu}_{0})}{(d-1)}\ J_{m}
\end{equation}\\
\\
By squaring eq.~\eqref{D903}, we obtain\\
\\
\begin{equation}\label{D904}
I^{\mu\nu}_{m}I_{m\,\mu\nu}=\frac{1}{d}\,J_{m}^{2}+\frac{d}{d-1}\,\bigg(H_{m}-\frac{1}{d}\ J_{m}\bigg)^{2}
\end{equation}
The reason for the split of terms in eq.~\eqref{D904} is the following. In the vacuum, due to Lorentz invariance, not only $I^{\mu\nu}_{m}$ is diagonal, but it is also proportional to the identity matrix:\
\\
\[
I^{v\,\mu\nu}_{m}=X_{m}\ \delta^{\mu\nu}
\]
\\
with\\
\\
\[
X_{m}=\frac{1}{d}\ \delta_{\mu\nu}\,I^{v\,\mu\nu}_{m}=\frac{1}{d}\ \int\frac{d^{d}K}{(2\pi)^{d}}\ \frac{K^{2}}{K^{2}(K^{2}+m^{2})}=\frac{1}{d}\ J_{m}^{v}
\]\\
\\
We thus have\\
\\
\begin{equation}
\Big(I^{\mu\nu}_{m}I_{m\,\mu\nu}\Big)\Big|_{T=0}=\frac{1}{d}\,J_{m}^{v\,2}
\end{equation}\\
\\
so that in the limit $T\to 0$ the second term in eq.~\eqref{D904} is actually equal to zero. In particular, this implies that\\
\\
\begin{equation}
H_{m}^{v}=\frac{1}{d}\ J_{m}^{v}
\end{equation}\\
\\
The thermal contribution to $H_{m}$ again can be evaluated with elementary methods. By definition, it equals\\
\\
\begin{equation}
H_{m}^{th}=2\ \int_{-\infty-i\varepsilon}^{+\infty-i\varepsilon}\frac{dK^{0}}{2\pi}\int \frac{d^{3}K}{(2\pi)^{3}}\ \frac{(K^{0})^{2}}{K^{2}\,(K^{2}+m^{2})}\ n_{\beta}(iK^{0})
\end{equation}\\
\\
To compute the $K^{0}$ integral, we can close the contour of integration the same way we did for $J_{m}^{th}$. This time the poles that fall into the contour are found at $K^{0}=-i\,|\vec{K}|$ and $K^{0}=-i\,\varepsilon_{m}(|\vec{K}|)$. The residues give\\
\\
\[
\int_{-\infty-i\varepsilon}^{+\infty-i\varepsilon}\frac{dK^{0}}{2\pi} \frac{(K^{0})^{2}}{K^{2}\,(K^{2}+m^{2})}\ n_{\beta}(iK^{0})=\frac{1}{2m^{2}}\ \Big(\varepsilon_{m}(|\vec{K}|)\,n_{\beta}\big(\varepsilon_{m}(|\vec{K}|)\big)-|\vec{K}|\,n_{\beta}\big(|\vec{K}|\big)\Big)
\]\\
\\
Hence we find that\\
\\
\begin{equation}
H_{m}^{th}=\frac{1}{2\pi^{2}m^{2}}\ \int_{0}^{+\infty}dk\ k^{2}\ \Big(\varepsilon_{m}(k)\,n_{\beta}\big(\varepsilon_{m}(k)\big)-k\,n_{\beta}\big(k\big)\Big)
\end{equation}
The contribution due to the last term in parentheses can be evaluated explicitly and is equal to $-\pi^{2}T^{4}/30m^{2}$. The contribution due to the first term may be expressed in terms of $K_{m}^{th}$ and $J_{m}^{th}$ by writing\\
\\
\[
k^{2}\varepsilon_{m}(k)=\frac{k^{2}(k^{2}+m^{2})}{\varepsilon_{m}(k)}=\frac{k^{4}+m^{2}\,k^{2}}{\varepsilon_{m}(k)}
\]\\
\\
Recalling expressions \eqref{D930} and \eqref{D90} for $K_{m}^{th}$ and $J_{m}^{th}$, we find\\
\\
\begin{equation}
H_{m}^{th}=-\frac{3}{m^{2}}\ K_{m}^{th}+ J_{m}^{th}-\frac{\pi^{2}T^{4}}{30m^{2}}
\end{equation}
\\
\\
\\
\\
\\
\addcontentsline{toc}{section}{E. Numerical tables}  \markboth{E. Numerical tables}{E. Numerical tables}
\section*{E. Numerical tables\index{Numerical tables}}

In this appendix we collect some of the numerical data that was used for the GEP analysis of sec. 3.3.2. In Table 2 we report the values of the two minima of the GEP and the associated value of the GEP as functions of the temperature. In the table, the second and third columns refer to the $m=0$ minimum, while the fourth and fifth columns refer to the $m=m_{0}$ minimum. Beyond the critical temperature $T_{c}\approx 0.35\ m_{0}$, the $m=m_{0}$ minimum disappears and the relative columns are left blank. In Table 3 we report the values of the optimal mass parameter $m(T)$ as a function of the temperature, together with the associated value of the GEP, the entropy density and enthalpy density. In the GEP approximation, the GEP $\F_{G}(m(T),T)$ is equal to the free energy density of the system; the entropy density $s(T)$ and enthalpy density $h(T)$ can be evaluated accordingly as $s(T)=-d\F_{G}(m(T),T)/dT$, $h(T)=Ts(T)$. The entropy density has actually been computed as a finite-difference derivative. The values in both tables are computed for $\alpha_{s}=1$.\\
\newpage
\begin{table}[H]
\centering
\vspace{6mm}
\caption{Minima and values of the thermal GEP as functions of the temperature for $\alpha_{s}=1$}
\vspace{4mm}
\resizebox{0.95\textwidth}{!}{
\begin{tabular}{ccccccc}
\hline\\
$T\,/\,m_{0}$ &  & $m_{\text{min}}\,/\,m_{0}$ & $\mathcal{F}_{G}(m_{\text{min}},\,T)\,/\,m_{0}^{4}$ & & $m_{\text{min}}\,/\,m_{0}$ & $\mathcal{F}_{G}(m_{\text{min}},\,T)\,/\,m_{0}^{4}$\\ \\ \hline
             &  &                &                 &  &                &                 \\ \hline
0.01 &  & 0.80744600E-02 & -0.18191944E-07 &  & 0.10000000E+01 & -0.12665148E-01 \\ \hline
0.02 &  & 0.17159180E-01 & -0.28708825E-06 &  & 0.10000000E+01 & -0.12665148E-01 \\ \hline
0.03 &  & 0.26811790E-01 & -0.14385173E-05 &  & 0.10000000E+01 & -0.12665148E-01 \\ \hline
0.04 &  & 0.36911320E-01 & -0.45079802E-05 &  & 0.10000000E+01 & -0.12665148E-01 \\ \hline
0.05 &  & 0.47397370E-01 & -0.10924048E-04 &  & 0.10000000E+01 & -0.12665148E-01 \\ \hline
0.06 &  & 0.58233400E-01 & -0.22498964E-04 &  & 0.99999997E+00 & -0.12665150E-01 \\ \hline
0.07 &  & 0.69395230E-01 & -0.41419720E-04 &  & 0.99999987E+00 & -0.12665155E-01 \\ \hline
0.08 &  & 0.80865910E-01 & -0.70239283E-04 &  & 0.99999937E+00 & -0.12665174E-01 \\ \hline
0.09 &  & 0.92633190E-01 & -0.11186777E-03 &  & 0.99999740E+00 & -0.12665238E-01 \\ \hline
0.10 &  & 0.10468796E+00 & -0.16956347E-03 &  & 0.99999137E+00 & -0.12665430E-01 \\ \hline
0.11 &  & 0.11702331E+00 & -0.24692350E-03 &  & 0.99997610E+00 & -0.12665937E-01 \\ \hline
0.12 &  & 0.12963385E+00 & -0.34787412E-03 &  & 0.99994288E+00 & -0.12667112E-01 \\ \hline
0.13 &  & 0.14251516E+00 & -0.47666046E-03 &  & 0.99987874E+00 & -0.12669556E-01 \\ \hline
0.14 &  & 0.15566342E+00 & -0.63783569E-03 &  & 0.99976604E+00 & -0.12674199E-01 \\ \hline
0.15 &  & 0.16907496E+00 & -0.83624940E-03 &  & 0.99958235E+00 & -0.12682379E-01 \\ \hline
0.16 &  & 0.18274584E+00 & -0.10770352E-02 &  & 0.99930056E+00 & -0.12695917E-01 \\ \hline
0.17 &  & 0.19667142E+00 & -0.13655971E-02 &  & 0.99888909E+00 & -0.12717178E-01 \\ \hline
0.18 &  & 0.21084578E+00 & -0.17075952E-02 &  & 0.99831207E+00 & -0.12749135E-01 \\ \hline
0.19 &  & 0.22526110E+00 & -0.21089298E-02 &  & 0.99752947E+00 & -0.12795415E-01 \\ \hline
0.20 &  & 0.23990684E+00 & -0.25757240E-02 &  & 0.99649696E+00 & -0.12860344E-01 \\ \hline
0.21 &  & 0.25476872E+00 & -0.31143054E-02 &  & 0.99516565E+00 & -0.12948990E-01 \\ \hline
0.22 &  & 0.26982756E+00 & -0.37311853E-02 &  & 0.99348140E+00 & -0.13067202E-01 \\ \hline
0.23 &  & 0.28505781E+00 & -0.44330374E-02 &  & 0.99138373E+00 & -0.13221650E-01 \\ \hline
0.24 &  & 0.30042586E+00 & -0.52266745E-02 &  & 0.98880428E+00 & -0.13419874E-01 \\ \hline
0.25 &  & 0.31588841E+00 & -0.61190244E-02 &  & 0.98566440E+00 & -0.13670333E-01 \\ \hline
0.26 &  & 0.33139084E+00 & -0.71171056E-02 &  & 0.98187178E+00 & -0.13982470E-01 \\ \hline
0.27 &  & 0.34686633E+00 & -0.82280045E-02 &  & 0.97731535E+00 & -0.14366787E-01 \\ \hline
0.28 &  & 0.36223621E+00 & -0.94588547E-02 &  & 0.97185767E+00 & -0.14834955E-01 \\ \hline
0.29 &  & 0.37741231E+00 & -0.10816822E-01 &  & 0.96532290E+00 & -0.15399945E-01 \\ \hline
0.30 &  & 0.39230210E+00 & -0.12309096E-01 &  & 0.95747690E+00 & -0.16076232E-01 \\ \hline
0.31 &  & 0.40681609E+00 & -0.13942891E-01 &  & 0.94799190E+00 & -0.16880080E-01 \\ \hline
0.32 &  & 0.42087711E+00 & -0.15725457E-01 &  & 0.93637894E+00 & -0.17830013E-01 \\ \hline
0.33 &  & 0.43442826E+00 & -0.17664090E-01 &  & 0.92183900E+00 & -0.18947619E-01 \\ \hline
0.34 &  & 0.44743812E+00 & -0.19766163E-01 &  & 0.90287362E+00 & -0.20259172E-01 \\ \hline
0.35 &  & 0.45990123E+00 & -0.22039135E-01 &  & 0.87586526E+00 & -0.21799702E-01 \\ \hline
0.36 &  & 0.47183459E+00 & -0.24490574E-01 &  & 0.82176090E+00 & -0.23631001E-01 \\ \hline
0.37 &  & 0.48327100E+00 & -0.27128163E-01 &  &                &                 \\ \hline
0.38 &  & 0.49425461E+00 & -0.29959707E-01 &  &                &                 \\ \hline
0.39 &  & 0.50483079E+00 & -0.32993135E-01 &  &                &                 \\ \hline
0.40 &  & 0.51504643E+00 & -0.36236495E-01 &  &                &                 \\ \hline
0.41 &  & 0.52494506E+00 & -0.39697954E-01 &  &                &                 \\ \hline
0.42 &  & 0.53456601E+00 & -0.43385789E-01 &  &                &                 \\ \hline
0.43 &  & 0.54394396E+00 & -0.47308389E-01 &  &                &                 \\ \hline
0.44 &  & 0.55310900E+00 & -0.51474243E-01 &  &                &                 \\ \hline
0.45 &  & 0.56208704E+00 & -0.55891946E-01 &  &                &                 \\ \hline
\end{tabular}
}
\end{table}
\begin{table}[H]
\centering
\vspace{6mm}
\resizebox{0.95\textwidth}{!}{
\begin{tabular}{ccccccc}
\hline\\
$T\,/\,m_{0}$ &  & $m_{\text{min}}\,/\,m_{0}$ & $\mathcal{F}_{G}(m_{\text{min}},\,T)\,/\,m_{0}^{4}$ & & $m_{\text{min}}\,/\,m_{0}$ & $\mathcal{F}_{G}(m_{\text{min}},\,T)\,/\,m_{0}^{4}$\\ \\ \hline
             &  &                &                 &  &                &                 \\ \hline
0.46 &  & 0.57090016E+00 & -0.60570187E-01 &  &                &                 \\ \hline
0.47 &  & 0.57956727E+00 & -0.65517751E-01 &  &                &                 \\ \hline
0.48 &  & 0.58810440E+00 & -0.70743513E-01 &  &                &                 \\ \hline
0.49 &  & 0.59652529E+00 & -0.76256440E-01 &  &                &                 \\ \hline
0.50 &  & 0.60484164E+00 & -0.82065585E-01 &  &                &                 \\ \hline
0.51 &  & 0.61306360E+00 & -0.88180088E-01 &  &                &                 \\ \hline
0.52 &  & 0.62119981E+00 & -0.94609169E-01 &  &                &                 \\ \hline
0.53 &  & 0.62925771E+00 & -0.10136213E+00 &  &                &                 \\ \hline
0.54 &  & 0.63724384E+00 & -0.10844836E+00 &  &                &                 \\ \hline
0.55 &  & 0.64516388E+00 & -0.11587732E+00 &  &                &                 \\ \hline
0.56 &  & 0.65302278E+00 & -0.12365856E+00 &  &                &                 \\ \hline
0.57 &  & 0.66082488E+00 & -0.13180167E+00 &  &                &                 \\ \hline
0.58 &  & 0.66857396E+00 & -0.14031637E+00 &  &                &                 \\ \hline
0.59 &  & 0.67627357E+00 & -0.14921241E+00 &  &                &                 \\ \hline
0.60 &  & 0.68392666E+00 & -0.15849962E+00 &  &                &                 \\ \hline
0.61 &  & 0.69153586E+00 & -0.16818793E+00 &  &                &                 \\ \hline
0.62 &  & 0.69910366E+00 & -0.17828730E+00 &  &                &                 \\ \hline
0.63 &  & 0.70663230E+00 & -0.18880777E+00 &  &                &                 \\ \hline
0.64 &  & 0.71412364E+00 & -0.19975948E+00 &  &                &                 \\ \hline
0.65 &  & 0.72157954E+00 & -0.21115259E+00 &  &                &                 \\ \hline
0.66 &  & 0.72900162E+00 & -0.22299735E+00 &  &                &                 \\ \hline
0.67 &  & 0.73639128E+00 & -0.23530408E+00 &  &                &                 \\ \hline
0.68 &  & 0.74374990E+00 & -0.24808315E+00 &  &                &                 \\ \hline
0.69 &  & 0.75107880E+00 & -0.26134500E+00 &  &                &                 \\ \hline
0.70 &  & 0.75837906E+00 & -0.27510012E+00 &  &                &                 \\ \hline
0.71 &  & 0.76565163E+00 & -0.28935909E+00 &  &                &                 \\ \hline
0.72 &  & 0.77289773E+00 & -0.30413252E+00 &  &                &                 \\ \hline
0.73 &  & 0.78011804E+00 & -0.31943109E+00 &  &                &                 \\ \hline
0.74 &  & 0.78731356E+00 & -0.33526554E+00 &  &                &                 \\ \hline
0.75 &  & 0.79448490E+00 & -0.35164668E+00 &  &                &                 \\ \hline
0.76 &  & 0.80163297E+00 & -0.36858534E+00 &  &                &                 \\ \hline
0.77 &  & 0.80875833E+00 & -0.38609245E+00 &  &                &                 \\ \hline
0.78 &  & 0.81586166E+00 & -0.40417897E+00 &  &                &                 \\ \hline
0.79 &  & 0.82294354E+00 & -0.42285592E+00 &  &                &                 \\ \hline
0.80 &  & 0.83000460E+00 & -0.44213436E+00 &  &                &                 \\ \hline
0.81 &  & 0.83704532E+00 & -0.46202543E+00 &  &                &                 \\ \hline
0.82 &  & 0.84406629E+00 & -0.48254031E+00 &  &                &                 \\ \hline
0.83 &  & 0.85106788E+00 & -0.50369022E+00 &  &                &                 \\ \hline
0.84 &  & 0.85805061E+00 & -0.52548645E+00 &  &                &                 \\ \hline
0.85 &  & 0.86501485E+00 & -0.54794032E+00 &  &                &                 \\ \hline
0.86 &  & 0.87196107E+00 & -0.57106322E+00 &  &                &                 \\ \hline
0.87 &  & 0.87888967E+00 & -0.59486658E+00 &  &                &                 \\ \hline
0.88 &  & 0.88580100E+00 & -0.61936187E+00 &  &                &                 \\ \hline
0.89 &  & 0.89269533E+00 & -0.64456061E+00 &  &                &                 \\ \hline
0.90 &  & 0.89957320E+00 & -0.67047438E+00 &  &                &                 \\ \hline
\end{tabular}
}
\end{table}
\newpage
\begin{table}[H]
\centering
\vspace{6mm}
\resizebox{0.95\textwidth}{!}{
\begin{tabular}{ccccccc}
\hline\\
$T\,/\,m_{0}$ &  & $m_{\text{min}}\,/\,m_{0}$ & $\mathcal{F}_{G}(m_{\text{min}},\,T)\,/\,m_{0}^{4}$ & & $m_{\text{min}}\,/\,m_{0}$ & $\mathcal{F}_{G}(m_{\text{min}},\,T)\,/\,m_{0}^{4}$\\ \\ \hline
             &  &                &                 &  &                &                 \\ \hline
0.91 &  & 0.90643482E+00 & -0.69711480E+00 &  &                &                 \\ \hline
0.92 &  & 0.91328043E+00 & -0.72449352E+00 &  &                &                 \\ \hline
0.93 &  & 0.92011044E+00 & -0.75262226E+00 &  &                &                 \\ \hline
0.94 &  & 0.92692514E+00 & -0.78151276E+00 &  &                &                 \\ \hline
0.95 &  & 0.93372472E+00 & -0.81117682E+00 &  &                &                 \\ \hline
0.96 &  & 0.94050950E+00 & -0.84162629E+00 &  &                &                 \\ \hline
0.97 &  & 0.94727972E+00 & -0.87287303E+00 &  &                &                 \\ \hline
0.98 &  & 0.95403561E+00 & -0.90492898E+00 &  &                &                 \\ \hline
0.99 &  & 0.96077750E+00 & -0.93780609E+00 &  &                &                 \\ \hline
1.00 &  & 0.96750549E+00 & -0.97151638E+00 &  &                &                 \\ \hline
\end{tabular}
}
\end{table}
\begin{table}[H]
\centering
\caption{Optimal mass parameter, value of the GEP, entropy and enthalpy density in the GEP approximation as functions of the temperature for $\alpha_{s}=1$}
\vspace{4mm}
\resizebox{0.95\textwidth}{!}{
\begin{tabular}{cccccc}
\hline\\
$T\,/\,m_{0}$ &  & $m(T)\,/\,m_{0}$ & $\mathcal{F}_{G}(m(T),\,T)\,/\,m_{0}^{4}$ & $s(T)\,/\,m_{0}^{3}$ & $h(T)\,/\,m_{0}^{4}$\\ \\ \hline
              &  &                  &                                           &                      &                      \\ \hline
0.01  &  & 0.10000000E+01   & -0.12665148E-01                           & 0.00000000E-07       & 0.00000000E-08       \\ \hline
0.02  &  & 0.10000000E+01   & -0.12665148E-01                           & 0.00000000E-07       & 0.00000000E-08       \\ \hline
0.03  &  & 0.10000000E+01   & -0.12665148E-01                           & 0.00000000E-07       & 0.00000000E-08       \\ \hline
0.04  &  & 0.10000000E+01   & -0.12665148E-01                           & 0.32000000E-07       & 0.12800000E-08       \\ \hline
0.05  &  & 0.10000000E+01   & -0.12665148E-01                           & 0.16000000E-06       & 0.80000000E-08       \\ \hline
0.06  &  & 0.99999997E+00   & -0.12665150E-01                           & 0.52800000E-06       & 0.31680000E-07       \\ \hline
0.07  &  & 0.99999987E+00   & -0.12665155E-01                           & 0.18880000E-05       & 0.13216000E-06       \\ \hline
0.08  &  & 0.99999937E+00   & -0.12665174E-01                           & 0.64000000E-05       & 0.51200000E-06       \\ \hline
0.09  &  & 0.99999740E+00   & -0.12665238E-01                           & 0.19232000E-04       & 0.17308800E-05       \\ \hline
0.10  &  & 0.99999137E+00   & -0.12665430E-01                           & 0.50640000E-04       & 0.50640000E-05       \\ \hline
0.11  &  & 0.99997610E+00   & -0.12665937E-01                           & 0.11747200E-03       & 0.12921920E-04       \\ \hline
0.12  &  & 0.99994288E+00   & -0.12667112E-01                           & 0.24443200E-03       & 0.29331840E-04       \\ \hline
0.13  &  & 0.99987874E+00   & -0.12669556E-01                           & 0.46432000E-03       & 0.60361600E-04       \\ \hline
0.14  &  & 0.99976604E+00   & -0.12674199E-01                           & 0.81803200E-03       & 0.11452448E-03       \\ \hline
0.15  &  & 0.99958235E+00   & -0.12682379E-01                           & 0.13537280E-02       & 0.20305920E-03       \\ \hline
0.16  &  & 0.99930056E+00   & -0.12695917E-01                           & 0.21261280E-02       & 0.34018048E-03       \\ \hline
0.17  &  & 0.99888909E+00   & -0.12717178E-01                           & 0.31957120E-02       & 0.54327104E-03       \\ \hline
0.18  &  & 0.99831207E+00   & -0.12749135E-01                           & 0.46280000E-02       & 0.83304000E-03       \\ \hline
0.19  &  & 0.99752947E+00   & -0.12795415E-01                           & 0.64929120E-02       & 0.12336533E-02       \\ \hline
0.20  &  & 0.99649696E+00   & -0.12860344E-01                           & 0.88646080E-02       & 0.17729216E-02       \\ \hline
\end{tabular}
}
\end{table}
\begin{table}[H]
\centering
\vspace{6mm}
\resizebox{0.95\textwidth}{!}{
\begin{tabular}{cccccc}
\hline\\
$T\,/\,m_{0}$ &  & $m(T)\,/\,m_{0}$ & $\mathcal{F}_{G}(m(T),\,T)\,/\,m_{0}^{4}$ & $s(T)\,/\,m_{0}^{3}$ & $h(T)\,/\,m_{0}^{4}$\\ \\ \hline
              &  &                  &                                           &                      &                      \\ \hline
0.21  &  & 0.99516565E+00   & -0.12948990E-01                           & 0.11821136E-01       & 0.24824386E-02       \\ \hline
0.22  &  & 0.99348140E+00   & -0.13067202E-01                           & 0.15444832E-01       & 0.33978630E-02       \\ \hline
0.23  &  & 0.99138373E+00   & -0.13221650E-01                           & 0.19822416E-01       & 0.45591557E-02       \\ \hline
0.24  &  & 0.98880428E+00   & -0.13419874E-01                           & 0.25045920E-01       & 0.60110208E-02       \\ \hline
0.25  &  & 0.98566440E+00   & -0.13670333E-01                           & 0.31213632E-01       & 0.78034080E-02       \\ \hline
0.26  &  & 0.98187178E+00   & -0.13982470E-01                           & 0.38431760E-01       & 0.99922576E-02       \\ \hline
0.27  &  & 0.97731535E+00   & -0.14366787E-01                           & 0.46816768E-01       & 0.12640527E-01       \\ \hline
0.28  &  & 0.97185767E+00   & -0.14834955E-01                           & 0.56499040E-01       & 0.15819731E-01       \\ \hline
0.29  &  & 0.96532290E+00   & -0.15399945E-01                           & 0.67628672E-01       & 0.19612315E-01       \\ \hline
0.30  &  & 0.95747690E+00   & -0.16076232E-01                           & 0.80384800E-01       & 0.24115440E-01       \\ \hline
0.31  &  & 0.94799190E+00   & -0.16880080E-01                           & 0.94993280E-01       & 0.29447917E-01       \\ \hline
0.32  &  & 0.93637894E+00   & -0.17830013E-01                           & 0.11176064E+00       & 0.35763405E-01       \\ \hline
0.33  &  & 0.92183900E+00   & -0.18947619E-01                           & 0.13115536E+00       & 0.43281269E-01       \\ \hline
0.34  &  & 0.90287362E+00   & -0.20259173E-01                           & 0.15405296E+00       & 0.52378006E-01       \\ \hline
0.35  &  & 0.45990123E+00   & -0.22039136E-01                           & 0.24514384E+00       & 0.85800344E-01       \\ \hline
0.36  &  & 0.47183459E+00   & -0.24490574E-01                           & 0.26375888E+00       & 0.94953197E-01       \\ \hline
0.37  &  & 0.48327100E+00   & -0.27128163E-01                           & 0.28315440E+00       & 0.10476713E+00       \\ \hline
0.38  &  & 0.49425461E+00   & -0.29959707E-01                           & 0.30334272E+00       & 0.11527023E+00       \\ \hline
0.39  &  & 0.50483079E+00   & -0.32993134E-01                           & 0.32433616E+00       & 0.12649110E+00       \\ \hline
0.40  &  & 0.51504643E+00   & -0.36236496E-01                           & 0.34614576E+00       & 0.13845830E+00       \\ \hline
0.41  &  & 0.52494506E+00   & -0.39697954E-01                           & 0.36878352E+00       & 0.15120124E+00       \\ \hline
0.42  &  & 0.53456601E+00   & -0.43385789E-01                           & 0.39226000E+00       & 0.16474920E+00       \\ \hline
0.43  &  & 0.54394396E+00   & -0.47308389E-01                           & 0.41658544E+00       & 0.17913174E+00       \\ \hline
0.44  &  & 0.55310900E+00   & -0.51474243E-01                           & 0.44177024E+00       & 0.19437891E+00       \\ \hline
0.45  &  & 0.56208704E+00   & -0.55891946E-01                           & 0.46782416E+00       & 0.21052087E+00       \\ \hline
0.46  &  & 0.57090016E+00   & -0.60570187E-01                           & 0.49475632E+00       & 0.22758791E+00       \\ \hline
0.47  &  & 0.57956727E+00   & -0.65517750E-01                           & 0.52257632E+00       & 0.24561087E+00       \\ \hline
0.48  &  & 0.58810440E+00   & -0.70743514E-01                           & 0.55129264E+00       & 0.26462047E+00       \\ \hline
0.49  &  & 0.59652529E+00   & -0.76256440E-01                           & 0.58091456E+00       & 0.28464813E+00       \\ \hline
0.50  &  & 0.60484164E+00   & -0.82065586E-01                           & 0.61145024E+00       & 0.30572512E+00       \\ \hline
0.51  &  & 0.61306360E+00   & -0.88180088E-01                           & 0.64290800E+00       & 0.32788308E+00       \\ \hline
0.52  &  & 0.62119981E+00   & -0.94609168E-01                           & 0.67529648E+00       & 0.35115417E+00       \\ \hline
0.53  &  & 0.62925771E+00   & -0.10136213E+00                           & 0.70862304E+00       & 0.37557021E+00       \\ \hline
0.54  &  & 0.63724384E+00   & -0.10844836E+00                           & 0.74289616E+00       & 0.40116393E+00       \\ \hline
0.55  &  & 0.64516388E+00   & -0.11587732E+00                           & 0.77812304E+00       & 0.42796767E+00       \\ \hline
0.56  &  & 0.65302278E+00   & -0.12365856E+00                           & 0.81431184E+00       & 0.45601463E+00       \\ \hline
0.57  &  & 0.66082488E+00   & -0.13180167E+00                           & 0.85146960E+00       & 0.48533767E+00       \\ \hline
0.58  &  & 0.66857396E+00   & -0.14031637E+00                           & 0.88960384E+00       & 0.51597023E+00       \\ \hline
0.59  &  & 0.67627357E+00   & -0.14921241E+00                           & 0.92872160E+00       & 0.54794574E+00       \\ \hline
0.60  &  & 0.68392666E+00   & -0.15849962E+00                           & 0.96882960E+00       & 0.58129776E+00       \\ \hline
0.61  &  & 0.69153586E+00   & -0.16818792E+00                           & 0.10099376E+01       & 0.61606194E+00       \\ \hline
0.62  &  & 0.69910366E+00   & -0.17828730E+00                           & 0.10520480E+01       & 0.65226976E+00       \\ \hline
0.63  &  & 0.70663230E+00   & -0.18880778E+00                           & 0.10951696E+01       & 0.68995685E+00       \\ \hline
0.64  &  & 0.71412364E+00   & -0.19975947E+00                           & 0.11393120E+01       & 0.72915968E+00       \\ \hline
0.65  &  & 0.72157954E+00   & -0.21115259E+00                           & 0.11844752E+01       & 0.76990888E+00       \\ \hline
\end{tabular}
}
\end{table}
\begin{table}[H]
\centering
\vspace{6mm}
\resizebox{0.95\textwidth}{!}{
\begin{tabular}{cccccc}
\hline\\
$T\,/\,m_{0}$ &  & $m(T)\,/\,m_{0}$ & $\mathcal{F}_{G}(m(T),\,T)\,/\,m_{0}^{4}$ & $s(T)\,/\,m_{0}^{3}$ & $h(T)\,/\,m_{0}^{4}$\\ \\ \hline
              &  &                  &                                           &                      &                      \\ \hline
0.66  &  & 0.72900162E+00   & -0.22299734E+00                           & 0.12306736E+01       & 0.81224458E+00       \\ \hline
0.67  &  & 0.73639128E+00   & -0.23530408E+00                           & 0.12779072E+01       & 0.85619782E+00       \\ \hline
0.68  &  & 0.74374990E+00   & -0.24808315E+00                           & 0.13261840E+01       & 0.90180512E+00       \\ \hline
0.69  &  & 0.75107880E+00   & -0.26134499E+00                           & 0.13755136E+01       & 0.94910438E+00       \\ \hline
0.70  &  & 0.75837906E+00   & -0.27510013E+00                           & 0.14258960E+01       & 0.99812720E+00       \\ \hline
0.71  &  & 0.76565163E+00   & -0.28935909E+00                           & 0.14773424E+01       & 0.10489131E+01       \\ \hline
0.72  &  & 0.77289773E+00   & -0.30413251E+00                           & 0.15298576E+01       & 0.11014975E+01       \\ \hline
0.73  &  & 0.78011804E+00   & -0.31943109E+00                           & 0.15834464E+01       & 0.11559159E+01       \\ \hline
0.74  &  & 0.78731356E+00   & -0.33526555E+00                           & 0.16381120E+01       & 0.12122029E+01       \\ \hline
0.75  &  & 0.79448490E+00   & -0.35164667E+00                           & 0.16938672E+01       & 0.12704004E+01       \\ \hline
0.76  &  & 0.80163297E+00   & -0.36858534E+00                           & 0.17507104E+01       & 0.13305399E+01       \\ \hline
0.77  &  & 0.80875833E+00   & -0.38609245E+00                           & 0.18086528E+01       & 0.13926627E+01       \\ \hline
0.78  &  & 0.81586166E+00   & -0.40417898E+00                           & 0.18676944E+01       & 0.14568016E+01       \\ \hline
0.79  &  & 0.82294354E+00   & -0.42285592E+00                           & 0.19278448E+01       & 0.15229974E+01       \\ \hline
0.80  &  & 0.83000460E+00   & -0.44213437E+00                           & 0.19891072E+01       & 0.15912858E+01       \\ \hline
0.81  &  & 0.83704532E+00   & -0.46202544E+00                           & 0.20514864E+01       & 0.16617040E+01       \\ \hline
0.82  &  & 0.84406629E+00   & -0.48254030E+00                           & 0.21149920E+01       & 0.17342934E+01       \\ \hline
0.83  &  & 0.85106788E+00   & -0.50369022E+00                           & 0.21796224E+01       & 0.18090866E+01       \\ \hline
0.84  &  & 0.85805061E+00   & -0.52548645E+00                           & 0.22453872E+01       & 0.18861252E+01       \\ \hline
0.85  &  & 0.86501485E+00   & -0.54794032E+00                           & 0.23122912E+01       & 0.19654475E+01       \\ \hline
0.86  &  & 0.87196107E+00   & -0.57106323E+00                           & 0.23803344E+01       & 0.20470876E+01       \\ \hline
0.87  &  & 0.87888967E+00   & -0.59486658E+00                           & 0.24495296E+01       & 0.21310908E+01       \\ \hline
0.88  &  & 0.88580100E+00   & -0.61936187E+00                           & 0.25198736E+01       & 0.22174888E+01       \\ \hline
0.89  &  & 0.89269533E+00   & -0.64456061E+00                           & 0.25913776E+01       & 0.23063261E+01       \\ \hline
0.90  &  & 0.89957320E+00   & -0.67047438E+00                           & 0.26640416E+01       & 0.23976374E+01       \\ \hline
0.91  &  & 0.90643482E+00   & -0.69711480E+00                           & 0.27378720E+01       & 0.24914635E+01       \\ \hline
0.92  &  & 0.91328043E+00   & -0.72449352E+00                           & 0.28128736E+01       & 0.25878437E+01       \\ \hline
0.93  &  & 0.92011044E+00   & -0.75262226E+00                           & 0.28890496E+01       & 0.26868161E+01       \\ \hline
0.94  &  & 0.92692514E+00   & -0.78151275E+00                           & 0.29664064E+01       & 0.27884220E+01       \\ \hline
0.95  &  & 0.93372472E+00   & -0.81117682E+00                           & 0.30449472E+01       & 0.28926998E+01       \\ \hline
0.96  &  & 0.94050950E+00   & -0.84162629E+00                           & 0.31246736E+01       & 0.29996867E+01       \\ \hline
0.97  &  & 0.94727972E+00   & -0.87287302E+00                           & 0.32055952E+01       & 0.31094273E+01       \\ \hline
0.98  &  & 0.95403561E+00   & -0.90492898E+00                           & 0.32877120E+01       & 0.32219578E+01       \\ \hline
0.99  &  & 0.96077750E+00   & -0.93780610E+00                           & 0.33710288E+01       & 0.33373185E+01       \\ \hline
1.00  &  & 0.96750549E+00   & -0.97151638E+00                           & 0.34555488E+01       & 0.34555488E+01       \\ \hline
\end{tabular}
}
\end{table}
\clearpage{}
\thispagestyle{empty}
\
\clearpage

\end{document}